\documentclass[aps,amsmath,amssymb,prx,10pt,showpacs,letterpaper,twocolumn]{revtex4-1}
\usepackage{graphicx,txfonts,color,tcolorbox}
\usepackage{bm}
\usepackage{mathrsfs}
\usepackage{comment}
\usepackage{mathtools}
\usepackage{hyperref}
\usepackage{epstopdf}
\usepackage{letltxmacro}
\usepackage{appendix}

\newcommand{\al}{\alpha}
\newcommand{\be}{\beta}
\newcommand{\g}{\gamma}
\newcommand{\de}{\delta}
\newcommand{\e}{\epsilon}
\newcommand{\z}{\zeta}

\newcommand{\thi}{\theta}
\newcommand{\io}{\iota}

\newcommand{\la}{\lambda}

\newcommand{\p}{\pi}
\newcommand{\ro}{\rho}
\newcommand{\s}{\sigma}
\newcommand{\y}{\upsilon}

\newcommand{\w}{\omega}
\newcommand{\W}{\Omega}
\newcommand{\De}{\Delta}
\newcommand{\G}{\Gamma}
\renewcommand{\S}{\Sigma}
\newcommand{\D}{\Delta}

\renewcommand{\y}{\psi}

\newcommand{\pd}{\partial}

\newcommand{\round}[1]{\left( #1 \right)}
\renewcommand{\square}[1]{\left[ #1 \right]}

\newcommand{\abs}[1]{\left| #1 \right|}

\newcommand{\beq}{\begin{equation}}
\newcommand{\eeq}{\end{equation}}
\newcommand{\Beq}{\begin{eqnarray}}
\newcommand{\Eeq}{\end{eqnarray}}
\newcommand{\bml}{\begin{multline}}

\newcommand{\bea}{\begin{align}}
\newcommand{\ena}{\end{align}}
\newcommand{\bsp}{\begin{split}}
\newcommand{\esp}{\end{split}}

\newcommand{\nn}{\nonumber}

\newcommand{\br}{{\boldsymbol r}}

\newcommand{\bS}{{\boldsymbol{S}}}

\newcommand{\bp}{{\boldsymbol p}}

\newcommand{\bi}{{\boldsymbol i}}
\newcommand{\bj}{{\boldsymbol j}}

\newcommand{\bk}{{\boldsymbol k}}

\newcommand{\bq}{{\boldsymbol q}}

\DeclareMathOperator{\diag}{Diag}

\newcommand{\sA}{\mathscr{A}}
\newcommand{\cH}{\mathcal{H}}
\newcommand{\cJ}{\mathcal{J}}

\newcommand{\cN}{\mathcal{N}}
\newcommand{\cM}{\mathcal{M}}

\newcommand{\bx}{\boldsymbol{x}}

\newcommand{\bn}{\boldsymbol{n}}

\newcommand{\ve}{\varepsilon}
\newcommand{\bs}{{\boldsymbol{s}}}
\newcommand{\bta}{{\boldsymbol{\tau}}}

\newcommand{\hu}{\hat{u}}

\newcommand{\hT}{T_K}

\newcommand{\bc}{\bar{c}}

\newcommand{\bnab}{\boldsymbol{\nabla}}

\newcommand{\ba}{\boldsymbol{a}}
\newcommand{\hbq}{\hat{\boldsymbol{q}}}
\newcommand{\hbk}{\hat{\boldsymbol{k}}}
\newcommand{\hbz}{\hat{\boldsymbol{z}}}

\makeatletter
\renewcommand*\env@matrix[1][\arraystretch]{%
  \edef\arraystretch{#1}%
  \hskip -\arraycolsep
  \let\@ifnextchar\new@ifnextchar
  \array{*\c@MaxMatrixCols c}}
\makeatother

\newcommand{\updownarrows}{\uparrow\mathrel{\mspace{-1mu}}\downarrow}
\newcommand{\downuparrows}{\downarrow\mathrel{\mspace{-1mu}}\uparrow}

\DeclareMathAlphabet{\mathpzc}{OT1}{pzc}{m}{it}
\DeclareMathOperator{\sinc}{sinc}

\begin{document}
\title{Probing quantum spin liquids in equilibrium using the inverse spin Hall effect}
\author{Joshua Aftergood}
\author{So Takei}
\affiliation{Department of Physics, Queens College of the City University of New York, Queens, NY 11367, USA}
\affiliation{Physics Doctoral Program, The Graduate Center of the City University of New York, New York, NY 10016, USA}
\date{\today}
%\pacs{show pacs here}

\begin{abstract}
We propose an experimental method utilizing a strongly spin-orbit coupled metal to quantum magnet bilayer that will probe quantum magnets lacking long range magnetic order, e.g., quantum spin liquids, via examination of the voltage noise spectrum in the metal layer. The bilayer is held in thermal and chemical equilibrium, and spin fluctuations arising across the single interface are converted into voltage fluctuations in the metal as a result of the inverse spin Hall effect.
%In thermal equilibrium, changes to the voltage noise in the metal are measurable as changes to the resistance via the fluctuation dissipation theorem.
We elucidate the theoretical workings of the proposed bilayer system, and provide precise predictions for the frequency characteristics of the enhancement to the ac electrical resistance measured in the metal layer for three candidate quantum spin liquid models. Application to the Heisenberg spin-$1/2$ kagom{\'e} lattice model should allow for the extraction of any spinon gap present. A quantum spin liquid consisting of fermionic spinons coupled to a $U(1)$ gauge field should cause subdominant $\W^{4/3}$ scaling of the resistance of the coupled metal. Finally, if the magnet is well-captured by the Kitaev model in the gapless spin liquid phase, then the proposed bilayer can extract the two-flux gap which arises in spite of the gapless spectrum of the fermions. We therefore show that spectral analysis of the ac resistance in the metal in a single interface, equilibrium bilayer can test the relevance of a quantum spin liquid model to a given candidate material.
%This proposed bilayer is therefore a diagnostic tool for discriminating the relevance of a spin liquid model to a particular candidate material due to these unique temperature scalings.
%We find that our setup should be able to extract the spinon gap in a material well characterized by the gapped spin-$1/2$ kagom{\'e} model, that the enhancement to the dc resistance measured across the metal layer should have $T^3$ temperature scaling when coupled to a Kitaev compound, and that the gauge fluctuation renormalization to the dc resistance measured across the metal layer when affixed to a compound hosting fermionic spinons coupled to a $U(1)$ gauge field should evince $T^{4/3}$ temperature scaling.
\end{abstract}
\maketitle

%\tableofcontents

%%%%%%%%%%%%%%%%%%%%%%%%%%%%%%%%%%%%%%%%%%%%%%%%%%%%%%%%%%%%%%
\section{Introduction}
Quantum spin liquids (QSLs) refer to intriguing states of quantum spin systems in which strong quantum fluctuations prevent spins from ordering down to zero temperature and the prototypical wavefunction exhibits extensive many-body entanglement~\cite{balentsNAT10,savaryRPP17}. They are endowed with fascinating physical properties like non-local excitations~\cite{kitaevPRL06,levinPRL06} and non-trivial topology~\cite{senthilPRB00,senthilPRL01}, and substantial pioneering work has been accomplished in the pursuit of a physical instantiation of this long-sought-after phase of matter. The most promising candidates to date include the mineral herbertsmithite~\cite{normanRMP16}, certain organic salt compounds~\cite{powellRPP11}, and the so-called Kitaev materials~\cite{takagiNRP19}, and their experimental studies have ranged from nuclear magnetic resonance~\cite{shimizuPRL03,olariuPRL08,fuSCI15,kitagawaNAT18}, to susceptibility and heat capacity investigations~\cite{bertPRB07,yamashitaNATP08b,yamashitaNATC11}, to thermal transport studies~\cite{yamashitaNATP08a,yamashitaSCI10,bourgeoishopeCM19,niCM19}, and to neutron scattering~\cite{heltonPRL07,devriesPRL09,heltonPRL10,banerjeeNATM16}. Theoretical works have shown that QSL ground states are realized in the antiferromagnetic $S=1/2$ Heisenberg model on the kagom{\'e} lattice~\cite{sachdevPRB92,yildirimPRB06,yanSCI11,jeschkePRB13}, in models that couple low-energy fermionic spin excitations to an emergent $U(1)$ gauge field~\cite{motrunichPRB05,leePRL05,navePRB07,leePRB08,leePRB09,sanyalPRB19}, as well as in the exactly solvable Kitaev model~\cite{kitaevAP06} in two~\cite{jackeliPRL09,knollePRL14} and three~\cite{mandalPRB09,nasuPRL14} dimensions. However, while the number and types of potential QSL models have proliferated, the experimental methods available for studying them have not~\cite{knolleAR19}. It is therefore important to the search for compounds possessing QSL ground states to propose previously unknown techniques by which to probe and categorize candidate materials.

The inverse spin Hall effect (ISHE)~\cite{saitohAPL06} is an essential component of the spintronics repertoire~\cite{sinovaRMP15}, and provides an opportunity to develop probes of unconventional quantum magnets. This phenomenon has found utility in detecting spin currents induced by thermal gradients~\cite{uchidaNAT08,uchidaAPL10,hirobeNATP16} and in performing non-local spin transport experiments through magnetic insulators~\cite{kajiwaraNAT10,cornelissenNATP15,goennenweinAPL15,liNATC16,hanNATM19}.
%In recent years, proposals to probe certain QSL states via nonequilibrium spin transport | and therefore spin Hall effects | have also been made~\cite{chenPRB13,chatterjeePRB15}. However, no extant probe of QSLs exploits the effect in thermal equilibrium.
The prime function of the ISHE is that it acts as a transducer between spin and charge current densities in metals with strong spin-orbit coupling.
%, and so, for e.g., thermal fluctuations arising in the spin sector of a heavy-element metal (e.g., Pt, Ta, W) can be transferred into electrical fluctuations.
One may therefore envisage a bilayer system in which a strongly spin-orbit coupled heavy-element metal (e.g., Pt, Ta, W) is deposited on top of a quantum magnet, i.e., consider coupling the metal to a spin dissipating subsystem that freezes out charge degrees of freedom to focus solely on the spin sector, while holding the entire bilayer in thermal and chemical equilibrium. The addition of a spin dissipating subsystem will result in equilibrium spin current fluctuations across the interface~\cite{callenPR51,kuboRPP66}, which, as a result of the ISHE, are converted into charge fluctuations inside the metal | charge fluctuations that may encode information about the microscopic structure of the quantum magnet. The QSL-to-normal-metal bilayer therefore presents a straightforward table-top setup that directly probes the low energy density of states of the QSL material. This type of bilayer system has garnered some study with respect to ferromagnetic insulators hosting magnons~\cite{kamraPRB16}; however, the utility of the ISHE as a noise conversion mechanism has not yet been explored in the context of exotic quantum magnets in general and the search for QSLs in particular.

%and is similar in that sense to heat capBcity experiments; there would, however, be no need to properly account for and remove the presence of phonons, as is the case when measuring heat capBcity, as phonons do not affect spin correlations. 
%We believe that this opens the door to previously unknown experimental possibilities.
%The prime function of the ISHE is that it acts as a transducer between the spin and charge current densities in metals with strong spin orbit coupling. The phenomenon therefore , and  However, no extant probe employs an equilibrium technique, where fluctuations arising in the spin sector of, e.g., a Pt sample can be transferred into electrical fluctuations via the ISHE.
%, and quantifying transport parameters of strongly spin orbit coupled metals~\cite{kimuraPRL07,castelAPL12,hahnPRB13,choSR15}. 
%The inverse spin Hall effect (ISHE)~\cite{saitohAPL06} is now an essential component of the spintronics repertoire~\cite{sinovaRMP15}, and provides an opportunity to develop a probe of exotic quantum magnets. 

In this work, we propose a relatively simple mechanism that can probe QSL ground states via equilibrium or near equilibrium measurements. We will develop the theory underpinning the aforementioned QSL-to-normal-metal bilayer that utilizes the QSL material as a spin dissipating subsystem, where spin dissipation via the QSL material results in spin noise generation due to the fluctuation dissipation theorem. The normal metal layer acts as a resistive element, and in the quantum limit, i.e., at very low temperatures, the power spectrum of the asymmetrized spin noise in the metal is entirely quantum in nature~\cite{clerkRMP10} and encodes information about the spin sector of the QSL layer. We show that the asymmetrized spin noise generated in this manner must affect the ac resistance measured in the normal metal, if the normal metal possesses strong spin-orbit coupling. We then use our theory to examine three QSL models: the $S=1/2$ nearest neighbor Heisenberg antiferromagnet on the kagom{\'e} lattice, a model consisting of gapless fermionic spin excitations coupled to an emergent $U(1)$ gauge field, and the bare Kitaev honeycomb model in the gapless spin liquid phase. 

We consider a gapped ground state in our approach to the kagom{\'e} lattice antiferromagnet owing to DMRG studies that indicate the presence of a gap~\cite{yanSCI11}, in addition to gapped out flux excitations, and utilize a bosonic parton mean-field theory~\cite{sachdevPRB92} when characterizing the emergent low energy spin excitations. In so doing, we show that our proposed bilayer system can successfully quantify any spinon gap present. 

Motivated by the fact that the slave-rotor representation~\cite{florensPRB04} of the Hubbard model on a triangular lattice results in stable QSL mean-field states comprised of a spinon Fermi surface coupled to a $U(1)$ gauge field~\cite{leePRL05} that may apply to existing candidate materials [e.g., $\kappa$-(BEDT-TTF)$_2$-Cu$_2$(CN)$_3$ and YbMgGaO$_4$], we examine this QSL model theoretically in our proposed bilayer system. We find that a QSL model coupling gapless fermions with a Fermi surface to an emergent $U(1)$ gauge field produces a subdominant $\W^{4/3}$ frequency correction to the ac resistance measured across the normal metal that our proposed system can educe. 

Finally, our treatment of the bare, gapless Kitaev model considers the isotropic point of the quantum spin liquid phase, where all bond interactions are equal in strength, and assumes a flux-free background | a valid assumption for temperatures lower than $1\%$ of the bond strength~\cite{nasuPRB15,rousochatzakisPRB19}. Under these considerations we find that when coupled to a normal metal, the presence of a Kitaev QSL will allow characterization of the two-flux gap energy that emerges in spite of the gapless nature of the fermion spectrum.

\section{Theory}
\begin{figure}
\includegraphics[width=0.9\linewidth]{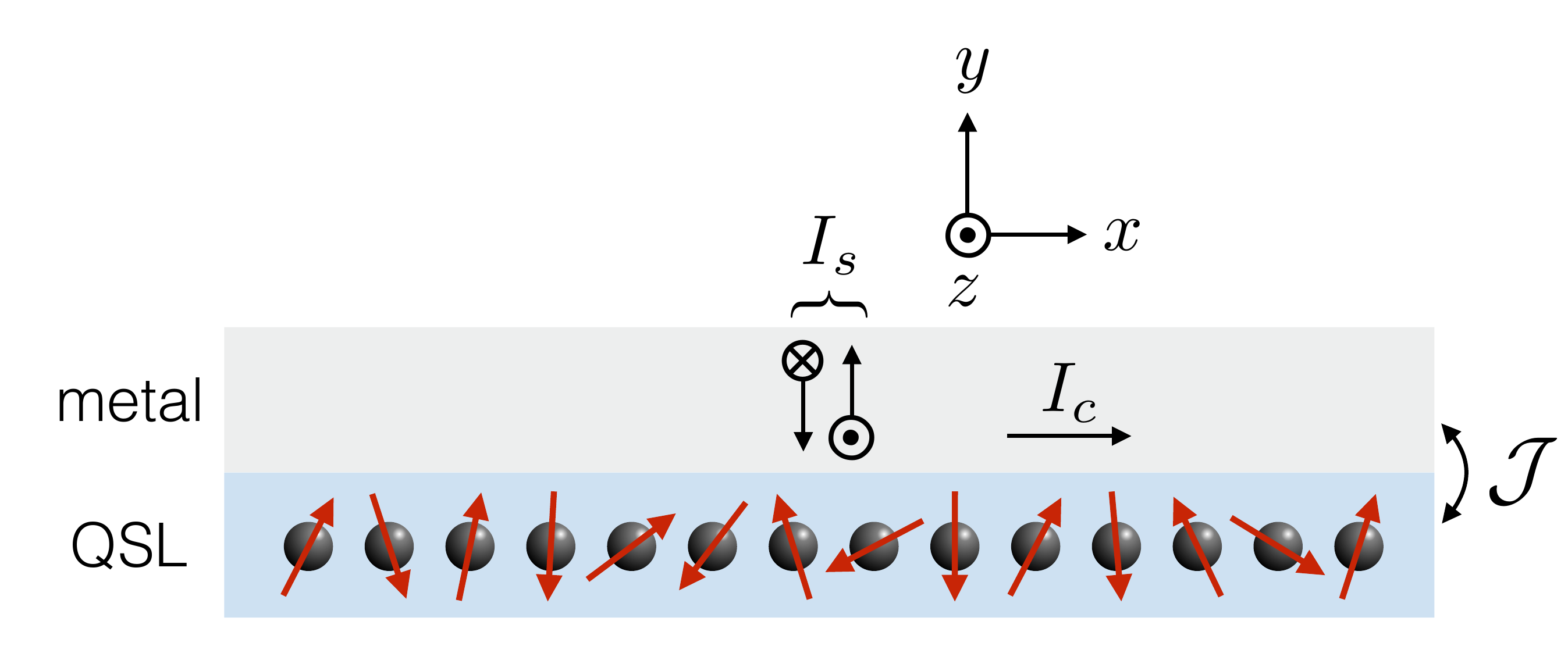}
\caption{(Color online) A cartoon of the proposed system: a quantum spin liquid to normal metal bilayer in equilibrium, where fluctuating spin current $I_s$ across the interface becomes fluctuating charge current $I_c$ (or fluctuating voltage) in the normal metal via the inverse spin Hall effect and results in measurable modifications to the resistance of the metal.}
\label{fig1}
\end{figure}

Let us consider coupling an insulating spin system (i.e., a quantum manget) to a strongly spin-orbit coupled heavy-element metal, as depicted in Fig.~\ref{fig1}. The addition of the spin system gives rise to increased spin dissipation, which, as a result of the fluctuation-dissipation theorem (FDT), contributes additional spin fluctuations inside the metal. If strong spin-orbit interactions are present in the conductor (as in, e.g., Pt, Ta, W~\cite{hallNBS68,desaiJPC84}), then FDT requires that spin fluctuations across the interface will give rise to voltage fluctuations in the conductor via the ISHE, i.e., $S=S_0+\de S$, where $S$ is the total voltage noise in the metal in thermal equilibrium, $S_0$ is the total noise present in the bare metal, and $\de S$ is the portion of the noise that arises due to the presence of the quantum magnet.
%The fluctuation-dissipation theorem (FDT) requires that dissipation in a quantity gives rise to fluctuations of that quantity. This is the starting point for the concept behind our proposal. 

%establishes a strict relationship between the resistance through a conductor and the ac voltage noise power spectrum in thermal equilibrium,
%\beq
%S_V(\W)=\frac{4\hbar\W R(\W)}{1 - e^{-\be \hbar \W}},
%\eeq
%where $S_V(\W)$ is the voltage noise power spectrum, $\be$ is the inverse temperature of the conductor, and $R(\W)$ is its electrical resistance. 

In the quantum limit | when the sample is cooled to as low a temperature as possible | background thermal noise in the metal is strongly suppressed, and
\beq
\label{fdt}
S(\W)=4\hbar\W R(\W)\thi(\W)\ .
\eeq
Here, the Heaviside step function $\thi(\W)$ signifies that only positive frequencies are possible in the quantum limit. This indicates that the metal is able to absorb spin quasiparticles from the coupled bath | i.e., the QSL | but is unable to emit them into the bath~\cite{clerkRMP10}. The voltage fluctuations present in the metal are thus entirely due to the presence of the QSL candidate material and any background quantum noise. 

We assume that in the quantum limit the base resistance is essentially constant in frequency range of interest and known in a given strongly spin-orbit coupled metal, and therefore any quantum background noise linear in frequency that arises can be accounted for and removed, exposing $\de S(\W)$.

In the upcoming sections, we provide the technical calculations necessary to extract the correction to the equilibrium voltage noise $\de S(\W)$ | the enhancement to the ac power spectrum of the voltage noise in a strongly spin-orbit coupled metal interfaced with a general quantum magnet possessing no long-ranged order | and then apply our results to some well-known QSL models. Characteristic features in the frequency distribution will then allow for discriminating between the QSL ground states of the various candidate materials via a relatively straightforward near-equilibrium measurements of the ac resistance present in the metal layer via Eq.~\eqref{fdt}. 

We therefore consider a bilayer system as shown in Fig.~\ref{fig1}, comprised of a thin normal metal layer (thickness $d$) affixed atop a QSL material. The interface is set at the $xz$ plane, and the QSL is treated as a 2d lattice of fixed quantum spins $\bS_\bi$. The particular lattice structures of the QSLs come into play later in the discussion. We assume that the entire heterostructure is thermalized to a temperature $T$ in order to eliminate the effects of nonequilibrium drives.

We assume the QSL and the normal metal are coupled via exchange interaction
\beq
H_{c} = -\cJ v_0\displaystyle\sum_{\bi}\bs(y=0,\br_\bi) \cdot \bS_{\bi}\ ,
\label{ham}
\eeq
where $v_0$ is the volume in the metal per spin of the QSL, $\cJ$ is the exchange constant, and $\br_\bi$ is a vector of the interfacial coordinates at the $y=0$ plane specifying the position of $\bS_\bi$. The local spin density of the metal is given by $\bs(\bx)=(1/2)\psi^\dag_s(\bx)\bta_{ss'}\psi_{s'}(\bx)$, where $\bta$ is the vector of Pauli matrices and $s$ labels the electron spin quantum number.

The interfacial spin current operator $I_s$ is defined by considering the total $z$ polarized spin entering the metal, i.e., 
\begin{multline}
\label{cur}
I_s(t)=(-\io)\frac{\cJ v_0}{2}\sum_{\bi}\ [s^-(y=0,\br_\bi,t)S^+_\bi(t)\\
-s^+(y=0,\br_\bi,t)S^-_\bi(t)]\ ,
\end{multline}
where $\io=\sqrt{-1}$. The spin current noise across the interface can then be computed to lowest non-trivial order in $\cJ$ using, e.g., the real-time Keldysh formalism and the result formally takes the form~\cite{joshiPRB18}
\begin{multline}
\label{iic}
\langle I_s(t)I_s(0)\rangle=\round{\frac{\cJ v_0}{2}}^2\sum_{\bi\bj}\langle s^-(y=0,\br_\bi,t)s^+(y=0,\br_\bj,0)\rangle\\
\times \square{\langle S^+_\bi(t)S^-_\bj(0)\rangle+\langle S^-_\bi(t)S^+_\bj(0)\rangle},
\end{multline}
where $\langle\cdots\rangle$ represents correlation functions with respect to the unperturbed Hamiltonian. The spectrum of the interfacial spin current noise is then defined via
\beq
S_s(\W)=\int_{-\infty}^\infty dt\ \langle I_s(t)I_s(0)\rangle e^{i\W t}\ .
\eeq
%We must take apBrt the $\left< T^\dag_\bi(t) T_\bj(0) \right>$ correlation functions in order to make progress. The result, due to the fact that $\left[ s^\pm,S^\mp \right] = \left[ s^\pm,S^\pm \right] = 0$, is
%\beq
%\left< T^\dag_\bi(t) T_\bj(0) \right> = \left< s^-(\bR_\bi,t)s^+(\bR_\bj,0) \right> \left< S^+_\bi(t)S^-_\bj(0) \right>, \nn
%\eeq
%and the $\left< T_\bi(t) T^\dag_\bj(0) \right>$ version is similar. Now, we first examine the electron density correlator. The electron wavefunction in real space, $\psi_{\s}(\bx)$, is
%\beq
%\psi_\s(\bx) =  \sqrt{\frac{2}{d_N \mathpzc{A}}} \displaystyle\sum_{\bk} e^{-ik_x x - ik_z z} \cos{(k_y y)} c_{\bk \s}, \nn
%\eeq
%where $\mathpzc{A}$ is the interfacial area of the interface, $d_N$ is the thickness of the Pt contact, $\s$ is the spin-1/2 index, and $c_{\bk \s}$ ($c^\dag_{\bk \s}$) is the electron annihilation (creation) operator that obeys the expected fermionic anti-commutation relations. Also note that the above wavefunction obeys the boundary condition at the interface $y=0$. The momenta are therefore $k_{x/z} = 2 \pi n_{x/z}/L$ and $k_y = \pi n_y/L$, with $n_{x/z} \e~\mathbb{Z}$ and $n_y\e~\mathbb{N}$.

The spin correlation function in the metallic sector can be readily computed, so we obtain a general expression for the equilibrium spin current noise spectrum at finite temperature (see, e.g., Ref.~\onlinecite{joshiPRB18}),
\begin{multline}
\label{noj}
S_s(\W,T)=2\io\round{\frac{\cJ v_0mk_F}{2\p^2\hbar}}^2\sum_{\bi \bj}\int_{-\infty}^\infty d\nu\ \frac{\nu-\W}{e^{\be\hbar(\nu-\W)}-1}\\
\times\sinc^2(k_F|\br_\bi - \br_\bj|)\square{\chi_{\bi \bj}^{+-}(\nu) + \chi_{\bi \bj}^{-+}(\nu)}\ ,
\end{multline}
where $k_F$ is the Fermi wavevector of the metal, and we have introduced the dynamical spin correlation function of the QSL,
\beq
\chi^{\mp \pm}_{\bi\bj}(\nu)\equiv -\io \int dt\ \langle S^\mp_\bi(t)S^\pm_\bj(0)\rangle e^{i\nu t}\ ,
\label{sus}
\eeq
to account for the portion of the noise that arises due to spin fluctuations in the QSL. Finally, for large Fermi wavevectors, i.e., $k_F\abs{\br_\bi - \br_\bj} \gg 1$ for all $\bi,\bj$, Eq.~\eqref{noj} can be approximated in spatially local terms, and we obtain
\begin{multline}
\label{noi}
S_s(\W,T)\approx2\io\round{\frac{\cJ v_0mk_F}{2\p^2\hbar}}^2\sum_{\bi}\int_{-\infty}^\infty d\nu\ \frac{\nu-\W}{e^{\be\hbar(\nu-\W)}-1}\\
\times\square{\chi_{\bi \bi}^{+-}(\nu) + \chi_{\bi \bi}^{-+}(\nu)}\ .
\end{multline}

We emphasize here that the derivation has so far assumed no particular form for the QSL Hamiltonian. Instead, Eq.~\eqref{noi} is constructed in order to apply to noise generated in a strongly spin-orbit coupled normal metal due to the proximity of a general quantum magnet (albeit with no long-ranged magnetic order), such that extracting a usable quantity depends only on characterizing $\chi^{\pm\mp}_{\bi\bi}(\nu)$ for a given QSL model. 

The ISHE finally converts the $z$ polarized spin current density in the $y$ direction into a charge current density along the $x$ direction (see Fig.~\ref{fig1}). The conversion ratio between the interfacial spin current fluctuations and the measurable voltage fluctuations may depend non-trivially on various properties of the metal, e.g., its spin Hall angle, geometry, and spin diffusion length. However, as shown in, e.g., Refs.~\cite{kamraPRB14,joshiPRB18}, these details may be lumped into a single phenomenological spin-to-voltage noise conversion constant $\Theta$ so that the ac voltage fluctuations generated by the quantum magnet can be expressed in terms of the spin fluctuations in Eq.~\eqref{noi} as
\beq
\de S(\W,T)=\Theta S_s(\W,T)\, .
\eeq
Therefore, what we have calculated in Eq.~\eqref{noi} is directly related to the ac voltage noise in the metal generated by the presence of the quantum magnet. We may then extract the modification to the resistance acquired by the presence of the quantum magnet from this quantity using Eq.~\eqref{fdt}.

%%%%%%%%%%%%%%%%%%%%%%%%%%%%%%%%%%%%%%%%%%%%%%%%%%%%%%%%%%%%%%
\section{Application to Quantum Spin Liquids}
\label{sthree}

We now consider three types of QSL models in the context of our proposed bilayer system: the antiferromagnetic Heisenberg $S=1/2$ kagom\'e lattice model (HKLM); a model involving a spinon Fermi surface coupled to an emergent $U(1)$ gauge field; and the Kitaev honeycomb model in the gapless spin liquid phase. We ultimately show that the bilayer system we propose will allow the extraction of characteristic frequency dependencies in each case. All three models above are believed to be of relevance for the descriptions of some QSL candidate materials. Detailed discussion of these connections will be presented in Sec.~\ref{discon}.
%First, we utilize the Schwinger boson~\cite{liuPRB19} approach to the HKLM that can account for a phase transition between QSL and N\'eel phases. Next, motivated by Ref.~\onlinecite{balentsCM19} and the search for observables that can identify signatures of gauge fields in spin liquid states, we perform a real-frequency, finite temperature calculation in our treatment of fermionic spinons with a Fermi surface coupled to an emergent $U(1)$ gauge field. Finally, we examine the Kitaev spin liquid as it is exactly solvable and therefore compelling theoretically. 

%This has made understanding the various phases of the Kitaev model
%---that is, when perturbatively including Heisenberg interactions and cross terms~\cite{songPRL16}---
%of pressing interest, including and, for our purposes, especially the gapless spin liquid phase.

%%%%%%%%%%%%%%%%%%%%%%%%%%%%%%%%%%%%%%%%%%%%%%%%%%%%%%%%%%%%%%
\subsection{Heisenberg antiferromagnet on the kagom\'e lattice}
\label{herbsmith}
The HKLM is an important prototypical model that supports a QSL ground state, and is believed to be a relevant model for the mineral compound ZnCu$_3$(OH)$_6$Cl$_2$ (herbertsmithite), one of the prime contenders for an experimental realization of a QSL. Recent neutron scattering experiments suggest that the relevant theoretical model for the QSL state observed in this compound consists of the so-called $Z_2$ spin liquids~\cite{sachdevPRB92,wangPRB06,punkNATP14}. These QSL states possess two types of excitations, i.e., the $S=1/2$ spinon and a gapped vortex, also known as a vison, corresponding to an emergent $Z_2$ gauge field~\cite{senthilPRB00}. It was shown recently that these visons can act as a momentum sink for the spinons and can flatten the dynamic spin structure factor~\cite{punkNATP14} in accordance with experimental observations of herbertsmithite. However, for the purposes of the heterostructure we are proposing in this work, we note that only local spin fluctuations are important and visons do not themselves carry spin. Therefore, we expect that visons should not significantly renormalize spin current fluctuations, and that the re-shuffling of the spin quasiparticle spectral weight in momentum space should not drastically affect local quantities. We therefore ignore the effects of the visons in this subsection, follow the calculations performed in Ref.~\onlinecite{sachdevPRB92} for the HKLM, and apply the result to the observable we are proposing as a probe of the QSL state.
%even above the vison gap energy, which is taken to be $\D_v \sim \D_s/2$~\cite{punkNATP14} as they must interact with spinons in order to have an effect, 

We begin with the Hamiltonian for the bare HKLM
\beq
H_{HKLM}=-\frac{J}{2}\sum_{\langle\bi,\bj\rangle}\bS_{\bi}\cdot\bS_{\bj}\ ,
\label{hkl}
\eeq
where $\langle\bi,\bj\rangle$ indicates nearest neighbor pairings and $J$ is the exchange coupling. Mean-field approaches to characterizing the HKLM ground states include bosonic representations~\cite{sachdevPRB92} of the gapped spinons and  fermionic representations, where a gap arises due to pairing in the Hamiltonian~\cite{luPRB11hc}. In what follows, we focus on the bosonic representation of the QSL phase as done in Ref.~\onlinecite{sachdevPRB92}.
\begin{figure}
\includegraphics[width=.7\linewidth]{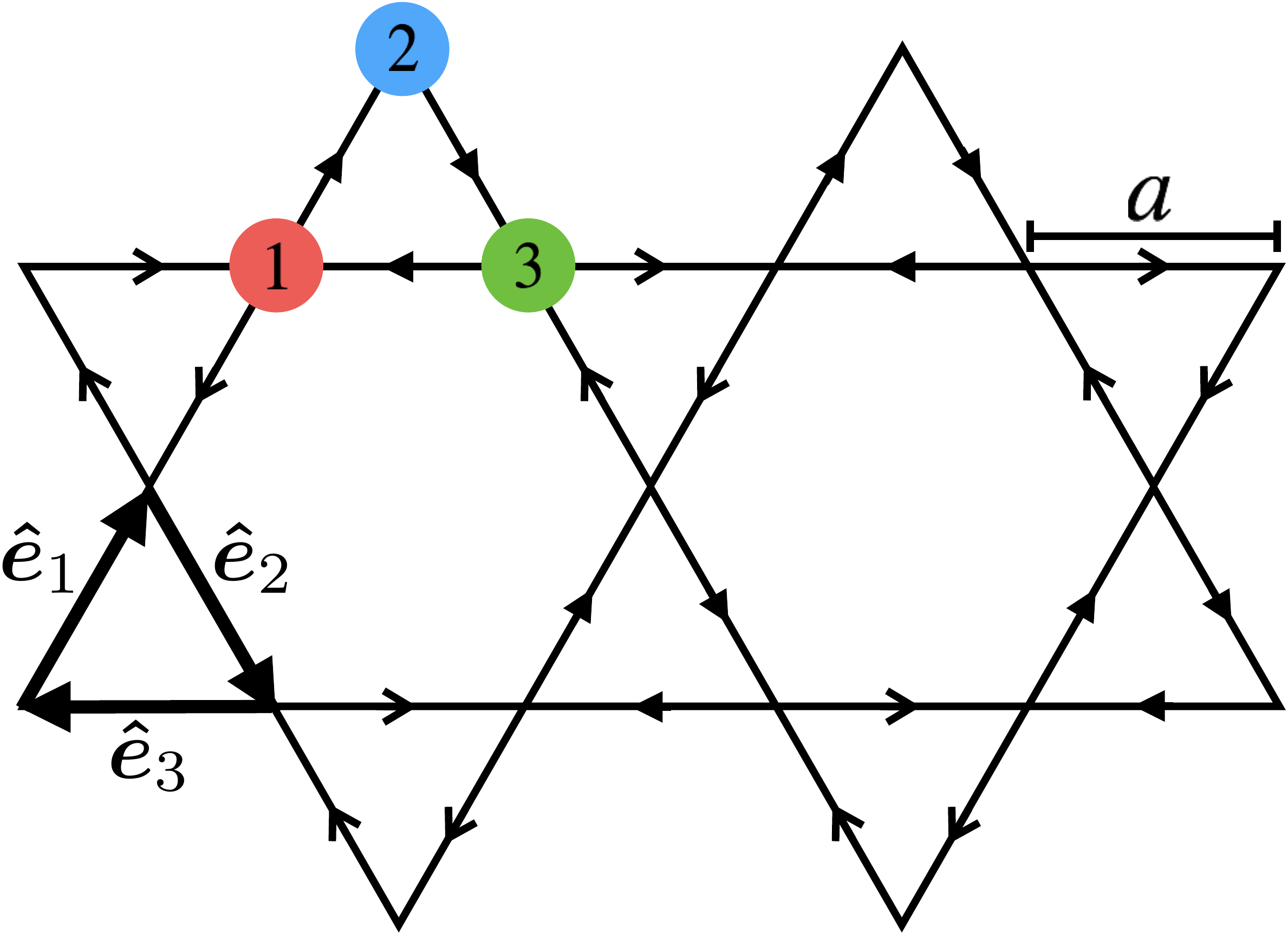}
\caption{(Color online) The kagom{\'e} lattice, with positions on a representative unit cell marked. The $\boldsymbol{\hat{e}}_i$ represent the unit vectors of a unit cell, and $a$ is the length of a bond. Arrows between sites correspond to the direction dependent $Q_{\bi\bj}$. Solid arrows around upward facing triangles depict  $Q_{\bi\bj}=Q_1$, and wire arrows around downward facing triangles depict $Q_{\bi\bj}=Q_2$.}
\label{fig2}
\end{figure}
%The goal is to compute the local dynamic spin susceptibility $\x_{\bi\bi}^{\pm\mp}(\nu)$, which is ultimately inserted into Eq.~\eqref{noi} for the evaluation of the spin current fluctuations. 

We begin with the standard Schwinger boson representation of the spin operators $\bS_{\bi}=(1/2)b^\dag_{\bi\s} \bta_{\s \s'} b_{\bi\s'}$, where $b_{\bi\s}$ represents a bosonic spinon, and then use the mean-field decoupling $Q_{\bi\bj}=(1/2)\langle\ve_{\s\s'}b_{\bi\s}b_{\bj\s'}\rangle$, where $\ve_{\s\s'}$ is the completely antisymmetric tensor of {\em SU}(2) and the field satisfies $Q_{\bi\bj}=-Q_{\bj\bi}$. Substituting this representation into Eq.~\eqref{hkl}, the mean-field decoupling results in the Hamiltonian
\begin{multline}
\label{hklm}
H_{HKLM}=-\frac{J}{2}\sum_{\langle\bi,\bj\rangle}\sum_{\s\s'}Q_{\bi\bj}\ve_{\s\s'}b^\dag_{\bi\s}b^\dag_{\bj\s'}+h.c.\\
+\la\sum_{\bi,\s}b^\dag_{\bi\s}b_{\bi\s}\ ,
\end{multline}
where $h.c.$ denotes the hermitian conjugate and $\la$ is a Lagrange multiplier that constrains the model to one boson per site.%, i.e., $\sum_\s b^\dag_{\bi\s}b_{\bi\s}=1$. 
%For ease of calculation, as we are primarily interested in illustrating the workings of our proposed heterostructure, we do not here consider Dzyaloshinskii-Moriya interactions~\cite{huhPRB10,doddsPRB13}. 
%which can be justified in the large-$N$ limit~\cite{sachdevPRB92}

Figure~\ref{fig2} shows a depiction of the kagom{\'e} lattice, where the arrows between sites in a unit cell correspond to the direction dependent $Q_{\bi\bj}$, i.e., $Q_1$ and $Q_2$ are the two distinct expectation values of $Q_{\bi\bj}$. In the figure, solid arrows correspond to $Q_{\bi\bj}=Q_1$ and wire arrows correspond to $Q_{\bi\bj}=Q_2$. The two possible locally stable QSL mean-field solutions~\cite{sachdevPRB92,wangPRB06,huhPRB10,punkNATP14} occur when $Q_1=Q_2 = Q$, corresponding to $\pi$ flux through the hexagonal plaquette, and $Q_1=-Q_2$, corresponding to no flux through the plaquette, both with $Q_1,Q_2\in\mathbb{R}$. Here, we will specifically consider the $\pi$ flux case and utilize mean-field theoretical values put forward in Ref.~\onlinecite{punkNATP14} for our purposes. 

Upon diagonalizing Eq.~\eqref{hklm}, we can solve for the total spin susceptibility $\chi(\nu)=\sum_{\bi}[\chi^{+-}_{\bi \bi}(\nu)+\chi^{-+}_{\bi \bi}(\nu)]$, and use it to derive the noise correction. We perform the finite temperature calculation in Appendix~\ref{apB}, and then take the $T\to0$ limit. Our resultant expression for the zero temperature ac noise power spectrum then becomes
\begin{multline}
\label{hsn0t}
S_s(\W,0)=\frac{4\pi}{N_u\hbar}\left( \frac{\cJ v_0mk_F}{2 \pi^2 \hbar} \right)^2 \sum_{\bk,\bq}\sum_{n,m,l} U^{\bk*}_{nm}U^\bk_{nm}U^{\bq*}_{\bar nl}U^{\bq}_{\bar nl}\\
\times(\hbar\W-\e_{\bk m}-\e_{\bq l})\thi(\hbar\W-\e_{\bk m}-\e_{\bq l})\ ,
\end{multline}
where $U^\bk_{nm}$ is the $6\times6$ matrix that diagonalizes Eq.~\eqref{hklm}, $\e_{\bk m}$ is the $m$-th energy eigenvalue of the diagonalized Hamiltonian, and $\thi(x)$ is the Heaviside theta function. Note that $\bar{n} = n+3$, the index $n=1,2,3$, and $m,l=1,\dots,6$. We now analyze this expression numerically.

\begin{figure}
\includegraphics[width=.7\linewidth]{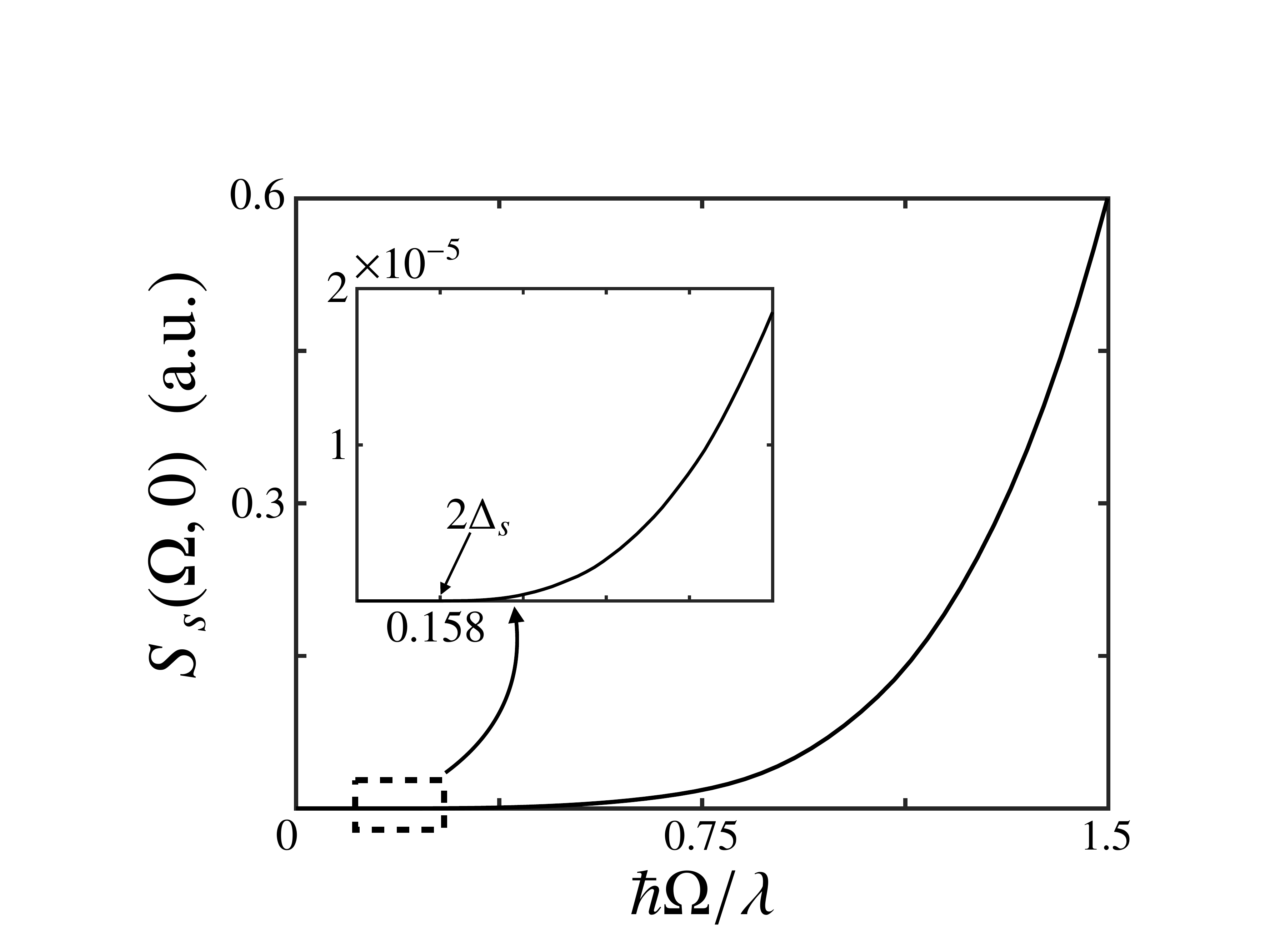}
\caption{A numerical plot showing $S_s(\W,0)$ in the HKLM as a function of ac frequency normalized by the mean-field coupling, $\la = 0.695 J$. The main plot shows the computed sum, and the inset zooms in on the low frequency regime. The graph is identically zero until a critical frequency of twice the gap energy is achieved, $\hbar\W_c = 2\D_s = 0.158\la$, at which point it becomes possible to create spinon pairs in the QSL and noise enhancement commences.}
\label{fig3}
\end{figure}

Figure~\ref{fig3} is a plot of Eq.~\eqref{hsn0t}, from which we find that, in the quantum limit, it is possible to extract an estimate of the spin gap using our proposed bilayer. We use the variable values proffered in Refs.~\onlinecite{chatterjeePRB15,punkNATP14} for the HKLM, taking as a reasonable estimate for the spin gap $\D_s = 0.055J$. In the HKLM, the gap energy can be found by setting $\bk = 0$ and finding the minimal eigenvalue~\cite{huhPRB10}, with the outcome $\D_s = \sqrt{\la^2 -12 J^2 Q^2}$. Therefore, rather than computing the values of $\la$ and $Q$ variationally, we set $\la = 0.695 J$ and tune $Q$ such that $\D_s = 0.055 J$, resulting in $Q = 0.2$. Note that $Q$ quantifies antiferromagnetic correlations and is restricted to $\abs{Q} \leq 1/\sqrt{2}$, at which point nearest neighbor spins form singlets and the model experiences a phase transition into an ordered phase. 

Importantly, the HKLM models a highly frustrated antiferromagnetic material, and therefore includes quantum fluctuations. At very low temperatures, quantum fluctuations are required in order to drive the interfacial equilibrium noise generated as a result of the mechanism we are considering. Figure~\ref{fig3} shows that no noise enhancement is expected until probing frequencies greater than the critical frequency $\hbar \W_c = 2\D_s$, at which point production of a pair of spinons can occur in the QSL and an enhancement in the noise begins to manifest. Therefore probing this region of the frequency range for a given material that is adequately modeled by the HKLM should allow for a direct estimate of the spin gap, $\D_s$, present in that material.

%This highlights our expectation that the bilayer system we have proposed as a mechanism for studying QSL candidate materials | and quantum magnets generally | can be used in the low temperature regime to extract the spin gap of the QSL material that can be modeled with the HKLM.

%%%%%%%%%%%%%%%%%%%%%%%%%%%%%%%%%%%%%%%%%%%%%%%%%%%%%%%%%%%%%%

\subsection{Spinon Fermi surface coupled to an emergent \\ $U(1)$ gauge field}
\label{fermu1}
Some QSL candidate materials, e.g., organic salt compounds~\cite{motrunichPRB05,leePRL05,navePRB07} and YbMgGaO$_4$~\cite{liPRB16,liuPRB16,liPRB17}, may be described as spinon Fermi seas, where the spin susceptibility is strongly renormalized by an emergent $U(1)$ gauge field. In this subsection, we first compute (using the Keldysh functional integral formalism) the thermal spin current fluctuations arising in the metal due to the presence of fermions in the QSL material, and second how the gauge field renormalization affects those thermal fluctuations in our proposed bilayer system at finite temperature. We then take the $T\to 0$ limit at the end and present the zero temperature ac frequency result.
%While the finite temperature calculation can also be performed using the more familiar Matsubara formalism (since the system is in thermal equilibrium), we opt for the Keldysh real-frequency calculation to avoid the need for analytic continuation at a later point. 

%We use the Keldysh functional integral formalism in this subsection to facilitate our calculation of the thermal spin current fluctuations.
Let us begin with the real-time action for $\cN$ flavors of fermionic spinons coupled to a compact $U(1)$ gauge field in $2+1$ dimensions~\cite{leePRB92,leePRL05,polchinskiNPB94}
\beq
\begin{aligned}
\label{fu1a}
S&=\sum_{\s}\int dtd\bx\ \Big\{\bar c_\s(t,\bx)(\io\hbar\pd_t+\mu)c_\s(t,\bx)\\
&\qquad-\frac{1}{2m_s}\bar c_\s(t,\bx)[-\io\hbar\bnab+\ba(t,\bx)]^2c_\s(t,\bx)\Big\}\ ,
\end{aligned}
\eeq
where $c_\s(t,\bx)$ is a spinon Grassmann field, $\s$ is the flavor index, $m_s$ is the spinon effective mass, $\mu$ is the spinon chemical potential, and $\ba(t,\bx)$ is the gauge field. The (real-frequency) retarded RPA propagator for the gauge fluctuations in the Coulomb gauge can be obtained by analytically continuing the standard result from the imaginary-time formalism~\cite{leePRB92,kimPRB94}
\begin{align}
D^{R}_{\xi\xi'}(\bq,\W)&=-\round{\de_{\xi\xi'}-\frac{q_\xi q_{\xi'}}{q^2}}\frac{1}{\cN} \frac{1}{\chi_dq^2 - \io \frac{E_{Fs}}{\pi\hbar^3}\frac{\W}{\varv_{Fs}q}}\ ,\nn\\
&\equiv-\round{\de_{\xi\xi'}-\frac{q_\xi q_{\xi'}}{q^2}}d^{R}_\bq(\W)\ ,
\label{RPA}
\end{align}
where $\xi,\xi'$ label the Cartesian components, $\varv_{Fs} = \hbar k_{Fs}/m_s$ is the spinon Fermi velocity, $k_{Fs}$ is the spinon Fermi wavevector, $E_{Fs}$ is the spinon Fermi energy, and $\chi_d = (24\hbar m_s)^{-1}$ is the Landau diamagnetic susceptibility of the fermions. Lastly, we define the interaction portion of the Keldysh action by placing Eq.~\eqref{fu1a} on the Keldysh contour and extracting the term dependent on the gauge field to first order,
\begin{multline}
S_{\rm int}=-\frac{1}{2\hbar\sqrt{\sA}}\int_{-\infty}^\infty dt\sum_{\bk \bk'\s}\sum_{\xi=x,z}\sum_{\eta=\pm}\eta\\
\times\varv^\xi_{\bk+\bk's}\bar c_{\bk \s}(t^\eta)a^{\xi}_{\bk - \bk'}(t^\eta)c_{\bk' \s}(t^\eta)\ ,
\label{kint}
\end{multline}
where $\eta=\pm$ represents the Keldysh time-loop forward and backward branches, $\varv_{\bk s}=\hbar\bk/m_s$, and $\sA$ is the total area of the QSL.

%%%%%%%%%%%%%%%%%%%%%%%%%%%%%%%%%%%%%%%%%%%%%%%%%%%%%%%%%%%%%%
\subsubsection{Noise correction due to bare susceptibility}
The spin operators of the QSL are represented using Abrikosov fermions, i.e., $\bS_{\bi \s} = (1/2) c^\dag_{\bi \s} \bta_{\s \s'} c_{\bi \s'}$, where the constraint of one fermion per site is enforced. Starting from the calculation of the bare bubble, without gauge field renormalization, we can find the uncorrected noise enhancement generated across the interface by first calculating the total susceptibility, $\chi^{(0)}(\nu) = \sum_\bi \big[ \chi^{+- (0)}_{\bi \bi}(\nu) + \chi^{-+ (0)}_{\bi \bi}(\nu) \big]$, given by
\begin{multline}
\label{chi0}
\chi^{(0)}(\nu) = -2 \io \sum_{\bi} \int dt~e^{\io \nu t}\\
\times\langle\hT \bar c_{\bi \downarrow}(t^-)c_{\bi \uparrow}(t^-)\bar c_{\bi \uparrow}(0^+) c_{\bi \downarrow}(0^+) \rangle\ ,
\end{multline}
where $T_K$ is the Keldysh time ordering operator. Equation~\eqref{chi0} quantifies the portion of the susceptibility that arises due simply to the presence of the fermions in the QSL material, without the additional corrections from emergent gauge photons, and evaluates to
\begin{align}
\label{chi1}
\chi^{(0)}(\nu)=-\io N\frac{a_s^2m_s^2}{\pi\hbar^2}\frac{\nu}{1-e^{-\be\hbar\nu}}\ ,
\end{align}
where $N$ is the total number of lattice points and $a_s$ is the area occupied by each QSL spin. %Therefore we present the finite temperature, ac voltage noise correction due to the fermions,
%\frac{-\io N k_F^2}{\pi \varv_F^2 \hbar^2} \int_{-\mu}^\infty d\e \int_{-\mu}^{\infty} d\e' &~\de{(\nu - (\e/\hbar - \e'/\hbar))} \nn \\ \times n_F(\e') \left[ 1 - n_F(\e) \right],
%$n_B(\nu) = ( e^{\be \hbar \nu} - 1 )^{-1}$ is the Bose-Einstein distribution, 
%\beq
%\begin{aligned}
%S_s^{(0)}(\W,T)&=\frac{4Na_s^2m_s^2}{\hbar^2}\left( \frac{\cJ v_0 mk_F}{2 \pi^2 \hbar}\right)^2\\
%&\qquad\times\int\frac{d\nu}{2\p}\frac{(\nu - \W)}{e^{\be \hbar (\nu - \W)}-1}\frac{\nu}{1-e^{-\be\hbar\nu}}\ .
%\end{aligned}
%\eeq
By inserting Eq.~\eqref{chi1} into Eq.~\ref{noi} and then moving to the zero temperature limit the zeroth order ac noise correction is therefore found to be
\beq
S_s^{(0)}(\W,0)=\frac{N}{3\p}\left( \frac{\cJ v_0 mk_F}{2\pi^2\hbar} \right)^2\round{\frac{m_sa_s}{\hbar}}^2 \W^3\, .
\eeq
%S_s^{(0)}(0,T) =\frac{4\p N}{3}\left( \frac{\cJ v_0 mk_F}{2\pi^2\hbar} \right)^2\round{\frac{m_sa_s}{\hbar}}^2 \left( \frac{k_B T}{\hbar} \right)^3\ .

%%%%%%%%%%%%%%%%%%%%%%%%%%%%%%%%%%%%%%%%%%%%%%%%%%%%%%%%%%%%%%
\subsubsection{Gauge field correction to the noise}
\begin{figure}
\includegraphics[width=.9\linewidth]{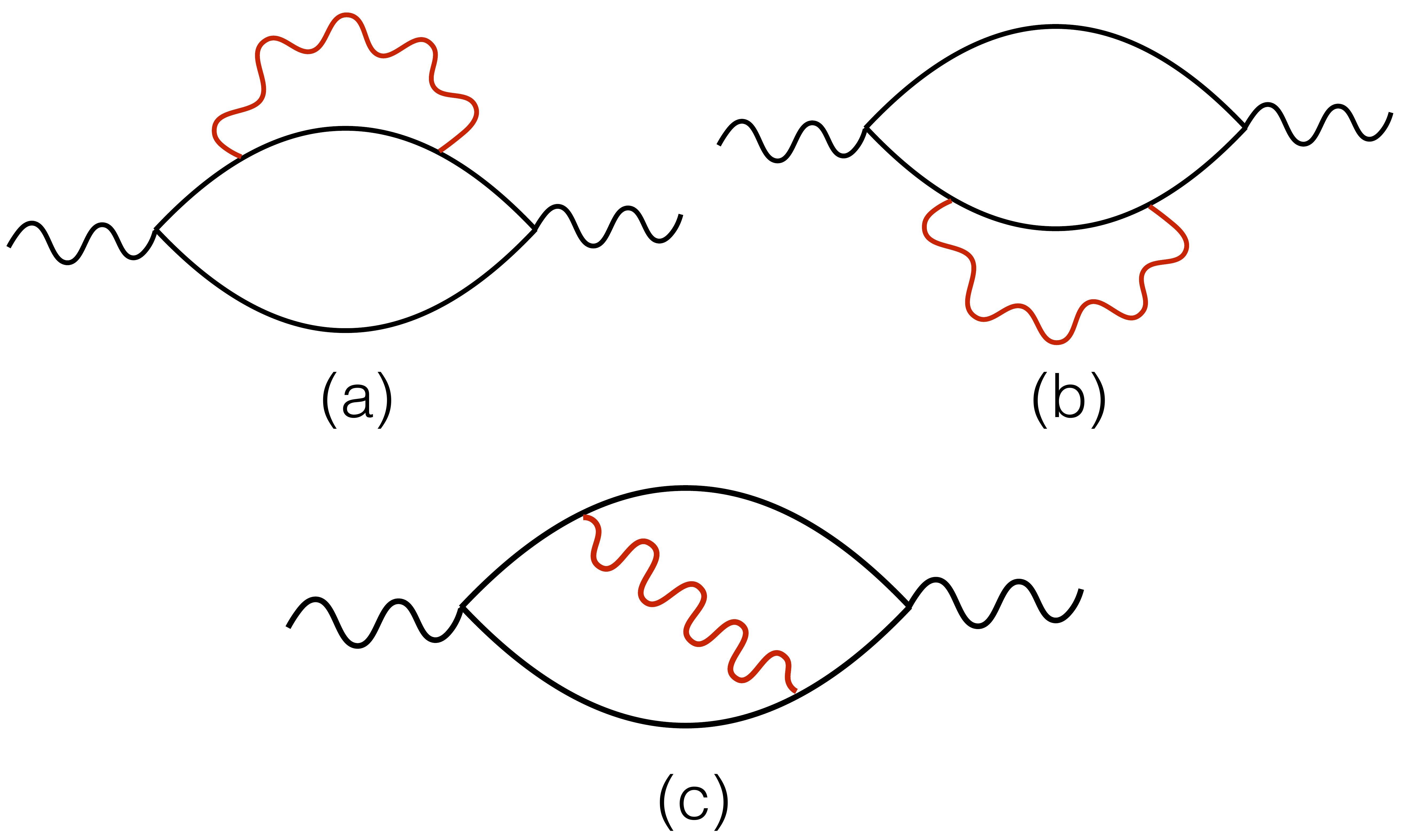}
\caption{Three diagrams contributing to the $\mathcal{O}(1/\cN)$ correction to the spin susceptibility generated by the presence of $U(1)$ gauge field fluctuations.}
\label{fig4}
\end{figure}

The next non-zero correction to the susceptibility arises due to gauge field renormalization and is given by the second order expansion using Eq.~\eqref{kint} as the perturbation, 
\begin{multline}
\label{u1int}
\chi^{\mp \pm (2)}_{\bi \bi}(\nu) = \frac{\io}{2} \int dt~e^{i \nu t}\\
\times\langle\hT \bar c_{\bi \downuparrows}(t^-)c_{\bi \updownarrows}(t^-)\bar c_{\bi \updownarrows}(0^+)c_{\bi \downuparrows}(0^+)S_{\rm int}^2\rangle\ .
\end{multline}
Expanding this gives five possible diagrams, only three of which contribute and must therefore be considered; these three are diagrammatically depicted in Fig.~\ref{fig4}. It is important when calculating gauge invariant quantities to sum over all three contributing diagrams, because only then do the divergences cancel exactly~\cite{kimPRB94,balentsCM19}. 

Accounting for the required diagrams (see Appendix~\ref{apC} for details of the calculation), the total susceptibility at two-loop order $\chi^{(2)}(\nu)=\sum_\bi[\chi^{+-(2)}_{\bi\bi}(\nu)+\chi^{-+(2)}_{\bi\bi}(\nu)]$ reads
\begin{multline}
\label{u1sus}
\chi^{(2)}(\nu)=\frac{N}{1-e^{-\be\hbar\nu}}\frac{m_{s}^2a_s^2}{2(2\pi)^2\hbar^4}\int \frac{d \w}{2 \pi} \int_0^\infty dz\,F(\nu,\w) \\
\times\left[ d^R_{-\bq}(-\w) - d^A_{-\bq}(-\w) \right]\round{1-z^{-1}\tan^{-1}z}\ ,
\end{multline}
where 
\begin{multline}
\label{u1f}
F(\nu,\w)=\int\frac{d\w'}{2\p}\square{\tanh\round{\tfrac{\hbar(\w'+\nu)}{2k_BT}}-\tanh\round{\tfrac{\hbar\w'}{2k_BT}}}\\
\times\Big[2\coth\round{\tfrac{\hbar\w}{2k_BT}}-\tanh\round{\tfrac{\hbar(\w'+\w+\nu)}{2k_BT}}-\tanh\round{\tfrac{\hbar(\w'+\w)}{2k_BT}}\Big]\ .
\end{multline}
We can then express the second order ac voltage noise correction as
\begin{multline}
\label{u1acnoi}
\de S_s^{(2)}(\W,T)=\io N \round{\frac{\cJ v_0mk_F}{2\p^2\hbar}}^2\frac{m_{s}^2a_s^2}{(2\p)^2\hbar^4}\\ 
\times\int d \nu\frac{\nu - \W}{e^{\be\hbar(\nu - \W)}-1}\int \frac{d \w}{2 \pi} \int_0^\infty dz\,\frac{F(\nu,\w)}{1-e^{-\be \hbar \nu}}\\
\times\left[d^R_{-\bq}(-\w) - d^A_{-\bq}(-\w) \right]\round{1-z^{-1}\tan^{-1}z}\ .
\end{multline}
%Thus we can express the total ac correction to the voltage noise across the heavy metal sample as $S_s(\W,T) = S_s^{(0)}(\W,T) + \de S_s^{(2)}(\W,T)$.

We now move into the zero temperature limit, which allows completion of each of the integrals. We expand the $q$ integral to lowest order in $\w$ for $\w \ll \W_{Fs}$, where $\W_{Fs}$ is the spinon Fermi frequency. Thus, the ac noise correction due to gauge fluctuations finally becomes
\begin{equation}
\de S^{(2)}_s(\W,0)=\frac{0.075}{\cN}\left( \frac{\W}{\W_{Fs}} \right)^{4/3}S_s^{(0)}(\W,0)\ .
\end{equation}
We can therefore see from the zeroth and second order calculations that the total ac voltage fluctuations generated across the normal metal sample when interfaced with a QSL material possessing an emergent $U(1)$ gauge field will have a $\W^3$ lowest order temperature dependence modified by a $\W^{4/3}$ subdominant component. The total ac noise correction generated is given by the sum of the two terms, such that the power spectrum of the enhancement becomes
\beq
\label{sga}
S_s(\W,0)=S_s^{(0)}(\W,0)\square{1+\frac{0.075}{\cN}\left(\frac{\W}{\W_{Fs}} \right)^{4/3}}\ ,
\eeq
when considering the regime $\W\ll \W_{Fs}$.
%and we note that $\W \ll \W_{Fs}$ should always pertain in the QSL phase.

That a noise enhancement accrues due to the presence of an emergent gauge field is physically understandable as the collective fermionic spin excitations present, i.e., the spinons, acquiring a higher scattering probability due to the added presence of gauge fluctuations. Additional scattering must result in additional noise generation. This enhancement should be extricable in the quantum limit using Eq.~\eqref{fdt}: first plotting $\de R(\W)/\W^2$ and then subtracting off the intercept reveals the bare $\W^{4/3}$ correction term characteristic of the gauge field renormalization calculated in Eq.~\eqref{sga}.
%This serves to enhance the voltage noise generated in the metal contact by the proximity of a quantum magnetic material well characterized by a model consisting of a spinon fermi sea that includes an emergent $U(1)$ gauge field. 

%%%%%%%%%%%%%%%%%%%%%%%%%%%%%%%%%%%%%%%%%%%%%%%%%%%%%%%%%%%%%%
\subsection{The Kitaev Honeycomb Model}
\label{dferm}
\begin{figure}
\includegraphics[width=.8\linewidth]{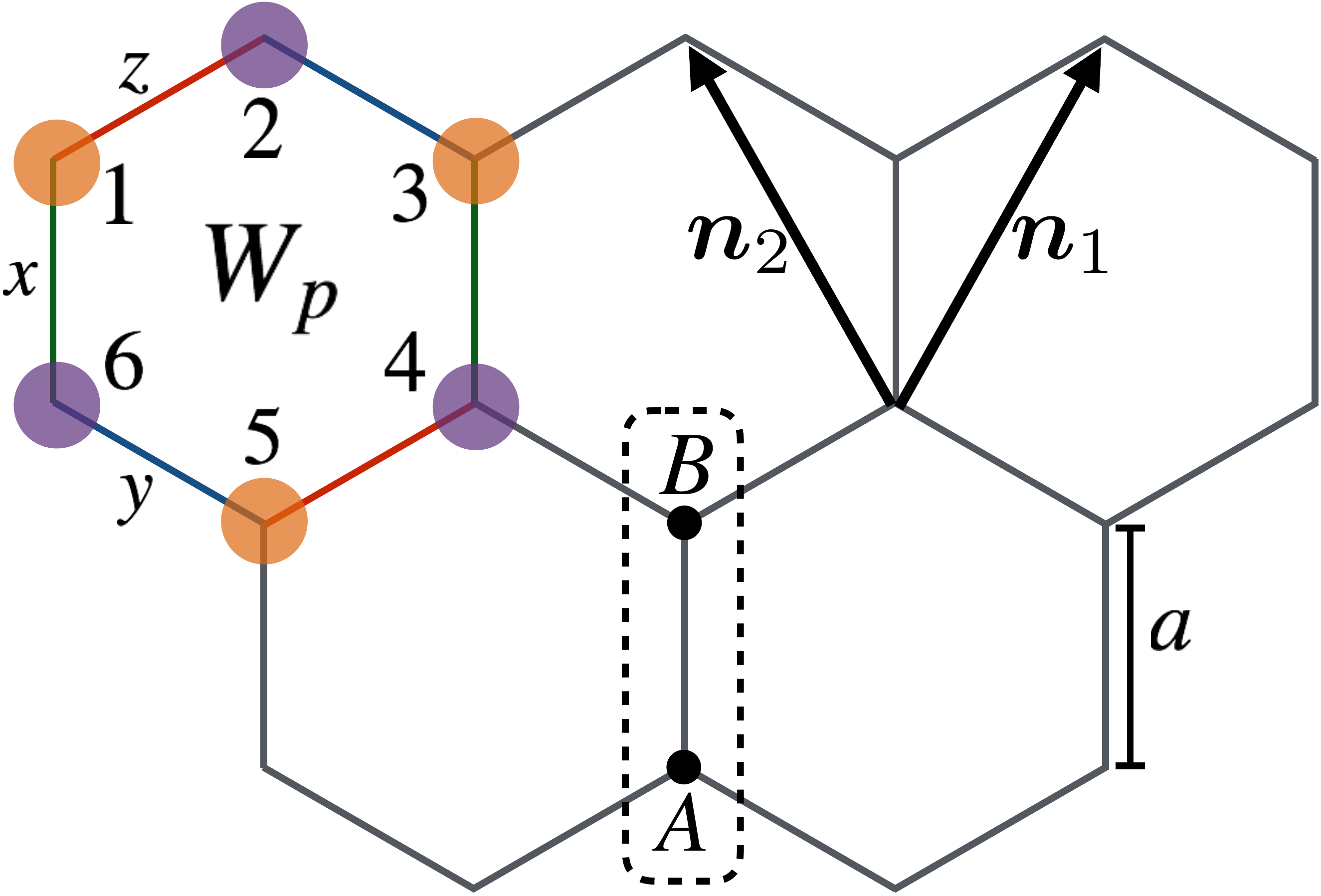}
\caption{(Color online) The Kitaev honeycomb lattice, where the two sub-lattices are shown by the purple (sub-lattice $A$) and orange (sub-lattice $B$) dots. Representative $x$-, $y$-, and $z$-links are depicted. The flux operator on plaquette $p$, $W_p$, is shown, $\boldsymbol{n}_1$ and $\boldsymbol{n}_2$ are the primitive lattice vectors, and $a$ is the lattice spacing. Also depicted is a representative unit cell enclosed in the dashed box.}
\label{fig5}
\end{figure}

A second example of a $Z_2$ QSL is the Kitaev model on the honeycomb lattice (see Fig.~\ref{fig5}), where exchange frustration arising due to the inability to simultaneously satisfy all Kitaev interactions along neighboring bonds can drive the system into a QSL phase. The Kitaev spin liquid is exactly solvable, and we select this example for consideration in our proposed system for that reason, in addition to the fact that there are potential material candidates available currently~\cite{kitagawaNAT18,banerjeeNATM16}. Furthermore, we restrict our investigation to the gapless phase, where the Kitaev exchange couplings are equal. However, while fermionic excitations are gapless, the emergent $Z_2$ gauge field is not, and this has the peculiar effect of generating an apparent gap in the fermionic sector~\cite{knollePRL14,knollePRB15}. The Kitaev model therefore provides something of a synthesis of the last two sections, as it is both a $Z_2$ and gapless theory. We give a short overview of the approach to solving the Kitaev model following Refs.~\onlinecite{kitaevAP06,hermannsAR18}, and then apply the solution to extracting an observable out of our proposed heterostructure.

The Kitaev model on the honeycomb lattice is given by~\cite{kitaevAP06}
\beq
\label{kh}
H_K=\sum_{\g,\langle\bi,\bj\rangle_\g}-K_\g S^\g_\bi S^\g_\bj\ , 
\eeq
where $\g=\{x,y,z\}$ represents the different nearest neighbor bond directions (see Fig.~\ref{fig5}) at each lattice point with interaction strength $K_\g$. In understanding the gapless spin liquid phase of the Kitaev model, parton mean-field theories have been proposed~\cite{kitaevAP06,knollePRB18} that characterize the emergent excitations as Dirac fermions~\cite{hermannsAR18} arising in tandem with an emergent gapped flux.

This can be seen by first representing the spin operators in terms of four Majorana fermions, i.e., $S^\g_\bi=\io f_{\bi\g} c_\bi$ with $\{f_{\bi\g},f_{\bi'\g'}\}=2\de_{\bi\bi'}\de_{\g\g'}$, $\{c_\bi,c_{\bi'}\}=2\de_{\bi\bi'}$, and $\{f_{\bi\g},c_{\bi'}\}=0$, which then gives $S^\g_\bi S^\g_\bj=-\io\hu_{\bi\bj}c_\bi c_\bj$, where $\hu_{\bi\bj}=\io\sum_\g f_{\bi\g}f_{\bj\g}$ is the bond operator. Noting that the bond operators commute with each other and with any bilinear operator containing $c_\bi$, it is possible to replace them with their eigenvalues $\pm 1$. One thus obtains $S^\g_\bi S^\g_\bj=\pm\io c_\bi c_\bj$, and $H_K$ becomes bilinear in the Majorana fermions. Second, we note that the so-called flux operator on a plaquette $W_p=f_{1x}f_{2y}f_{3z}f_{4x}f_{5y}f_{6z}$, where the subscript $p$ labels the plaquette number, commutes with the Hamiltonian and is therefore an integral of motion. When a plaquette has an even number of bonds, as Fig.~\ref{fig5} makes clear is the case for the honeycomb lattice, its eigenvalues are $\pm 1$. Finally, we note that the spin representation in terms of four Majorana fermions enlarges the Hilbert space from two to four, and must therefore be constrained in order to recover the physical Hilbert space. This constraint is enforced via a projection operator $P_\bi=(1/2)(1+f_{\bi x}f_{\bi y}f_{\bi z}c_\bi)$ for each site $\bi$, which requires that the initial spin algebra be conserved. Effectively, what has been done is to reduce the initial Hamiltonian to a noninteracting Dirac fermion hopping Hamiltonian in a static $Z_2$ gauge field, where the choice of values for $\hu_{\bi\bj}$ amounts to fixing a gauge, and the gauge invariant quantities are the plaquette operators $W_p$.

A theorem by Lieb~\cite{liebPRL94} guarantees that if the number of sites per plaquette is $2$ mod $4$, then the ground state is in the flux-free sector, i.e., where it is possible to set $\hu_{\bi\bj}=1$. The ground state is therefore described by the free Majorana hopping Hamiltonian,
\beq
H_K=\io K \sum_{\langle\bi,\bj\rangle}c_{\bi}c_{\bj}\ ,
\label{kith}
\eeq
where $K_x=K_y=K_z\equiv K$, and the subscripts $\bi$ and $\bj$ are now labeling $A$ and $B$ sites respectively. 

Equation~\eqref{kith} can be expressed in terms of complex fermions using a standard procedure~\cite{knollePRL14}. We take $\br$ as a unit cell coordinate, comprised of an $A$ site and a $B$ site as shown in Fig.~\ref{fig5}, and combine the two Majorana species into a single complex fermion species through $b_\br=(c_{A\br} + \io c_{B\br})/2$. Putting the system on a torus with $N_u$ unit cells, and changing to reciprocal space with $b_{\br}=N_u^{-1/2} \sum_{\bk} e^{\io \bk \br} b_{\bk}$, Eq.~\eqref{kith} becomes
\beq
H_K = \sum_{\bk} \begin{pmatrix} b^\dag_\bk & b_{-\bk} \end{pmatrix} \begin{pmatrix} \ve_{\bk} & \io\D_{\bk} \\ -\io\D_{\bk} & -\ve_{\bk} \end{pmatrix} \begin{pmatrix} b_{\bk} \\ b^\dag_{-\bk} \end{pmatrix}\ ,
\label{kitd}
\eeq
%\beq
%H_K=\sum_{\bk \in B.Z.}2\abs{s_\bk}\left(a^\dag_\bk a_\bk-\frac{1}{2}\right)\ ,
%\label{kitd}
%\eeq
where $\ve_\bk = {\rm Re}{(s_\bk)}$, $\D_\bk = {\rm Im}(s_\bk)$, $s_\bk=K(1+e^{\io\bk\bn_1}+e^{\io\bk\bn_2})$, and the primitive vectors $\bn_1$ and $\bn_2$ are defined in Fig.~\ref{fig5}. Thus the ground state energy is $E_0 = -\sum_{\bk} \abs{s_\bk}$~\cite{kitaevAP06,knollePRL14}. %Our goal is to use Eq.~\eqref{kitd} to derive the susceptibility in order to extract the noise correction emerging in our proposed bilayer system when the QSL material is a gapless Kitaev spin liquid compound.
%and performing a Bogoliubov transformation $b_{\bk} = \cos{\thi_\bk} a_\bk + \io \sin{\thi_\bk} a^\dag_{-\bk}$, where $\tan{2 \thi_\bk} = -{\rm Im}(s_\bk)/{\rm Re}{(s_\bk)}$~\cite{knollePRL14}, Eq.~\eqref{kith} diagonalizes to

The isotropic Kitaev spin liquid offers a unique opportunity in that only on-site and nearest neighbor spin correlations contribute~\cite{baskaranPRL07}. Therefore we use Eq.~\eqref{noj} in calculating the noise and include both on-site and nearest neighbor contributions under the assumption that the envelope $\sinc^2(k_F|\br_\bi-\br_\bj|)$ does not noticeably vary over the lattice constant $a$. We then find that the total susceptibility $\chi(\nu)=\sum_{\bi\bj}[\chi_{\bi \bj}^{+-}(\nu) + \chi_{\bi \bj}^{-+}(\nu)]$ becomes
\begin{multline}
\chi(\nu)=-8\io N_u\int dt~\Big(\langle S^x_{A0}(t) S^x_{A0}(0) \rangle\\
+\langle S^x_{A0}(t) S^x_{B0}(0) \rangle \Big)\,e^{\io \nu t}\, ,
\label{kits}
\end{multline}
where we invoke the translation invariance of the isotropic spin liquid to select the $\br = 0$ unit cell.

The above two dynamical spin correlation functions can be written as~\cite{baskaranPRL07}
\begin{align}
\label{sxsx}
\begin{aligned}
\big< S^x_{A0}(t) S^x_{A0}(0) \big> &= \big< e^{\io H_K t/\hbar} c_{A0} e^{-\io (H_K + V_x)t/\hbar} c_{A0} \big>_K\, \\
\big< S^x_{A0}(t) S^x_{B0}(0) \big> &= -\io \big< e^{\io H_K t/\hbar} c_{A0} e^{-\io (H_K + V_x)t/\hbar} c_{B0} \big>_K\ , 
\end{aligned}
\end{align}
where the subscript $K$ for the expectation values indicates they are taken with respect to the ground state of $H_K$, and the bond potential reads $V_x = -2\io K c_\bi c_\bj$ with $\bi$ and $\bj$, again, representing the $A$ and its adjacent $x$-bond connected $B$ site, respectively~\cite{baskaranPRL07,knollePRL14}. 

The form of the correlators in Eq.~\eqref{sxsx} can be understood by taking into consideration the effect of operator $S^\g_{\bi}$ on an eigenstate. In addition to adding a single Majorana fermion $c_\bi$ at site $\bi$, the spin operator also adds one $\p$ flux each to the two plaquettes sharing the $\g$-bond emanating from $\bi$. The bond potential $V_x$ represents the insertion of this flux pair, and Eq.~\eqref{sxsx} calculates the dynamic rearrangement of the Majorana fermions at time $t$ following the sudden appearance of the fluxes at $t=0$. Incidentally, Baskaran {\em et al}. showed that this quench problem is equivalent to an exactly solvable x-ray edge problem~\cite{baskaranPRL07}. 
%We see from Eq.~\eqref{kits} that the potential $V_x$ on the $x$ bond at site $A0$ enters as a quench that suddenly changes the bond fermion number by 1, which is equivalent to injecting one flux each in the plaquettes adjacent to the altered bond. 

It was later shown by Knolle {\em et al}.~\cite{knollePRL14} that due to a vanishing Majorana density of states in the Kitaev model, slowly switching on the bond potential in the infinite past and then switching it off in the infinite future | what may be called the adiabatic approximation | replicates the quench dynamics in the low energy limit. Thus, as we desire a probe of the low energy density of states, we perform the calculation under this approximation, meaning we re-write Eqs.~\eqref{sxsx} so that the correlation functions are taken with respect to the Hamiltonian $H_x = H_K + V_x$, i.e., the Hamiltonian with a flux pair present, or equivalently one flipped $x$-bond. As a result, we evaluate 
\begin{align}
\label{skxx}
\begin{aligned}
\big< S^x_{A0}(t) S^x_{A0}(0) \big> &\approx e^{\io E_0 t/\hbar}\big< c_{A0}e^{-\io H_x t/\hbar}c_{A0} \big>_x \\
\big< S^x_{A0}(t) S^x_{B0}(0) \big> &\approx -\io e^{\io E_0 t/\hbar}\big< c_{A0} e^{-\io H_x t/\hbar} c_{B0} \big>_x \, ,
\end{aligned}
\end{align} 
where the subscript $x$ on the correlators explicitly indicates that they are now taken with respect to the ground state of $H_x$. Time-evolving the Majorana operators in Eqs.~\eqref{skxx} in the Heisenberg picture $c_\bi(t) = e^{\io H_x t/\hbar}c_\bi e^{-\io H_x t/\hbar}$, we write
\begin{align}
\label{skxxe}
\begin{aligned}
\big< S^x_{A0}(t) S^x_{A0}(0) \big> &= e^{-\io \D_F t/\hbar}\big< c_{A0}(t)c_{A0}(0) \big>_x \\
\big< S^x_{A0}(t) S^x_{B0}(0) \big> &= -\io e^{-\io \D_F t/\hbar}\big< c_{A0}(t)c_{B0}(0) \big>_x \, ,
\end{aligned}
\end{align} 
where the two-flux gap energy $\D_F \approx 0.26K$ is the energy required to insert the fluxes~\cite{kitaevAP06}.

%using the real-time (real-frequency) Keldysh diagrammatic formalism. Here, the Keldysh $s$-matrix reads
%\beq
%\label{vbp}
%S_K(-\infty,-\infty)=\exp\curly{-\frac{\io}{\hbar} \int_{c_K} dt'~V_x(t')}\ ,
%\eeq
%the time integrals are performed over the Keldysh time-loop contour $c_K$, and $\hT$ is the time ordering operator on the contour. The superscripts on the time arguments $+$ and $-$ label the forward and backward portions of the Keldysh contour, respectively. We perform the calculation connecting Eq.~\eqref{skxx} and Eq.~\eqref{susk} explicitly in Appendix~\ref{apD}. The total susceptibility, Eq.~\eqref{kits}, then becomes
%\begin{multline}
%\label{susk}
%\chi(\nu)=32\p\io\hbar KN\frac{1}{N_u^2}\sum_{\bp\bq}\cos(\bq \boldsymbol{n}_2+2\thi_\bq)n_F(-\nu)\\
%\times\frac{|s_\bq|}{|s_\bp|^2-|s_\bq|^2}\left[\de(\hbar\nu-2|s_\bq|)\right.+\de(\hbar\nu+2|s_\bq|)\\
%-\de(\hbar\nu-2|s_\bp|)-\left.\de(\hbar\nu+2|s_\bp|)\right]\, ,
%\end{multline}
%where $n_F(\nu)=(e^{\be\hbar\nu}+1)^{-1}$ is the Fermi-Dirac distribution.

%\begin{align}
%\label{skxxb}
%\begin{aligned}
%\big< S^x_{A0}(t) S^x_{A0}(0) \big> &= e^{-\io \D_F t} \square{\big< b_0(t) b^\dag_0(0) \big>_x + \big< b^\dag_0(t) b_0(0) \big>_x} \\ 
%\big< S^x_{A0}(t) S^x_{B0}(0) \big> &= e^{-\io \D_F t}\square{\big< b_0(t) b^\dag_0(0) \big>_x - \big< b^\dag_0(t) b_0(0) \big>_x}
%\end{aligned}
%\end{align}

Replacing the Majorana operators in the correlation functions with the complex fermions $b_\bk$ allows us to re-write the total susceptibility Eq.~\eqref{kits} as
\beq
\chi(\nu)=-16\io N_u\int dt e^{\io(\nu-\D_F/\hbar)t} \big<b_0(t) b^\dag_0(0) \big>_x\ .
\eeq
The last correlator can be obtained by solving a Dyson equation, details of which can be found in Appendix~\ref{apD}. Utilizing the result, we produce (at zero temperature)
\beq
\label{kitsf}
\chi(\nu)=\frac{16\hbar\thi(\hbar\nu-\D_F)[g_{11}^R(\hbar\nu-\De_F)-g_{11}^A(\hbar\nu-\De_F)]}{\square{1+\frac{4K}{N_u}g_{11}^R(\hbar\nu - \D_F)}\square{1+\frac{4K}{N_u}g_{11}^A(\hbar\nu - \D_F)}}\ ,
\eeq
where the local retarded Green function $g^R_{11}(\w)$ is derived in Appendix~\ref{apD} and is given by
\beq
\label{rgf}
g_{11}^R(x) = \sum_\bk\frac{x+2\ve_\bk}{\round{x+\io0^+}^2-4\abs{s_\bk}^2}=g_{11}^{A*}(x)\ .
\eeq
The ac noise correction generated in the metal by proximity to a gapless Kitaev spin liquid can now be calculated by utilizing Eq.~\eqref{kitsf} in Eq.~\eqref{noi} in the quantum limit. We find
\begin{multline}
\label{noik}
S_s(\W,0)=\frac{32\io}{\hbar}\left( \frac{\cJ v_0 mk_F}{2 \pi^2 \hbar} \right)^2\int d\ve(\hbar\W-\ve)\thi(\hbar\W-\ve)\\
\times\frac{\thi(\ve-\D_F)[g_{11}^R(\ve-\De_F)-g_{11}^A(\ve-\De_F)]}{\square{1+\frac{4K}{N_u}g_{11}^R(\ve-\D_F)}\square{1+\frac{4K}{N_u}g_{11}^A(\ve-\D_F)}}\ ,
\end{multline}
which can be measured electrically as ac resistance via Eq.~\eqref{fdt}. %We will examine this result numerically in order to extract the low temperature behavior. \frac{2|s_\bq|}{\sinh{2\be|s_\bq|}}-\frac{2|s_\bp|}{\sinh{2\be|s_\bp|}}
\begin{figure}
\includegraphics[width=.7\linewidth]{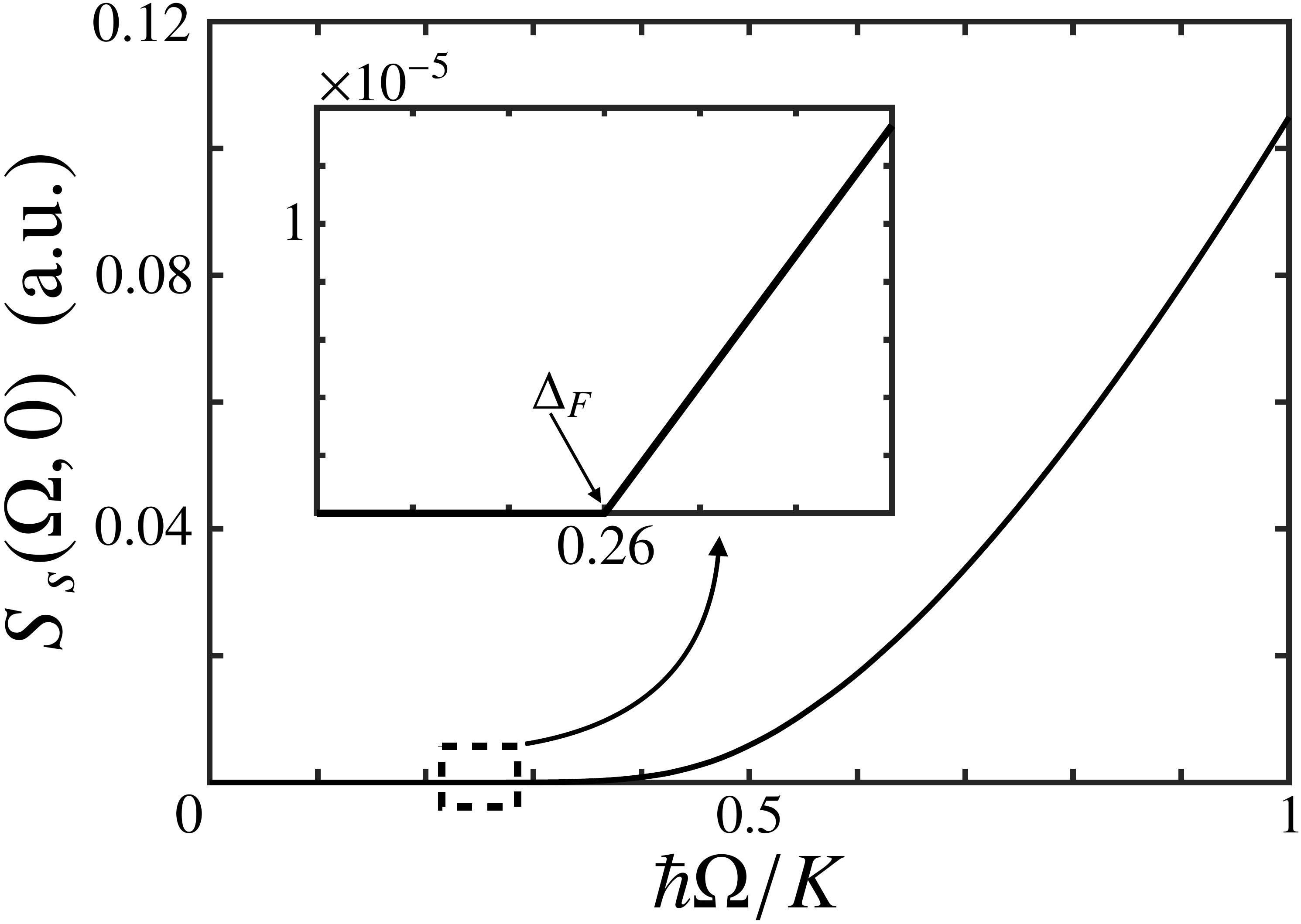}
\caption{A numerical plot showing $S_s(\W,0)$ for the Kitaev model as a function of ac frequency normalized by the isotropic interaction strength, $K$. The main plot shows the computed sum, and the inset zooms in on the low frequency regime near the gap energy. The graph is identically zero until a critical frequency equal to the gap energy is achieved, $\hbar\W_c = \D_F \approx 0.26 K$, at which point it becomes possible to inject flux pairs into the QSL and noise enhancement commences. Here, $N_u=4\times10^4$ has been used.}
\label{fig6}
\end{figure}

Figure~\ref{fig6} depicts our numerical evaluation of Eq.~\eqref{noik} in the main plot, with the inset showing a zoomed in plot of the region immediately adjacent to the critical frequency $\W_c=\D_F/\hbar$ on either side. The figure displays our primary result for the gapless Kitaev spin liquid compound, namely, that even in the ostensibly gapless case the ac noise correction generated in the metal is suppressed until frequencies greater than the two-flux gap, $\D_F$. Equivalently, via the FDT, measurement of the ac resistance enhancement in the metal near the critical frequency $\W_c$ should expose the two-flux gap when measured utilizing the proposed heterostructure. 

That a completely suppressed region occurs seems odd given that the fermion spectrum $s_\bk$ is gapless, however while inserting a fermion into the lattice requires no extra energy input, it also requires the addition of a flux in each of the adjacent plquettes. These fluxes are gapped, and so an indirect gap appears even for the fermions. This outcome has been noted previously~\cite{knollePRL14,knollePRB15}, however the proposed bilayer system offers a method for quantifying the flux gap experimentally.

\section{Discussion and Conclusions}
\label{discon}
Equation~\eqref{noi} is a general relationship that quantifies the noise enhancement present in a normal metal adjacent to an insulating magnet lacking long ranged magnetic order as a result of the coupling to that magnet, i.e., as a result of spin fluctuations across the interface. We assume that in the quantum limit the base ac resistance of a particular strongly spin-orbit coupled metal layer is known and constant, and as a result that background quantum noise can be accounted for and removed.
%In the quantum limit | when the sample is cooled to as low a temperature as possible | background thermal noise in the metal is strongly suppressed, and the voltage fluctuations present in the metal are entirely due to the presence of the QSL candidate material and any background quantum noise. We assume that for temperatures in the quantum limit the base resistance is essentially constant and known in a given strongly spin orbit coupled normal metal, e.g., Pt~\cite{hallNBS68}, Ta, and W~\cite{desaiJPC84}, and as a result of Eq.~\eqref{fdt} any quantum background noise linear in frequency that arises can be accounted for and removed.
This exposes the portion of the voltage noise power spectrum that quantifies the presence of the affixed quantum magnet, or more specifically the QSL candidate material. So, by measuring the ac frequency dependence of the spin fluctuations in the metal layer, it should be possible to compare the various QSL candidate materials in order to examine gapped and gapless models, and the low energy spin excitations of these exotic phases of matter in general. Now we will connect the models we have examined to real materials.%: herbertsmithite, $\kappa$-type organic compounds, and Kitaev compounds.

%It connects to resistance through the normal metal in the linear response regime via the FDT, as shown in Eq.~\eqref{fdt}.

The compound herbertsmithite, ZnCu$_3$(OH)$_6$Cl$_2$, evinces no long range magnetic order down to temperatures of at least $50$ mK, which is four orders of magnitude lower than its exchange coupling $J/k_B \sim 200$ K~\cite{heltonPRL07}, suggesting the presence of a QSL phase. Additionally, the Cu$^{2+}$ ions form a perfect kagom{\'e} lattice, and therefore herbertsmithite is modeled, to a first approximation, by the HKLM. Experimental treatments of herbertsmithite include NMR~\cite{olariuPRL08,fuSCI15}, neutron scattering~\cite{devriesPRL09,heltonPRL10}, and susceptibility~\cite{bertPRB07} studies, most of which indicate that ground state excitations should be gapless. However, numerical DMRG studies show that the ground state of herbertsmithite should be a gapped spin liquid with $Z_2$ topological order~\cite{yanSCI11}, and the more recent NMR measurements indicate a gapped ground state as well~\cite{fuSCI15}. Additional complications arise due to the fact that the low energy spectrum appears to be dominated by impurity spins~\cite{hanPRB16} that occur as a result of Cu$^{2+}$ ions replacing some non-magnetic Zn$^{2+}$ in the transition metal sites between kagom{\'e} layers~\cite{freedmanJACS10}. We have shown that our proposed heterostructure can address some of these issues. For instance, the presence of a gap energy will be indicated by a lack of frequency scaling in the voltage noise power spectrum measured in the heavy metal layer in the quantum limit up to a critical frequency of $\hbar\W_c = 2\D_s$. Technical issues notwithstanding, herbertsmithite remains one of the most promising QSL candidate material on offer, and our proposed technique adds a tool for further investigation.

The primary organic salt candidate QSL material that has been modeled as a spinon fermi sea coupled to an emergent $U(1)$ gauge field is $\kappa$-(BEDT-TTF)$_2$-Cu$_2$(CN)$_3$, henceforth $\kappa$-ET~\cite{leePRL05,motrunichPRB05}. It is an antiferromagnetic weak Mott insulator, where structural dimers possessing a single spin-$1/2$ degree of freedom arise and form an approximately triangular lattice that becomes geometrically frustrated when cooled. The primary experimental evidence of QSL like behavior comes from thermal and NMR measurements; in the specific heat versus temperature, for example, a large linear term is observed~\cite{yamashitaNATP08a,yamashitaNATP08b,yamashitaNATC11}, and NMR measurements show a lack of long range magnetic ordering down to temperatures of $32$ mK, approximately $4$ orders of magnitude lower than the exchange coupling of $J/k_B \sim 250$ K~\cite{shimizuPRL03}. The necessity of working at very low temperatures when probing the low energy density of states of $\kappa$-ET and similar compounds has caused some doubt as to the reliability of specific heat measurements, however, due to the difficulty of properly accounting for nuclear contributions. Nevertheless, other triangular lattice compounds exist, and the recent discoveries of possible QSL phases in YbMgGaO$_4$~\cite{kimchiPRX18} and NaYbO$_2$~\cite{leiPRB19}, neither of which order down to temperatures of at least $70$ mK, provide new opportunities to study previously unknown QSL candidate materials | opportunities the proposed heterostructure is intended to capitalize upon by providing a clear prediction for the frequency scaling of the modification to the ac resistance enhancement in the metal. We have shown that in its QSL state, a compound that can be modeled as a spinon fermi sea coupled to a $U(1)$ gauge field should evince a correction to the ac resistance measured across a coupled heavy metal that goes as $\W^{4/3}$, which would therefore shed light on the low energy density of states directly. The heterostructure proposed here would avoid the issue of contamination by signals due to nuclear spins entirely, and provide a powerful tool for probing the low energy excitations of more recently discovered candidate QSL materials as well.

Finally, we mention the iridate honeycomb materials~\cite{chaloupkaPRL13} and $\al$-RuCl$_3$~\cite{plumbPRB14} in the context of Kitaev compounds accessible to our bilayer technique. Initially, iridates like $\al$-Na$_2$IrO$_3$ and $\al$-Li$_2$IrO$_3$ were realized to have Ir$^{4+}$ ions, with effective $j=1/2$, arranged in a honeycomb lattice. While these particular compounds were found to order at low temperatures~\cite{liuPRB11,choiPRL12,kubotaPRB15}, this class of materials does exhibit bond-directional Kitaev interactions~\cite{chunNAT15}. Thus the discovery of a variant that does not magnetically order, H$_3$LiIr$_2$O$_6$, is interesting~\cite{kitagawaNAT18}; in particular, it would be an excellent test of this material to see if the measured resistance enhancement in a coupled heavy metal film in fact behaves as we predict should be the case for Kitaev spin liquid compounds. A second material, $\al$-RuCl$_3$, is also known to evince similar structural properties to the iridates, with magnetic Ru$^{3+}$ ions forming a honeycomb lattice. As with most iridates, $\al$-RuCl$_3$ orders at low temperatures~\cite{majumderPRB15}, though interesting Kitaev spin liquid-like behavior occurs before ordering~\cite{banerjeeNATM16}. However, while $\al$-RuCl$_3$ exhibits Kitaev-like physics prior to freezing, questions have arisen as to whether or not Kitaev physics are actually responsible for that appearance, rather than more conventional physics from the perspective of magnon-like excitations~\cite{winterNATC17}. We have shown that the proposed bilayer could perhaps address this controversy directly, as the presence of Kitaev spin liquid physics in the $\al$-RuCl$_3$ layer should result in characteristic suppression of the ac resistance enhancement in the coupled metal, particularly suppression below any flux gap present. We therefore believe that our proposed QSL-to-metal bilayer could be helpful to the search for a material exhibiting a true Kitaev spin liquid state.

%The third and final QSL model we have under consideration is the Kitaev honeycomb model. The Kitaev model has been the subject of much study due to the fact that it is exactly solvable in the spin liquid phase, in addition to its possible applicability as the basis for a minimal model for various materials, including Na$_2$IrO$_3$~\cite{chaloupkaPRL13}, Li$_2$IrO$_3$~\cite{chaloupkaPRL13}, and $\al$-RuCl$_3$~\cite{plumbPRB14}. It is now clear that these compounds in fact exhibit long range magnetic ordering at low temperatures~\cite{liuPRB11,choiPRL12,kubotaPRB15,majumderPRB15}. While this is unfortunate, and no true Kitaev spin liquid materials have yet been found, the search is ongoing, and it is hoped that one or all of these materials might somehow be driven into a QSL state by, e.g., field, pressure, or chemical modifications~\cite{savaryRPP16}. The Kitaev model itself is therefore interesting and worthy of study in the context of our proposed system, and so we include it here.

In conclusion, we have proposed a heterostructure composed of a QSL candidate material overlaid with a strongly spin-orbit coupled heavy metal film as a viable probe of QSL ground states that can perhaps alleviate some of the currently outstanding controversies. Our proposal marries concepts from spintronics and quantum magnetism in order to use the well-understood physics of equilibrium noise in a new context, namely the search for, and categorization of, QSL ground states in various materials. The theory advanced here indicates that an equilibrium measurement, i.e., a measurement taken within the linear response regime, of the ac resistance in the normal metal layer will provide information about the low energy density of states of the QSL material under observation due to a connection via the FDT and Eq.~\eqref{fdt} to the noise power spectrum we have calculated. The system we propose will have wide ranging applicability to probing insulating quantum magnets in general, and should provide an interesting method of examining and categorizing QSL candidate materials in particular. We show that the proposed bilayer should be able to extract any gap energy present in the HKLM, that fermions coupled to an emergent $U(1)$ gauge field should see a sub-dominant frequency dependent correction arising in the ac resistance that goes as $\W^{4/3}$, and that the bare Kitaev model in the gapless spin liquid phase should nevertheless evince gapped behavior extricable through ac resistance measurements across the coupled metal layer. 

We have restricted our analysis to bulk conversion effects in the metal layer, so in future works it could be interesting to consider interfacial effects that may affect conversion; in this vein examining different metal compounds like CuBi~\cite{niimiPRL12} with low intrinsic spin-orbit coupling could be illuminating. Additionally, it would be interesting to expand this analysis to include extensions to the QSL models such as Dzyaloshinskii-Moriya interactions, or Heisenberg and $\G$ terms in the Kitaev model, and especially to include disorder in the analysis, as disorder appears to be highly influential when attempting to discriminate between QSL and non-QSL ground states.

\acknowledgments
We thank Gerrit E. W. Bauer and Yaroslav Tserkovnyak for discussions, and acknowledge support from The Research Foundation of The City University of New York, Fund \#90922-07 10.

%%%%%%%%%%%%%%%%%%%%%%%%%%%%%%%%%%%%%%%%%%%%%%%%
\appendix

\section{The Heisenberg kagom{\'e} lattice model calculation}
\label{apB}

The three basis vectors for the unit cell shown in Fig.~\ref{fig2} are
\beq
\boldsymbol{\hat{e}}_1 = \frac{a}{2} \left( 1,\sqrt{3} \right)\ ,\ \ \boldsymbol{\hat{e}}_2 = \frac{a}{2} \left( 1,-\sqrt{3} \right)\ ,\ \ \boldsymbol{\hat{e}}_3 = a \left( -1,0 \right)\ , 
\eeq
where $a$ is the lattice constant. These unit vectors allow us to Fourier transform the Hamiltonian into momentum space, where we write the momenta as $k_i = \boldsymbol{\hat{e}}_i \cdot \bk$. Equation~\eqref{hklm} then reads
\beq
H_{HKLM} = \sum_{\bk} \Psi^\dag_\bk\begin{pmatrix} \la & \mathrm{C}^\dag_\bk \\ \mathrm{C}_\bk & \la \end{pmatrix}\Psi_\bk\ ,
\label{coma}
\eeq
where $\Psi(\bk) = [b_1(\bk),b_2(\bk),b_3(\bk),b^\dag_1(-\bk),b^\dag_2(-\bk),b^\dag_3(-\bk)]^T$ is the vector of particle-hole boson operators for each of the three sites of the unit cell, and the matrix $\mathrm{C}_\bk$ is a traceless $3\times3$ matrix with components
\begin{align}
c_{12} &= JQ_1 e^{ik_1} + JQ_2 e^{-ik_1}=-c_{21}^*\nn \\
c_{23} &= JQ_1 e^{ik_2} + JQ_2 e^{-ik_2}=-c_{32}^*\nn\\
c_{31} &= JQ_1 e^{ik_3} + JQ_2 e^{-ik_3}=-c_{13}^*\ .\nn
\end{align}
Here, $Q_1,Q_2\in\mathbb{R}$ are the two distinct expectation values of $Q_{ij}$ (see Fig.~\ref{fig2}), and as covered in the main text we explicitly consider the regime $Q_1 = Q_2 = Q$. 

Equation~\eqref{coma} can be diagonalized with a unitary matrix $U$, leading to 
\beq
H_{HKLM} = \sum_{\bk} \g^\dag_\bk D \g_\bk
\eeq 
with the rotated boson operators $\g_\bk = U^\dag \Psi_\bk$. The diagonalized matrix $D = \diag\round{\e_1,\e_2,\e_3,\e_1,\e_2,\e_3}$, where $\e_1=\la$ and
\beq
\e_2=\e_3= \sqrt{\la^2- 4J^2Q^2[\cos^2\round{k_1} + \cos^2\round{k_2} + \cos^2\round{k_3}]}\ .\nn
\eeq

To calculate the total spin susceptibility $\chi(\nu)=\sum_{\bi}[\chi^{+-}_{\bi \bi}(\nu)+\chi^{-+}_{\bi \bi}(\nu)]$, we re-express the sum over all lattice sites $\bi$ by reformulating it as $\sum_\bi = N_u \sum_{n=1}^3$, where $N_u$ is the number of unit cells and $n$ indexes each position within a single unit cell. We then obtain
%Under this consideration, we diagonalize Eq.~\eqref{hklm} while maintaining the commutation relations for the boson operators,  for details. 
\begin{multline}
\label{herbx}
\chi(\nu)=-\io\frac{2\pi\hbar N_u}{N_u^2}\sum_{\bk,\bq}\sum_{n=1}^3\sum_{m,l=1}^6 U^{\bk*}_{nm}U^\bk_{nm}U^{\bq*}_{\bar nl}U^{\bq}_{\bar nl} \\
\times \Big[ \de(\hbar\nu+\xi_{\bk m}+\xi_{\bq l}) \cN_m(\bk) \cN_l(\bq) \\
+ \de(\hbar\nu-\xi_{\bk m}-\xi_{\bq l}) \cM_m(\bk) \cM_l(\bq) \Big]\ ,
\end{multline}
% $V^{\bk\bq}_{nml}=U^{\bk*}_{nm}U^\bk_{nm}U^{\bq*}_{\bar nl}U^{\bq}_{\bar nl}$,
where $U^\bk_{ml}$ is the $6\times6$ matrix that diagonalizes Eq.~\eqref{hklm}, $\bar n=n+3$, and $\xi_{\bk}=(\e_{1}, \e_{2}, \e_{3}, -\e_{1}, -\e_{2}, -\e_{3})$ is the vector of energy eigenvalues with the last three negated. We also introduce vectors of distribution functions to display this more compactly,
\begin{align}
\cN(\bk) &= \left[ n_B(\e_1), n_B(\e_2), n_B(\e_3), \bar n_B(\e_1), \bar n_B(\e_2), \bar n_B(\e_3) \right] \nn\\
\cM(\bk) &= \left[\bar n_B(\e_1), \bar n_B(\e_2), \bar n_B(\e_3), n_B(\e_1), n_B(\e_2), n_B(\e_3)\right]\ , \nn
\end{align}
where $n_B(x)=(e^{\be x}-1)^{-1}$ and $\bar n_B=1+n_B$. 

With this, we can express the ac, finite temperature noise created by the proximate QSL by using Eq.~\eqref{herbx} in Eq.~\eqref{noi},
\begin{multline}
\label{hsnoi}
S_s(\W,T)=\frac{4\pi}{\hbar N_u}\left( \frac{\cJ v_0mk_F}{2 \pi^2 \hbar} \right)^2 \sum_{\bk,\bq}\sum_{n,m,l} U^{\bk*}_{nm}U^\bk_{nm}U^{\bq*}_{\bar nl}U^{\bq}_{\bar nl}\\
\times \left[ \frac{-\xi_{\bk m}-\xi_{\bq l}-\hbar\W}{e^{-\be(\xi_{\bk m}+\xi_{\bq l}+\hbar\W)}-1}\right. \cN_m(\bk) \cN_l(\bq) \\
+ \left.\frac{\xi_{\bk m}+\xi_{\bq l}-\hbar\W}{e^{\be(\xi_{\bk m}+\xi_{\bq l}-\hbar\W)}-1} \cM_m(\bk) \cM_l(\bq)\right].
\end{multline}
The zero temperature limit of this result gives Eq.~\eqref{hsn0t}.

\begin{widetext}
\section{$U(1)$ gauge field correction to the susceptibility}
\label{apC}
In this Appendix we examine the steps involved in proceeding from Eq.~\eqref{u1int} to Eq.~\eqref{u1sus}. When considering a single gauge propagator, there are three contributing diagrams as shown in Fig.~\ref{fig4}. Figure~\ref{fig4}a and~\ref{fig4}b, where the gauge propagator does not cross the particle-hole bubble, are self-energy corrections to a fermion line, and Fig.~\ref{fig4}c, where the propagator does cross the bubble, is a vertex correction. Maintaining the gauge invariance of the susceptibility requires that all three diagrams are included in the calculation~\cite{kimPRB94,balentsCM19}.

We begin this section with the action given in Eq.~\eqref{fu1a} of the main text, and place it explicitly on the Keldysh contour in the Keldysh basis. We write the components of the action zeroth and first order in the gauge field as
\begin{align}
S_{0} &=  \int dt \int dt' \sum_{\bk \s} \begin{pmatrix}[1.4] \bc^1_{\bk \s}(t) & \bc^2_{\bk \s}(t) \end{pmatrix} \begin{pmatrix}[1.4] g^R_{\bk \s}(t-t') & g^K_{\bk \s}(t-t') \\ 0 & g^A_{\bk \s}(t-t') \end{pmatrix}^{-1} \begin{pmatrix}[1.4] c^1_{\bk \s}(t') \\ c^2_{\bk \s}(t') \end{pmatrix}\ , \\
S_{1} &= -\frac{1}{2\hbar \sqrt{2 \sA}} \int dt \sum_{\bk \bk' \s \xi} \varv^\xi_{\bk + \bk' s} \begin{pmatrix}[1.4] \bc^1_{\bk \s}(t) & \bc^2_{\bk \s}(t) \end{pmatrix} \begin{pmatrix}[1.4] a^{\xi,c}_{\bk-\bk'}(t) & a^{\xi,q}_{\bk - \bk'}(t) \\ a^{\xi,q}_{\bk - \bk'}(t) & a^{\xi,c}_{\bk-\bk'}(t) \end{pmatrix} \begin{pmatrix}[1.4] c^1_{\bk' \s}(t) \\ c^2_{\bk' \s}(t) \end{pmatrix}\ ,
\end{align}
where $\varv_{\bk s} = \hbar \bk / m_s$. We have split the gauge field into its classical and quantum components denoted by the $c$ and $q$ superscripts, and split the fermion fields into their 1 and 2 components in accordance with Ref.~\onlinecite{kamenevBOOK11}. Here, $S_1$ is the RAK basis equivalent to what we have called $S_{int}$ in the main text.

Expanding Eq.~\eqref{u1int}, and restricting ourselves to terms with a single gauge propagator, it is possible to write three expressions of the form $\chi^{(2)}_n(\bp,\nu) = \chi^{+- (2)}_n(\bp,\nu) + \chi^{-+ (2)}_n(\bp,\nu)$, each corresponding to one of the diagrams in Fig.~\ref{fig4}:
\begin{align}
\chi^{(2)}_a(\bp,\nu) = \frac{1}{4 \hbar^2 \sA^2} \sum_{\bk \bq} \sum_{\xi \xi'} \sum_{\eta_1 \eta_2}&\int \frac{d \W}{2 \pi} \int \frac{d \w}{2 \pi} \eta_1 \eta_2 D^{\eta_1 \eta_2}_{\xi \xi'}(-\bq,-\w) \varv^\xi_{2\bk + 2\bp + \bq s} \varv^{\xi'}_{2\bk + 2\bp + \bq s} \nn \\
&~~~~~~~~~~~~~~~~~~~~~~~~~~~~~~\times \square{g^{- \eta_1}_{\bk + \bp}(\W + \nu) g^{\eta_1 \eta_2}_{\bk + \bp + \bq}(\W + \nu + \w) g^{\eta_2 +}_{\bk + \bp}(\W + \nu) g^{+-}_{\bk}(\W)}, \label{chia} \\
\chi^{(2)}_b(\bp,\nu) = \frac{1}{4 \hbar^2 \sA^2} \sum_{\bk \bq} \sum_{\xi \xi'} \sum_{\eta_1 \eta_2}&\int \frac{d \W}{2 \pi} \int \frac{d \w}{2 \pi} \eta_1 \eta_2 D^{\eta_1 \eta_2}_{\xi \xi'}(-\bq,-\w) \varv^\xi_{2\bk + 2\bp + \bq s} \varv^{\xi'}_{2\bk + 2\bp + \bq s}\nn \\
&~~~~~~~~~~~~~~~~~~~~~~~~~~~~~~\times \square{g^{- +}_{\bk}(\W + \nu) g^{+ \eta_1}_{\bk + \bp}(\W) g^{\eta_1 \eta_2}_{\bk + \bp + \bq}(\W + \w) g^{\eta_2 -}_{\bk + \bp}(\W)},\label{chib} \\
\chi^{(2)}_c(\bp,\nu) = \frac{1}{4 \hbar^2 \sA^2} \sum_{\bk \bq} \sum_{\xi \xi'} \sum_{\eta_1 \eta_2}&\int \frac{d \W}{2 \pi} \int \frac{d \w}{2 \pi} \eta_1 \eta_2 D^{\eta_1 \eta_2}_{\xi \xi'}(-\bq,-\w) \varv^\xi_{2\bk + \bq s} \varv^{\xi'}_{2\bk + 2\bp + \bq s}\nn \\
&~~~~~~~~~~~~~~~~~~~~~~~~~~~~~~\times \square{g^{- \eta_1}_{\bk + \bp}(\W + \nu) g^{\eta_1 +}_{\bk + \bp + \bq}(\W + \nu + \w) g^{+ \eta_2}_{\bk + \bq}(\W + \w) g^{\eta_2 -}_{\bk}(\W)}. \label{chic}
\end{align}
Here $\eta_1,\eta_2 = \pm$ represent Keldysh indices for the forward and backward contours, and bare $\pm$ superscripts on the Green functions also represent the externally fixed forward and backward Keldysh indices. The gauge propagator, which can be derived from the Matsubara formalism in the Coulomb gauge by analytic continuation of the zero temperature result, is given as
\beq
D^{R,A}_{\xi \xi'}(\bq,\w) = -\round{\de_{\xi \xi'} - \frac{ q_\xi q_{\xi'}}{q^2}} d^{R,A}_{\bq}(\w),~~{\rm with}~~d^{R,A}_{\bq}(\w) = \frac{1}{N} \frac{1}{\chi_d q^2 \mp \io \frac{\w E_F}{\varv_F q \pi \hbar^3}},
\eeq
where $E_F$ is the Fermi energy, and $\chi_d = (24\p \hbar m_s)^{-1}$ is the Landau diamagnetic susceptibility of the fermions.

We now introduce the fermion self-energy in order to represent Eqs.~\eqref{chia} and~\eqref{chib} in terms of the self-energy. Note at the outset that because the system we consider is in equilibrium, only the retarded spinon self-energy need be explicitly constructed; the advanced self-energy can be derived from the retarded term. The retarded one-loop self-energy can be calculated by considering the second-order correction to the retarded Green function using $S_1$ as a perturbation
\beq
g^{R(2)}_{\bk \s}(t,t') = \frac{\io}{2} \left< c^1_{\bk \s}(t) \bc^1_{\bk \s}(t') S^2_1 \right>\ .
\eeq
Extracting the self energy is then a matter of expanding this quantity and enforcing causality, resulting in
\beq
\S^R_{\bk \s}(t,t') = \frac{2 \io}{(4 \hbar)^2 \sA} \sum_{\bq} \sum_{\xi \xi'} \square{g^R_{\bk+\bq \s}(t-t') D^K_{\xi \xi'}(-\bq,t-t') + g^K_{\bk+\bq \s}(t-t') D^R_{\xi \xi'}(-\bq,t-t')} \varv^\xi_{2\bk + \bq s} \varv^{\xi'}_{2\bk + \bq s}\ . 
\eeq
Completing the $\xi$ and $\xi'$ sums and transforming to frequency space, we have
\beq
\label{se1}
\S^R_{\bk \s}(\W) = -\frac{\io}{2 \sA} \int \frac{d\w}{2\p} \sum_{\bq} \round{\frac{\abs{\bk \times \hbq}}{m_s}}^{2} \square{g^R_{\bk+\bq \s}(\W+\w) d^K_{-\bq}(-\w) + g^K_{\bk+\bq \s}(\W+\w) d^R_{-\bq}(-\w)}\ . 
\eeq
In general we expect $d_{-\bq}(-\w)$ to be dominated by small $\bq$, due to the form of the propagator. By selecting the coordinates $\bq = q_{\parallel} \hbk + \hbz \times \hbk q_\perp$ we can express the dispersion as
\beq
\z_{\bk+\bq} \approx \e_\bk - \mu + \frac{\hbar^2 q_\perp^2}{2 m_s} + \hbar \varv_{Fs} q_\parallel \ ,
\eeq
with $\e_\bk = \hbar^2 k^2/2m_s$. From this, we further expect to find momentum scaling as $q_\parallel \sim q^2_\perp \ll q_\perp$, i.e., where $\bq$ is essentially normal to $\bk$, which allows us to replace the momentum dependence of the gauge propagator with $q_\perp$ and write $\abs{\bk \times \hbq}^2 \approx k_{Fs}^2$ for momentum near the spinon Fermi surface. Therefore, by explicitly substituting in the spinon Green functions, Eq.~\eqref{se1} becomes
\begin{multline}
\S^R_{\bk \s}(\W) \approx -\frac{\io k_{Fs}^2 a_s}{2 m_s^2} \int \frac{d\w}{2\pi} \int \frac{dq_\perp}{2\pi} \int \frac{dq_\parallel}{2\pi} \\
\times \square{\frac{d^K_{-q_\perp}(-\w)}{\W + \w - \frac{\z_{\bk}}{\hbar} - \frac{\e_{q_\perp}}{\hbar} - \varv_{Fs} q_\parallel + \io \de} - 2\pi \io \de{\left(\W + \w - \frac{\z_{\bk}}{\hbar} - \frac{\e_{q_\perp}}{\hbar} - \varv_{Fs} q_\parallel \right)} \tanh{\round{\frac{\hbar(\W+\w)}{2k_B T}}} d^R_{-q_\perp}(-\w) } \ .
\end{multline}
At this point it is clear that the $q_\parallel$ integral poses no trouble, so we complete it and find
\beq
\label{sef}
\S^R_{\bk \s}(\W) \approx \frac{\varv_{Fs}a_s}{2 \hbar^2} \int \frac{d\w}{2\pi} \int \frac{dq_\perp}{2\pi} \square{-\frac{1}{2}\coth{\round{\frac{\hbar \w}{2k_B T}}} \square{d^R_{q_\perp}(-\w) - d^A_{q_\perp}(-\w)} + \tanh{\round{\frac{\hbar(\W+\w)}{2k_B T}}} d^R_{q_\perp}(-\w)}\ .
\eeq
Note that from this expression for the self-energy we can see that the $\bk$ dependence drops out entirely.

%Finally, by noting that we are primarily concerned with ${\rm Im}{\S^R_{\bk \s}(\W)}$ and evaluating further, we obtain
%\beq
%\label{self}
%\S^R_{\bk \s}(\W) = \frac{1}{\cN \chi_d \sA} \int_0^{\infty} \frac{d\w}{2\p} \sum_{\bq} \round{\frac{\abs{\bk \times \hbq}}{m_s}}^{2} \frac{ \frac{12 \w k_F^2}{\varv_F q} }{ q^4 + \round{\frac{12 \w k_F^2}{\varv_F q}}^2 } \curly{ \frac{n_B(\w) + n_F(\W + \w)}{\W + \w - \frac{\z_{\bk + \bq}}{\hbar} + \io \de} + \frac{1 + n_B(\w) - n_F(\W - \w)}{\W - \w - \frac{\z_{\bk + \bq}}{\hbar} + \io \de} },
%\eeq
%where $n_F$ and $n_B$ are the Fermi-Dirac and Bose-Einstein distributions respectively, and $\z_\bk = \e_\bk - \mu$ with $\e_\bk = \hbar^2 k^2/2m_s$ the dispersion and $\mu$ the Fermi surface.

Evaluating the sums over the Keldysh indices in the self-energy correction terms, Eqs.~\eqref{chia} and~\eqref{chib}, allows us to rewrite the expressions for diagrams $a$ and $b$ in terms of spinon Green's functions and the self energy:
\begin{align}
\chi^{(2)}_a(\bp,\nu) &= -\io \frac{1}{1-e^{-\be\hbar\nu}} \frac{1}{2 \sA} \sum_{\bk} \int \frac{d\W}{2 \p} \square{\tanh{\round{\frac{\hbar \W}{2 k_B T}}} - \tanh{\round{\frac{\hbar (\W + \nu)}{2 k_B T}}}} \nn \\
			       &~~~~~~~~~~\times \square{g^R_\bk(\W) - g^A_\bk(\W)} \square{ g^R_{\bk + \bp}(\W + \nu)\S^R_{\bk + \bp}(\W + \nu)g^R_{\bk + \bp}(\W + \nu) - g^A_{\bk + \bp}(\W + \nu)\S^A_{\bk + \bp}(\W + \nu)g^A_{\bk + \bp}(\W + \nu) }, \\
\chi^{(2)}_b(\bp,\nu) &= - \io \frac{1}{1-e^{-\be\hbar\nu}} \frac{1}{2 \sA} \sum_{\bk} \int \frac{d\W}{2 \p} \square{\tanh{\round{\frac{\hbar \W}{2 k_B T}}} - \tanh{\round{\frac{\hbar (\W + \nu)}{2 k_B T}}}} \nn \\
			       &~~~~~~~~~~\times \square{g^R_{\bk+\bp}(\W+\nu) - g^A_{\bk+\bp}(\W+\nu)} \square{ g^R_{\bk}(\W)\S^R_{\bk}(\W)g^R_{\bk}(\W) - g^A_{\bk}(\W)\S^A_{\bk}(\W)g^A_{\bk}(\W) }.
\end{align}
In the end we must sum together all three diagrams in order to maintain gauge invariance, which we begin by combining these two terms into $\chi^{(2)}_{ab}(\bp,\nu) = \chi^{(2)}_a(\bp,\nu) + \chi^{(2)}_b(\bp,\nu)$, and use the fact that any integral over only retarded or advanced terms vanishes,
%\beq
%\int d\z_{\bk}~g^R_{\bk}(\W) g^R_{\bk+\bp}(\W+\nu)\S^R_{\bk+\bp}(\W+\nu)g^R_{\bk+\bp}(\W+\nu) = \int d\z_{\bk}~g^A_{\bk}(\W) g^A_{\bk+\bp}(\W+\nu)\S^A_{\bk+\bp}(\W+\nu)g^A_{\bk+\bp}(\W+\nu) =0
%\eeq
to write $\chi^{(2)}_{ab}(\bp,\nu)$ compactly. The outcome is
\begin{multline}
\chi^{(2)}_{ab}(\bp,\nu) = \io \frac{m_s a_s}{2 \hbar} \frac{1}{1-e^{-\be\hbar\nu}} \int \frac{d\W}{2\p} \int^{\infty}_{-\mu} \frac{d\z_\bk}{2 \p} \int \frac{d\thi_\bk}{2\p} \square{\tanh{\round{\frac{\hbar \W}{2k_B T}}} - \tanh{\round{\frac{\hbar(\W + \nu)}{2k_B T}}}} \\
\times \square{ \frac{ g_\bk^A(\W) \square{\S_{\bk+\bp}^R(\W+\nu) - \S_{\bk}^A(\W) } g_{\bk+\bp}^R{(\W+\nu)} }{ \nu + \frac{\z_\bk}{\hbar} - \frac{\z_{\bk+\bp}}{\hbar} + \io \de } + \frac{ g_\bk^R(\W) \square{\S_{\bk+\bp}^A(\W+\nu) - \S_{\bk}^R(\W) } g_{\bk+\bp}^A{(\W+\nu)} }{ \nu + \frac{\z_\bk}{\hbar} - \frac{\z_{\bk+\bp}}{\hbar} - \io \de}}\ ,
\end{multline}
where we have represented the $\bk$ integral in polar coordinates, with $\thi_\bk$ the angle between $\bk$ and $\bp$, and performed a change of variables from $k$ to $\z_\bk$. Here it is important to point out two things: first, the Fermi surface $\mu$ is the largest energy scale in the problem, and so we extend the lower bound of the $\z_\bk$ integral to $-\infty$. Second, note that $\z_{\bk+\bp}-\z_{\bk} = \e_{\bp} + \hbar \varv_{Fs}p\cos{\thi_\bk} \approx \hbar \varv_{Fs}p\cos{\thi_\bk}$, where the last step is due to the expectation that small $p$ dominates, which means the $\z_\bk$ dependence drops out of the denominator in both terms. Thus there is no $\z_\bk$ dependence in either the denominator or the self-energies, and so we can complete the $\z_\bk$ integral by simply integrating over the spinon Green functions; this is the reason terms that only have retarded or advanced Green functions in them vanish. The result is
\begin{multline}
\chi^{(2)}_{ab}(\bp,\nu) = -\frac{ m_s a_s}{2 \hbar} \frac{1}{1-e^{-\be\hbar\nu}} \int \frac{d\W}{2\p} \int \frac{d\thi_\bk}{2\p} \\
\times \square{\tanh{\round{\frac{\hbar \W}{2k_B T}}} - \tanh{\round{\frac{\hbar(\W + \nu)}{2k_B T}}}} \square{ \frac{\S_{\bk+\bp}^R(\W+\nu) - \S_{\bk}^A(\W)}{ \round{\nu - \varv_{Fs}p\cos{\thi_\bk} + \io \de}^2} - \frac{ \S_{\bk+\bp}^A(\W+\nu) - \S_{\bk}^R(\W)}{ \round{\nu - \varv_{Fs}p\cos{\thi_\bk} - \io \de}^2}}\ .
\end{multline}
We can now substitute in the self energies represented in Eq.~\eqref{sef} to present the final step in combining these two diagrams,
\begin{align}
\label{chiab}
\chi^{(2)}_{ab}(\bp,\nu) = - \frac{ k_{Fs} a_s^2}{2\hbar^2} & \frac{1}{1-e^{-\be\hbar\nu}} \int \frac{d\W}{2\p} \int \frac{d\w}{2\pi} \int \frac{d\thi_\bk}{2\p} \int \frac{dq_\perp}{2\pi} \square{\tanh{\round{\frac{\hbar(\W + \nu)}{2k_B T}}} - \tanh{\round{\frac{\hbar \W}{2k_B T}}}} \nn \\
&\times \left\{ \frac{ \coth{\round{\frac{\hbar \w}{2k_B T}}} \square{d^R_{-q_\perp}(-\w) - d^A_{-q_\perp}(-\w)} - \tanh{\round{\frac{\hbar(\W + \nu + \w)}{2k_B T}}}d^R_{-q_\perp}(-\w) + \tanh{\round{\frac{\hbar(\W + \w)}{2k_B T}}}d^A_{-q_\perp}(-\w)}{ \round{\nu - \varv_{Fs}p\cos{\thi_\bk} + \io \de}^2 } \right. \nn \\
&+  \left. \frac{ \coth{\round{\frac{\hbar \w}{2k_B T}}} \square{d^R_{-q_\perp}(-\w) - d^A_{-q_\perp}(-\w) } + \tanh{\round{\frac{\hbar(\W + \nu + \w}{2k_B T}}}d^A_{-q_\perp}(-\w) - \tanh{\round{\frac{\hbar(\W + \w)}{2k_B T}}}d^R_{-q_\perp}(-\w)  }{ \round{\nu - \varv_{Fs}p\cos{\thi_\bk} - \io \de}^2 }   \right\}\ .
\end{align}

The third diagram corresponds to a vertex correction. We expect that $p \ll k$ dominates, and recall that the form of the gauge propagator restricts us to small $q$ as well. We express $\bq$ in components perpendicular and parallel to $\bk$ as above, and note that, via the same approximation as in Ref.~\onlinecite{balentsCM19}, near the Fermi surface we can write
\beq
-\sum_{\xi \xi'} \round{\de_{\xi \xi'} - \frac{ q_\xi q_{\xi'}}{q^2}} \varv^\xi_{2\bk + \bq s} \varv^{\xi'}_{2\bk + 2\bp + \bq s} \approx -\frac{4 \hbar^2 k_{Fs}^2}{m_s^2} \left( \frac{q_\perp^2}{q^2} + \frac{q_\parallel^2}{q^2} - \frac{q_\parallel q_\perp}{q^2} \right) \approx -4 \varv_{Fs}^2 \ .
\eeq
Above, the first approximation is from recalling that $p\ll k$ and the second is a result of $q_\parallel \sim q^2_\perp \ll q_\perp$. Therefore, summing over $\xi$ and $\xi'$ and expanding the Keldysh indices, Eq.~\eqref{chic} becomes
\begin{align}
\chi_{c}^{(2)}(\bp,\nu) &= -\frac{ \varv_F^2}{2 \hbar^2 \sA^2} \frac{1}{1-e^{-\be\hbar\nu}} \sum_{\bk \bq} \int \frac{d \W}{2 \pi} \int \frac{d \w}{2 \pi} \square{\tanh{\round{ \frac{\hbar(\W + \nu)}{2 k_B T} }} - \tanh{\round{ \frac{\hbar \W}{2 k_B T} }}} \nn \\
				 &\times \left\{ g^R_{\bk+\bp}(\W + \nu) g^R_{\bk+\bp+\bq}(\W + \nu + \w) g^A_{\bk+\bq}(\W + \w) g^A_{\bk}(\W) \left[ \coth{\round{\frac{ \hbar \nu }{ 2 k_B T }}} \left[ d_{-q_\perp}(-\w) - d_{-q_\perp}(-\w) \right] \right. \right. \nn \\
				 &~~~~~~~~~~~~~~~~~~~~- \left. d^R_{-q_\perp}(-\w) \tanh{\round{\frac{ \hbar (\W + \nu + \w) }{ 2 k_B T }}} + d^A_{-q_\perp}(-\w) \tanh{\round{\frac{ \hbar (\W + \w)}{ 2 k_B T }}}  \right] \nn \\
				 &+ g^A_{\bk+\bp}(\W + \nu) g^A_{\bk+\bp+\bq}(\W + \nu + \w) g^R_{\bk+\bq}(\W + \w) g^R_{\bk}(\W) \left[ \coth{\round{\frac{ \hbar \nu }{ 2 k_B T }}} \left[ d_{-q_\perp}(-\w) - d_{-q_\perp}(-\w) \right] \right.\nn \\
				 &~~~~~~~~~~~~~~~~~~~~+ \left. \left. d^A_{-q_\perp}(-\w) \tanh{\round{\frac{ \hbar (\W + \nu + \w) }{ 2 k_B T }}} - d^R_{-q_\perp}(-\w) \tanh{\round{\frac{ \hbar (\W + \w)}{ 2 k_B T }}} \right] \right\}\ .
\end{align}
Representing the $\bk$ integral in polar coordinates, where again the angle is with respect to the momentum $\bp$, and then performing a change of variables from the modulus $k$ to $\z_\bk$, we can complete the $q_\parallel$ and $\z_\bk$ integrals. Note that the only quantities that contain these variables now are the spinon Green's functions, and those only occur only in specific pairings. We find
\begin{align}
\int \frac{d\z_\bk}{2\pi} \int \frac{dq_\parallel}{2\pi} g^R_{\bk+\bp}(\W + \nu) g^R_{\bk+\bp+\bq}(\W + \nu + \w) g^A_{\bk+\bq}(\W + \w) g^A_{\bk}(\W) &= \frac{-k_{Fs}/v^2_{Fs}}{\Big( \nu - \varv_{Fs} p \cos{\thi_\bk} + \io \de \Big) \left( \nu - \varv_{Fs} p \cos{\thi_\bk} - \frac{\hbar}{m_s} q_\perp p \sin{\thi_k} + \io \de \right)} \nn \\
\int \frac{d\z_\bk}{2\pi} \int \frac{dq_\parallel}{2\pi} g^A_{\bk+\bp}(\W + \nu) g^A_{\bk+\bp+\bq}(\W + \nu + \w) g^R_{\bk+\bq}(\W + \w) g^R_{\bk}(\W) &= \frac{-k_{Fs}/v^2_{Fs}}{\Big( \nu - \varv_{Fs} p \cos{\thi_\bk} - \io \de \Big) \left( \nu - \varv_{Fs} p \cos{\thi_\bk} - \frac{\hbar}{m_s} q_\perp p \sin{\thi_k} - \io \de \right)} \nn
\end{align}
which we can use in $\chi_c^{(2)}(\bp,\nu)$. The result is
\begin{align}
\label{chicf}
\chi_{c}^{(2)}(\bp,\nu) &= \frac{k_{Fs} a^2_s}{2 \hbar^2} \frac{1}{1-e^{-\be\hbar\nu}} \int \frac{d \W}{2 \pi} \int \frac{d \w}{2 \pi} \int \frac{d \thi_\bk}{2\pi} \int \frac{d q_\perp}{2\pi} \square{\tanh{\round{ \frac{\hbar(\W + \nu)}{2 k_B T} }} - \tanh{\round{ \frac{\hbar \W}{2 k_B T} }}} \nn \\
				 &\times \left[ \frac{ \coth{\round{\frac{ \hbar \nu }{ 2 k_B T }}} \left[ d^R_{-q_\perp}(-\w) - d^A_{-q_\perp}(-\w) \right] - d^R_{-q_\perp}(-\w) \tanh{\round{\frac{ \hbar (\W + \nu + \w) }{ 2 k_B T }}} + d^A_{-q_\perp}(-\w) \tanh{\round{\frac{ \hbar (\W + \w)}{ 2 k_B T }}} }{ \Big( \nu - \varv_{Fs} p \cos{\thi_\bk} + \io \de \Big) \left( \nu - \varv_{Fs} p \cos{\thi_\bk} - \frac{\hbar}{m_s} q_\perp p \sin{\thi_k} + \io \de \right) } \right. \nn \\
				 &+ \left. \frac{ \coth{\round{\frac{ \hbar \nu }{ 2 k_B T }}} \left[ d^R_{-q_\perp}(-\w) - d^A_{-q_\perp}(-\w) \right] + d^A_{-q_\perp}(-\w) \tanh{\round{\frac{ \hbar (\W + \nu + \w) }{ 2 k_B T }}} - d^R_{-q_\perp}(-\w) \tanh{\round{\frac{ \hbar (\W + \w)}{ 2 k_B T }}} }{ \Big( \nu - \varv_{Fs} p \cos{\thi_\bk} - \io \de \Big) \left( \nu - \varv_{Fs} p \cos{\thi_\bk} - \frac{\hbar}{m_s} q_\perp p \sin{\thi_k} - \io \de \right) } \right]\ ,
\end{align}
which we must combine with Eq.~\eqref{chiab} in order to maintain gauge invariance.

In order to express the combined term more concisely, we perform the angular integral and introduce the function
\begin{align}
I^\pm(q) &= \int \frac{d\thi_\bk}{2\p} \frac{1}{x - \cos{\thi_\bk} \pm \io \de} \frac{1}{x - \cos{\thi_\bk} - \frac{q}{k_F} \sin{\thi_k} \pm \io \de} \nn \\
		       &= \frac{ \abs{x} }{ \round{x \pm \io \de }^2 \sqrt{\round{x \pm \io \de }^2 - \round{\round{\frac{q}{k_F}}^2 + 1}} }\ ,
\end{align}
where $x = \nu / (\varv_{Fs}p)$ and we have substituted $q_\perp \to q$ for notational convenience. Therefore the sum of Eq.~\eqref{chiab} and Eq.~\eqref{chicf} becomes
\begin{align}
\label{chii}
\chi^{(2)}(\bp,\nu) &= \frac{ k_{Fs} a_s^2}{2\hbar^2} \frac{1}{(\varv_{Fs} p)^2} \frac{1}{1-e^{-\be\hbar\nu}} \int \frac{d\W}{2\p} \int \frac{d\w}{2\p} \int \frac{dq}{2\p} \square{\tanh{\round{\frac{\hbar(\W + \nu)}{2 k_B T}}} - \tanh{\round{\frac{\hbar \W}{2 k_B T}}}} \nn \\
			   &\times \left\{ \coth{\round{\frac{\hbar \W}{2 k_B T}}} \square{d^R_{-q}(-\w) - d^A_{-q}(-\w)} \round{I^+(q) + I^-(q) - I^+(0) - I^-(0)} \right. \nn \\
			   &- \tanh{\round{\frac{\hbar(\W + \nu + \w)}{2 k_B T}}} \square{d^R_{-q}(-\w)\round{I^+(q) - I^+(0)} - d^A_{-q}(-\w)\round{I^-(q) - I^-(0)}} \nn \\
			   &- \left.\tanh{\round{\frac{\hbar(\W + \w)}{2 k_B T}}} \square{d^R_{-q}(-\w)\round{I^-(q) - I^-(0)} - d^A_{-q}(-\w)\round{I^+(q) - I^+(0)}} \right\}\ .
\end{align}
We need to connect to Eq.~\eqref{u1sus}, so we must obtain the momentum integrated susceptibility: $\chi^{(2)}(\nu) = (2\pi)^{-2} \int d\bp~\chi^{(2)}(\bp,\nu)$. Only the angular term $I^{\pm}(q)$ carries external momentum, however, so the momentum integral can be carried out:
\beq
\int_0^\infty \frac{pdp}{p^2} I^{\pm}(q) = \int_0^\infty \frac{dx}{x} I^{\pm}(q) = - \frac{tan^{-1}\round{q_\perp / k_{Fs}}}{q /k_{Fs}}\, .
\eeq
At this point note that the integrand in Eq.~\eqref{chii} is even in $q$, so it is possible to restrict the $q$ integral to half the domain. If we recall that there is an external sum over all lattice points, $\sum_{\bi} = N$, then because the quantum spin liquid is isotropic we have for the final result
\beq
\chi^{(2)}(\nu)=\frac{1}{1-e^{-\be\hbar\nu}}\frac{Nm_{s}^2a_s^2}{2(2\pi)^2\hbar^4}\int \frac{d \w}{2 \pi} \int_0^\infty \frac{dq}{k_{Fs}} F(\nu,\w) \left[ d^R_{-q}(-\w) - d^A_{-q}(-\w) \right] \left[1-\frac{\tan^{-1}(q/k_{Fs})}{q/k_{Fs}}\right]\, ,
\eeq
with
\beq
F(\nu,\w)=\int\frac{d\W}{2\p}\square{\tanh\round{\tfrac{\hbar(\W+\nu)}{2k_BT}}-\tanh\round{\tfrac{\hbar\W}{2k_BT}}} \Big[2\coth\round{\tfrac{\hbar\w}{2k_BT}}-\tanh\round{\tfrac{\hbar(\W+\w+\nu)}{2k_BT}}-\tanh\round{\tfrac{\hbar(\W+\w)}{2k_BT}}\Big]\, .
\eeq
These are Eqs.~\eqref{u1sus} and~\eqref{u1f} from the main text.

\end{widetext}

\section{The Kitaev model calculation}
\label{apD}
In this section we will explicitly show our calculation of the retarded Green function for the Kitaev model calculation in Sec~\ref{dferm}. The unperturbed Hamiltonian for the $b$ fermions reads
\beq
\label{bham}
\cH_0 = \sum_\bq \begin{pmatrix} b^\dag_\bq & b_{-\bq} \end{pmatrix} \begin{pmatrix} \ve_\bq & -\D_\bq \\ -\D^*_\bq & -\ve_\bq \end{pmatrix} \begin{pmatrix} b_\bq \\ b^\dag_{-\bq} \end{pmatrix}\, ,
\eeq
where $\ve_\bq = K(1 + \cos{\bq\cdot \bn_1} + \cos{\bq\cdot \bn_2})$, $\D_\bq =-\io K (\sin{\bq\cdot \bn_1} + \sin{\bq\cdot \bn_2})$, and $\bn_1,\bn_2$ are the real space lattice vectors as defined in the main text. The local potential across the $x$-bond of the $\br=0$ unit cell is given by
\beq
\label{lpot}
V_x = ub^\dag_0 b_0 - \frac{u}{2} = \frac{u}{N_u}\sum_{\bp \bq} b^\dag_\bp b_\bq - \frac{u}{2}\, ,
\eeq
where $u=-4K$ and $N_u$ is the number of unit cells in the lattice. Then the full Hamiltonian including the local potential is $H_x = \cH_0 + V_x$, for which we want to find the Green functions.

We take the equation of motion approach, so we require the commutation relations for $b_\bk$,
\beq
\label{eom}
[H_x,b_{\bk}] = -2\ve_\bk b_\bk + 2 \D_\bk b^\dag_{-\bk} - \frac{u}{N_u} \sum_{\bk} b_\bk\ .
\eeq
We now examine the retarded matrix Green functions in the Nambu basis, which we express as $G^R_{\bp \bq}(t,0) = \thi(t)[ G^>_{\bp \bq}(t,0) - G^<_{\bp \bq}(t,0)] $ (note that capitalized $G$ Green functions will be reserved for matrices, and lower-case $g$ for components). The greater-than matrix Green functions is
\beq
\label{gtgf}
G^>_{\bp \bq}(t,0) = -\io \begin{pmatrix}[1.5] \langle b_\bp(t) b^\dag_{-\bq}(0)\rangle & \langle b_\bp(t) b_\bq(0)\rangle \\ \langle b^\dag_{-\bp}(t) b^\dag_{-\bq}(0)\rangle & \langle b^\dag_{-\bp}(t) b_\bq(0)\rangle \end{pmatrix}\, ,
\eeq
and the lesser-than Green function matrix can be found by anti-commuting the fermions in each component. Using Eq.~\eqref{eom}, we then find 
%Taking the time derivative of the retarded Green function matrix we find
%\begin{multline}
%\io \pd_t G^R_{\bp \bq}(t,0) = \io \de(t)\round{G^>_{\bp \bq}(t,0)-G^<_{\bp \bq}(t,0)}\de_{\bp \bq} \\
%+ \io \thi(t)\round{\pd_t G^>_{\bp \bq}(t,0)-\pd_t G^<_{\bp \bq}(t,0)}\, .
%\end{multline}
%We apply Heisenberg's equation of motion on each element of the matrices under time derivatives, and replace them using . Thus 
\begin{multline}
\io \pd_t G^R_{\bp \bq}(t,0) = \io \de(t) \de_{\bp \bq} + \frac{2}{\hbar}H_0(\bp) G^R_{\bp \bq}(t,0) \\+ \frac{u}{\hbar N} \s_z \sum_{\bp} G^R_{\bp \bq}(t,0)\, ,
\end{multline}
where $H_0(\bp)$ is the coefficient matrix of $\cH_0$, and $\s_z$ is the $z$-component Pauli matrix. Fourier transforming to frequency space
\beq
\label{rgf}
G^R_{\bp \bq}(\w) = G^{R,0}_{\bp} \de_{\bp \bq} + \frac{u}{\hbar N_u} G^{R,0}_{\bp}(\w) \s_z \sum_{\bk} G^R_{\bk \bq}(\w)\, ,
\eeq
where the unperturbed Green function matrix for the $b_\bk$ fermions reads
\beq
G^{R,0}_{\bp}(\w) = \frac{1}{\cN_\bp(\w)} \begin{pmatrix}[1.5] \w + \frac{2}{\hbar}\ve_\bp + \io \de & \frac{2}{\hbar}\D_\bp \\ \frac{2}{\hbar} \D^*_\bp & \w - \frac{2}{\hbar} + \io \de \end{pmatrix}\, ,
\eeq
and $\cN_\bp(\w) = (\w + \io \de)^2 - (2/\hbar)^2 |s_\bp|^2$. When summing over the momentum index, we can see that the off diagonal elements of $G^{R,0}_{\bp}(\w)$ vanish because $\D_\bp$ is an odd function of $\bp$, and the Brillouin zone is symmetric about the origin. This is the reason there are no anomalous Green functions of the form $\langle b^\dag_0(t) b^\dag_0(t')\rangle$ | or the complex conjugate | as mentioned in the main text. Thus we have $\sum_\bp G^{R,0}_\bp(\w) = \diag\{g^R_{11}(\w),g^R_{22}(\w)\}$, where the elements are
\beq
\label{gfco}
g^R_{11,22}(\w) = \sum_\bp \frac{\w \pm \frac{2}{\hbar}\ve_\bp + \io\de}{\cN_\bp(\w)}\, .
\eeq
We then sum over $\bp$ in Eq.~\eqref{rgf} in order to rewrite the full Green function in terms of unperturbed Green functions as
\beq
\label{fugf}
\sum_\bp G^R_{\bp \bq}(\w) = \square{1 - \frac{u}{\hbar N_u} \round{\sum_\bp G^{R,0}_\bp(\w)} \s_z}^{-1} G^{R,0}_\bq(\w)\, .
\eeq
Finally, by substituting this expression back into Eq.~\eqref{rgf}
\beq
\label{frgf}
\sum_{\bp \bq} G^R_{\bp \bq}(\w) = \begin{pmatrix}[1.5] \frac{g^R_{11}(\w)}{1 - \frac{u}{\hbar N_u}g^R_{11}(\w)} & 0 \\ 0 &\frac{g^R_{22}(\w)}{1 + \frac{u}{\hbar N_u}g^R_{22}(\w)} \end{pmatrix}\, .
\eeq

The correlation function we require is $\langle b_0(t) b^\dag_0(0)\rangle_x = (1/N_u)\sum_{\bp \bq}\langle b_\bp(t) b^\dag_\bq(0)\rangle_x$, which in frequency space is $\io g^>_{11}(\w)$. In equilibrium at the quantum limit, $G^>(\w) = \thi(\w)[G^R(\w) - G^A(\w)]$, and therefore we have 
\beq
g_{11}^>(\w) = \thi(\w) \round{\frac{g^R_{11}(\w)}{1 - \frac{u}{\hbar N_u}g^R_{11}(\w)} - \frac{g^A_{11}(\w)}{1 - \frac{u}{\hbar N_u}g^A_{11}(\w)}}\, .\nn
\eeq
This is the expression used to compute the susceptibility in Eq.~\eqref{kits}.


\begin{thebibliography}{92}%
\makeatletter
\providecommand \@ifxundefined [1]{%
 \@ifx{#1\undefined}
}%
\providecommand \@ifnum [1]{%
 \ifnum #1\expandafter \@firstoftwo
 \else \expandafter \@secondoftwo
 \fi
}%
\providecommand \@ifx [1]{%
 \ifx #1\expandafter \@firstoftwo
 \else \expandafter \@secondoftwo
 \fi
}%
\providecommand \natexlab [1]{#1}%
\providecommand \enquote  [1]{``#1''}%
\providecommand \bibnamefont  [1]{#1}%
\providecommand \bibfnamefont [1]{#1}%
\providecommand \citenamefont [1]{#1}%
\providecommand \href@noop [0]{\@secondoftwo}%
\providecommand \href [0]{\begingroup \@sanitize@url \@href}%
\providecommand \@href[1]{\@@startlink{#1}\@@href}%
\providecommand \@@href[1]{\endgroup#1\@@endlink}%
\providecommand \@sanitize@url [0]{\catcode `\\12\catcode `\$12\catcode
  `\&12\catcode `\#12\catcode `\^12\catcode `\_12\catcode `\%12\relax}%
\providecommand \@@startlink[1]{}%
\providecommand \@@endlink[0]{}%
\providecommand \url  [0]{\begingroup\@sanitize@url \@url }%
\providecommand \@url [1]{\endgroup\@href {#1}{\urlprefix }}%
\providecommand \urlprefix  [0]{URL }%
\providecommand \Eprint [0]{\href }%
\providecommand \doibase [0]{http://dx.doi.org/}%
\providecommand \selectlanguage [0]{\@gobble}%
\providecommand \bibinfo  [0]{\@secondoftwo}%
\providecommand \bibfield  [0]{\@secondoftwo}%
\providecommand \translation [1]{[#1]}%
\providecommand \BibitemOpen [0]{}%
\providecommand \bibitemStop [0]{}%
\providecommand \bibitemNoStop [0]{.\EOS\space}%
\providecommand \EOS [0]{\spacefactor3000\relax}%
\providecommand \BibitemShut  [1]{\csname bibitem#1\endcsname}%
\let\auto@bib@innerbib\@empty
%</preamble>
\bibitem [{\citenamefont {Balents}(2010)}]{balentsNAT10}%
  \BibitemOpen
  \bibfield  {author} {\bibinfo {author} {\bibfnamefont {Leon}\ \bibnamefont
  {Balents}},\ }\bibfield  {title} {\enquote {\bibinfo {title} {Spin liquids in
  frustrated magnets},}\ }\href {http://dx.doi.org/10.1038/nature08917}
  {\bibfield  {journal} {\bibinfo  {journal} {Nature}\ }\textbf {\bibinfo
  {volume} {464}},\ \bibinfo {pages} {199--208} (\bibinfo {year}
  {2010})}\BibitemShut {NoStop}%
\bibitem [{\citenamefont {Savary}\ and\ \citenamefont
  {Balents}(2016)}]{savaryRPP17}%
  \BibitemOpen
  \bibfield  {author} {\bibinfo {author} {\bibfnamefont {Lucile}\ \bibnamefont
  {Savary}}\ and\ \bibinfo {author} {\bibfnamefont {Leon}\ \bibnamefont
  {Balents}},\ }\bibfield  {title} {\enquote {\bibinfo {title} {Quantum spin
  liquids: a review},}\ }\href@noop {} {\bibfield  {journal} {\bibinfo
  {journal} {Rep. Prog. Phys.}\ }\textbf {\bibinfo {volume} {80}} (\bibinfo
  {year} {2016})}\BibitemShut {NoStop}%
\bibitem [{\citenamefont {Kitaev}\ and\ \citenamefont
  {Preskill}(2006)}]{kitaevPRL06}%
  \BibitemOpen
  \bibfield  {author} {\bibinfo {author} {\bibfnamefont {Alexei}\ \bibnamefont
  {Kitaev}}\ and\ \bibinfo {author} {\bibfnamefont {John}\ \bibnamefont
  {Preskill}},\ }\bibfield  {title} {\enquote {\bibinfo {title} {Topological
  entanglement entropy},}\ }\href {\doibase 10.1103/PhysRevLett.96.110404}
  {\bibfield  {journal} {\bibinfo  {journal} {Phys. Rev. Lett.}\ }\textbf
  {\bibinfo {volume} {96}},\ \bibinfo {pages} {110404} (\bibinfo {year}
  {2006})}\BibitemShut {NoStop}%
\bibitem [{\citenamefont {Levin}\ and\ \citenamefont {Wen}(2006)}]{levinPRL06}%
  \BibitemOpen
  \bibfield  {author} {\bibinfo {author} {\bibfnamefont {Michael}\ \bibnamefont
  {Levin}}\ and\ \bibinfo {author} {\bibfnamefont {Xiao-Gang}\ \bibnamefont
  {Wen}},\ }\bibfield  {title} {\enquote {\bibinfo {title} {Detecting
  topological order in a ground state wave function},}\ }\href {\doibase
  10.1103/PhysRevLett.96.110405} {\bibfield  {journal} {\bibinfo  {journal}
  {Phys. Rev. Lett.}\ }\textbf {\bibinfo {volume} {96}},\ \bibinfo {pages}
  {110405} (\bibinfo {year} {2006})}\BibitemShut {NoStop}%
\bibitem [{\citenamefont {Senthil}\ and\ \citenamefont
  {Fisher}(2000)}]{senthilPRB00}%
  \BibitemOpen
  \bibfield  {author} {\bibinfo {author} {\bibfnamefont {T.}~\bibnamefont
  {Senthil}}\ and\ \bibinfo {author} {\bibfnamefont {Matthew P.~A.}\
  \bibnamefont {Fisher}},\ }\bibfield  {title} {\enquote {\bibinfo {title}
  {${Z}_{2}$ gauge theory of electron fractionalization in strongly correlated
  systems},}\ }\href {\doibase 10.1103/PhysRevB.62.7850} {\bibfield  {journal}
  {\bibinfo  {journal} {Phys. Rev. B}\ }\textbf {\bibinfo {volume} {62}},\
  \bibinfo {pages} {7850--7881} (\bibinfo {year} {2000})}\BibitemShut {NoStop}%
\bibitem [{\citenamefont {Senthil}\ and\ \citenamefont
  {Fisher}(2001)}]{senthilPRL01}%
  \BibitemOpen
  \bibfield  {author} {\bibinfo {author} {\bibfnamefont {T.}~\bibnamefont
  {Senthil}}\ and\ \bibinfo {author} {\bibfnamefont {Matthew P.~A.}\
  \bibnamefont {Fisher}},\ }\bibfield  {title} {\enquote {\bibinfo {title}
  {Fractionalization in the cuprates: Detecting the topological order},}\
  }\href {\doibase 10.1103/PhysRevLett.86.292} {\bibfield  {journal} {\bibinfo
  {journal} {Phys. Rev. Lett.}\ }\textbf {\bibinfo {volume} {86}},\ \bibinfo
  {pages} {292--295} (\bibinfo {year} {2001})}\BibitemShut {NoStop}%
\bibitem [{\citenamefont {Norman}(2016)}]{normanRMP16}%
  \BibitemOpen
  \bibfield  {author} {\bibinfo {author} {\bibfnamefont {M.~R.}\ \bibnamefont
  {Norman}},\ }\bibfield  {title} {\enquote {\bibinfo {title} {Colloquium:
  Herbertsmithite and the search for the quantum spin liquid},}\ }\href
  {\doibase 10.1103/RevModPhys.88.041002} {\bibfield  {journal} {\bibinfo
  {journal} {Rev. Mod. Phys.}\ }\textbf {\bibinfo {volume} {88}},\ \bibinfo
  {pages} {041002} (\bibinfo {year} {2016})}\BibitemShut {NoStop}%
\bibitem [{\citenamefont {Powell}\ and\ \citenamefont
  {McKenzie}(2011)}]{powellRPP11}%
  \BibitemOpen
  \bibfield  {author} {\bibinfo {author} {\bibfnamefont {B~J}\ \bibnamefont
  {Powell}}\ and\ \bibinfo {author} {\bibfnamefont {Ross~H}\ \bibnamefont
  {McKenzie}},\ }\bibfield  {title} {\enquote {\bibinfo {title} {Quantum
  frustration in organic mott insulators: from spin liquids to unconventional
  superconductors},}\ }\href {\doibase 10.1088/0034-4885/74/5/056501}
  {\bibfield  {journal} {\bibinfo  {journal} {Rep. Prog. Phys.}\ }\textbf
  {\bibinfo {volume} {74}},\ \bibinfo {pages} {056501} (\bibinfo {year}
  {2011})}\BibitemShut {NoStop}%
\bibitem [{\citenamefont {Takagi}\ \emph {et~al.}(2019)\citenamefont {Takagi},
  \citenamefont {Takayama}, \citenamefont {Jackeli}, \citenamefont
  {Khaliullin},\ and\ \citenamefont {Nagler}}]{takagiNRP19}%
  \BibitemOpen
  \bibfield  {author} {\bibinfo {author} {\bibfnamefont {Hidenori}\
  \bibnamefont {Takagi}}, \bibinfo {author} {\bibfnamefont {Tomohiro}\
  \bibnamefont {Takayama}}, \bibinfo {author} {\bibfnamefont {George}\
  \bibnamefont {Jackeli}}, \bibinfo {author} {\bibfnamefont {Giniyat}\
  \bibnamefont {Khaliullin}}, \ and\ \bibinfo {author} {\bibfnamefont
  {Stephen~E.}\ \bibnamefont {Nagler}},\ }\bibfield  {title} {\enquote
  {\bibinfo {title} {Concept and realization of kitaev quantum spin liquids},}\
  }\href {\doibase 10.1038/s42254-019-0038-2} {\bibfield  {journal} {\bibinfo
  {journal} {Nature Reviews Physics}\ }\textbf {\bibinfo {volume} {1}},\
  \bibinfo {pages} {264--280} (\bibinfo {year} {2019})}\BibitemShut {NoStop}%
\bibitem [{\citenamefont {Shimizu}\ \emph {et~al.}(2003)\citenamefont
  {Shimizu}, \citenamefont {Miyagawa}, \citenamefont {Kanoda}, \citenamefont
  {Maesato},\ and\ \citenamefont {Saito}}]{shimizuPRL03}%
  \BibitemOpen
  \bibfield  {author} {\bibinfo {author} {\bibfnamefont {Y.}~\bibnamefont
  {Shimizu}}, \bibinfo {author} {\bibfnamefont {K.}~\bibnamefont {Miyagawa}},
  \bibinfo {author} {\bibfnamefont {K.}~\bibnamefont {Kanoda}}, \bibinfo
  {author} {\bibfnamefont {M.}~\bibnamefont {Maesato}}, \ and\ \bibinfo
  {author} {\bibfnamefont {G.}~\bibnamefont {Saito}},\ }\bibfield  {title}
  {\enquote {\bibinfo {title} {Spin liquid state in an organic mott insulator
  with a triangular lattice},}\ }\href {\doibase 10.1103/PhysRevLett.91.107001}
  {\bibfield  {journal} {\bibinfo  {journal} {Phys. Rev. Lett.}\ }\textbf
  {\bibinfo {volume} {91}},\ \bibinfo {pages} {107001} (\bibinfo {year}
  {2003})}\BibitemShut {NoStop}%
\bibitem [{\citenamefont {Olariu}\ \emph {et~al.}(2008)\citenamefont {Olariu},
  \citenamefont {Mendels}, \citenamefont {Bert}, \citenamefont {Duc},
  \citenamefont {Trombe}, \citenamefont {de~Vries},\ and\ \citenamefont
  {Harrison}}]{olariuPRL08}%
  \BibitemOpen
  \bibfield  {author} {\bibinfo {author} {\bibfnamefont {A.}~\bibnamefont
  {Olariu}}, \bibinfo {author} {\bibfnamefont {P.}~\bibnamefont {Mendels}},
  \bibinfo {author} {\bibfnamefont {F.}~\bibnamefont {Bert}}, \bibinfo {author}
  {\bibfnamefont {F.}~\bibnamefont {Duc}}, \bibinfo {author} {\bibfnamefont
  {J.~C.}\ \bibnamefont {Trombe}}, \bibinfo {author} {\bibfnamefont {M.~A.}\
  \bibnamefont {de~Vries}}, \ and\ \bibinfo {author} {\bibfnamefont
  {A.}~\bibnamefont {Harrison}},\ }\bibfield  {title} {\enquote {\bibinfo
  {title} {$^{17}\mathrm{O}$ NMR study of the intrinsic magnetic susceptibility
  and spin dynamics of the quantum kagome antiferromagnet
  ${\mathrm{ZnCu}}_{3}(\mathrm{OH}{)}_{6}{\mathrm{Cl}}_{2}$},}\ }\href
  {\doibase 10.1103/PhysRevLett.100.087202} {\bibfield  {journal} {\bibinfo
  {journal} {Phys. Rev. Lett.}\ }\textbf {\bibinfo {volume} {100}},\ \bibinfo
  {pages} {087202} (\bibinfo {year} {2008})}\BibitemShut {NoStop}%
\bibitem [{\citenamefont {Fu}\ \emph {et~al.}(2015)\citenamefont {Fu},
  \citenamefont {Imai}, \citenamefont {Han},\ and\ \citenamefont
  {Lee}}]{fuSCI15}%
  \BibitemOpen
  \bibfield  {author} {\bibinfo {author} {\bibfnamefont {Mingxuan}\
  \bibnamefont {Fu}}, \bibinfo {author} {\bibfnamefont {Takashi}\ \bibnamefont
  {Imai}}, \bibinfo {author} {\bibfnamefont {Tian-Heng}\ \bibnamefont {Han}}, \
  and\ \bibinfo {author} {\bibfnamefont {Young~S.}\ \bibnamefont {Lee}},\
  }\bibfield  {title} {\enquote {\bibinfo {title} {Evidence for a gapped
  spin-liquid ground state in a kagome heisenberg antiferromagnet},}\
  }\href@noop {} {\bibfield  {journal} {\bibinfo  {journal} {Science}\ }\textbf
  {\bibinfo {volume} {350}},\ \bibinfo {pages} {655--658} (\bibinfo {year}
  {2015})}\BibitemShut {NoStop}%
\bibitem [{\citenamefont {Kitagawa}\ \emph {et~al.}(2018)\citenamefont
  {Kitagawa}, \citenamefont {Takayama}, \citenamefont {Matsumoto},
  \citenamefont {Kato}, \citenamefont {Takano}, \citenamefont {Kishimoto},
  \citenamefont {Bette}, \citenamefont {Dinnebier}, \citenamefont {Jackeli},\
  and\ \citenamefont {Takagi}}]{kitagawaNAT18}%
  \BibitemOpen
  \bibfield  {author} {\bibinfo {author} {\bibfnamefont {K.}~\bibnamefont
  {Kitagawa}}, \bibinfo {author} {\bibfnamefont {T.}~\bibnamefont {Takayama}},
  \bibinfo {author} {\bibfnamefont {Y.}~\bibnamefont {Matsumoto}}, \bibinfo
  {author} {\bibfnamefont {A.}~\bibnamefont {Kato}}, \bibinfo {author}
  {\bibfnamefont {R.}~\bibnamefont {Takano}}, \bibinfo {author} {\bibfnamefont
  {Y.}~\bibnamefont {Kishimoto}}, \bibinfo {author} {\bibfnamefont
  {S.}~\bibnamefont {Bette}}, \bibinfo {author} {\bibfnamefont
  {R.}~\bibnamefont {Dinnebier}}, \bibinfo {author} {\bibfnamefont
  {G.}~\bibnamefont {Jackeli}}, \ and\ \bibinfo {author} {\bibfnamefont
  {H.}~\bibnamefont {Takagi}},\ }\bibfield  {title} {\enquote {\bibinfo {title}
  {A spin--orbital-entangled quantum liquid on a honeycomb lattice},}\ }\href
  {https://doi.org/10.1038/nature25482} {\bibfield  {journal} {\bibinfo
  {journal} {Nature}\ }\textbf {\bibinfo {volume} {554}},\ \bibinfo {pages}
  {341 EP --} (\bibinfo {year} {2018})}\BibitemShut {NoStop}%
\bibitem [{\citenamefont {Bert}\ \emph {et~al.}(2007)\citenamefont {Bert},
  \citenamefont {Nakamae}, \citenamefont {Ladieu}, \citenamefont {L'H\^ote},
  \citenamefont {Bonville}, \citenamefont {Duc}, \citenamefont {Trombe},\ and\
  \citenamefont {Mendels}}]{bertPRB07}%
  \BibitemOpen
  \bibfield  {author} {\bibinfo {author} {\bibfnamefont {F.}~\bibnamefont
  {Bert}}, \bibinfo {author} {\bibfnamefont {S.}~\bibnamefont {Nakamae}},
  \bibinfo {author} {\bibfnamefont {F.}~\bibnamefont {Ladieu}}, \bibinfo
  {author} {\bibfnamefont {D.}~\bibnamefont {L'H\^ote}}, \bibinfo {author}
  {\bibfnamefont {P.}~\bibnamefont {Bonville}}, \bibinfo {author}
  {\bibfnamefont {F.}~\bibnamefont {Duc}}, \bibinfo {author} {\bibfnamefont
  {J.-C.}\ \bibnamefont {Trombe}}, \ and\ \bibinfo {author} {\bibfnamefont
  {P.}~\bibnamefont {Mendels}},\ }\bibfield  {title} {\enquote {\bibinfo
  {title} {Low temperature magnetization of the $s=\frac{1}{2}$ kagome
  antiferromagnet
  $\mathrm{Zn}{\mathrm{Cu}}_{3}{(\mathrm{O}\mathrm{H})}_{6}{\mathrm{Cl}}_{2}$},}\
  }\href {\doibase 10.1103/PhysRevB.76.132411} {\bibfield  {journal} {\bibinfo
  {journal} {Phys. Rev. B}\ }\textbf {\bibinfo {volume} {76}},\ \bibinfo
  {pages} {132411} (\bibinfo {year} {2007})}\BibitemShut {NoStop}%
\bibitem [{\citenamefont {Yamashita}\ \emph
  {et~al.}(2008{\natexlab{a}})\citenamefont {Yamashita}, \citenamefont
  {Nakata}, \citenamefont {Kasahara}, \citenamefont {Sasaki}, \citenamefont
  {Yoneyama}, \citenamefont {Kobayashi}, \citenamefont {Fujimoto},
  \citenamefont {Shibauchi},\ and\ \citenamefont {Matsuda}}]{yamashitaNATP08b}%
  \BibitemOpen
  \bibfield  {author} {\bibinfo {author} {\bibfnamefont {Minoru}\ \bibnamefont
  {Yamashita}}, \bibinfo {author} {\bibfnamefont {Norihito}\ \bibnamefont
  {Nakata}}, \bibinfo {author} {\bibfnamefont {Yuichi}\ \bibnamefont
  {Kasahara}}, \bibinfo {author} {\bibfnamefont {Takahiko}\ \bibnamefont
  {Sasaki}}, \bibinfo {author} {\bibfnamefont {Naoki}\ \bibnamefont
  {Yoneyama}}, \bibinfo {author} {\bibfnamefont {Norio}\ \bibnamefont
  {Kobayashi}}, \bibinfo {author} {\bibfnamefont {Satoshi}\ \bibnamefont
  {Fujimoto}}, \bibinfo {author} {\bibfnamefont {Takasada}\ \bibnamefont
  {Shibauchi}}, \ and\ \bibinfo {author} {\bibfnamefont {Yuji}\ \bibnamefont
  {Matsuda}},\ }\bibfield  {title} {\enquote {\bibinfo {title}
  {Thermal-transport measurements in a quantum spin-liquid state of the
  frustrated triangular magnet $\kappa$-(BEDT-TTF)$_2$Cu$_2$(CN)$_3$},}\ }\href
  {https://doi.org/10.1038/nphys1134} {\bibfield  {journal} {\bibinfo
  {journal} {Nature Physics}\ }\textbf {\bibinfo {volume} {5}},\ \bibinfo
  {pages} {44 EP --} (\bibinfo {year} {2008}{\natexlab{a}})}\BibitemShut
  {NoStop}%
\bibitem [{\citenamefont {Yamashita}\ \emph {et~al.}(2011)\citenamefont
  {Yamashita}, \citenamefont {Yamamoto}, \citenamefont {Nakazawa},
  \citenamefont {Tamura},\ and\ \citenamefont {Kato}}]{yamashitaNATC11}%
  \BibitemOpen
  \bibfield  {author} {\bibinfo {author} {\bibfnamefont {Satoshi}\ \bibnamefont
  {Yamashita}}, \bibinfo {author} {\bibfnamefont {Takashi}\ \bibnamefont
  {Yamamoto}}, \bibinfo {author} {\bibfnamefont {Yasuhiro}\ \bibnamefont
  {Nakazawa}}, \bibinfo {author} {\bibfnamefont {Masafumi}\ \bibnamefont
  {Tamura}}, \ and\ \bibinfo {author} {\bibfnamefont {Reizo}\ \bibnamefont
  {Kato}},\ }\bibfield  {title} {\enquote {\bibinfo {title} {Gapless spin
  liquid of an organic triangular compound evidenced by thermodynamic
  measurements},}\ }\href {http://dx.doi.org/10.1038/ncomms1274} {\bibfield
  {journal} {\bibinfo  {journal} {Nature Comm.}\ }\textbf {\bibinfo {volume}
  {2}},\ \bibinfo {pages} {275} (\bibinfo {year} {2011})}\BibitemShut {NoStop}%
\bibitem [{\citenamefont {Yamashita}\ \emph
  {et~al.}(2008{\natexlab{b}})\citenamefont {Yamashita}, \citenamefont
  {Nakazawa}, \citenamefont {Oguni}, \citenamefont {Oshima}, \citenamefont
  {Nojiri}, \citenamefont {Shimizu}, \citenamefont {Miyagawa},\ and\
  \citenamefont {Kanoda}}]{yamashitaNATP08a}%
  \BibitemOpen
  \bibfield  {author} {\bibinfo {author} {\bibfnamefont {Satoshi}\ \bibnamefont
  {Yamashita}}, \bibinfo {author} {\bibfnamefont {Yasuhiro}\ \bibnamefont
  {Nakazawa}}, \bibinfo {author} {\bibfnamefont {Masaharu}\ \bibnamefont
  {Oguni}}, \bibinfo {author} {\bibfnamefont {Yugo}\ \bibnamefont {Oshima}},
  \bibinfo {author} {\bibfnamefont {Hiroyuki}\ \bibnamefont {Nojiri}}, \bibinfo
  {author} {\bibfnamefont {Yasuhiro}\ \bibnamefont {Shimizu}}, \bibinfo
  {author} {\bibfnamefont {Kazuya}\ \bibnamefont {Miyagawa}}, \ and\ \bibinfo
  {author} {\bibfnamefont {Kazushi}\ \bibnamefont {Kanoda}},\ }\bibfield
  {title} {\enquote {\bibinfo {title} {Thermodynamic properties of a spin-1/2
  spin-liquid state in a {$[$}kappa{$]$}-type organic salt},}\ }\href
  {http://dx.doi.org/10.1038/nphys942} {\bibfield  {journal} {\bibinfo
  {journal} {Nature Phys.}\ }\textbf {\bibinfo {volume} {4}},\ \bibinfo {pages}
  {459--462} (\bibinfo {year} {2008}{\natexlab{b}})}\BibitemShut {NoStop}%
\bibitem [{\citenamefont {Yamashita}\ \emph {et~al.}(2010)\citenamefont
  {Yamashita}, \citenamefont {Nakata}, \citenamefont {Senshu}, \citenamefont
  {Nagata}, \citenamefont {Yamamoto}, \citenamefont {Kato}, \citenamefont
  {Shibauchi},\ and\ \citenamefont {Matsuda}}]{yamashitaSCI10}%
  \BibitemOpen
  \bibfield  {author} {\bibinfo {author} {\bibfnamefont {Minoru}\ \bibnamefont
  {Yamashita}}, \bibinfo {author} {\bibfnamefont {Norihito}\ \bibnamefont
  {Nakata}}, \bibinfo {author} {\bibfnamefont {Yoshinori}\ \bibnamefont
  {Senshu}}, \bibinfo {author} {\bibfnamefont {Masaki}\ \bibnamefont {Nagata}},
  \bibinfo {author} {\bibfnamefont {Hiroshi~M.}\ \bibnamefont {Yamamoto}},
  \bibinfo {author} {\bibfnamefont {Reizo}\ \bibnamefont {Kato}}, \bibinfo
  {author} {\bibfnamefont {Takasada}\ \bibnamefont {Shibauchi}}, \ and\
  \bibinfo {author} {\bibfnamefont {Yuji}\ \bibnamefont {Matsuda}},\ }\bibfield
   {title} {\enquote {\bibinfo {title} {Highly mobile gapless excitations in a
  two-dimensional candidate quantum spin liquid},}\ }\href {\doibase
  10.1126/science.1188200} {\bibfield  {journal} {\bibinfo  {journal}
  {Science}\ }\textbf {\bibinfo {volume} {328}},\ \bibinfo {pages} {1246--1248}
  (\bibinfo {year} {2010})}\BibitemShut {NoStop}%
\bibitem [{\citenamefont {{Bourgeois-Hope}}\ \emph {et~al.}()\citenamefont
  {{Bourgeois-Hope}}, \citenamefont {{Lalibert{\'e}}}, \citenamefont
  {{Lefran{\c{c}}ois}}, \citenamefont {{Grissonnanche}}, \citenamefont
  {{Ren{\'e} de Cotret}}, \citenamefont {{Gordon}}, \citenamefont {{Kato}},
  \citenamefont {{Taillefer}},\ and\ \citenamefont
  {{Doiron-Leyraud}}}]{bourgeoishopeCM19}%
  \BibitemOpen
  \bibfield  {author} {\bibinfo {author} {\bibfnamefont {P.}~\bibnamefont
  {{Bourgeois-Hope}}}, \bibinfo {author} {\bibfnamefont {F.}~\bibnamefont
  {{Lalibert{\'e}}}}, \bibinfo {author} {\bibfnamefont {E.}~\bibnamefont
  {{Lefran{\c{c}}ois}}}, \bibinfo {author} {\bibfnamefont {G.}~\bibnamefont
  {{Grissonnanche}}}, \bibinfo {author} {\bibfnamefont {S.}~\bibnamefont
  {{Ren{\'e} de Cotret}}}, \bibinfo {author} {\bibfnamefont {R.}~\bibnamefont
  {{Gordon}}}, \bibinfo {author} {\bibfnamefont {R.}~\bibnamefont {{Kato}}},
  \bibinfo {author} {\bibfnamefont {L.}~\bibnamefont {{Taillefer}}}, \ and\
  \bibinfo {author} {\bibfnamefont {N.}~\bibnamefont {{Doiron-Leyraud}}},\
  }\href@noop {} {\bibinfo  {journal} {arXiv:1904.10402}\ }\BibitemShut
  {NoStop}%
\bibitem [{\citenamefont {{Ni}}\ \emph {et~al.}()\citenamefont {{Ni}},
  \citenamefont {{Pan}}, \citenamefont {{Huang}}, \citenamefont {{Zeng}},
  \citenamefont {{Yu}}, \citenamefont {{Cheng}}, \citenamefont {{Wang}},
  \citenamefont {{Kato}},\ and\ \citenamefont {{Li}}}]{niCM19}%
  \BibitemOpen
\bibfield  {journal} {  }\bibfield  {author} {\bibinfo {author} {\bibfnamefont
  {J.~M.}\ \bibnamefont {{Ni}}}, \bibinfo {author} {\bibfnamefont {B.~L.}\
  \bibnamefont {{Pan}}}, \bibinfo {author} {\bibfnamefont {Y.~Y.}\ \bibnamefont
  {{Huang}}}, \bibinfo {author} {\bibfnamefont {J.~Y.}\ \bibnamefont {{Zeng}}},
  \bibinfo {author} {\bibfnamefont {Y.~J.}\ \bibnamefont {{Yu}}}, \bibinfo
  {author} {\bibfnamefont {E.~J.}\ \bibnamefont {{Cheng}}}, \bibinfo {author}
  {\bibfnamefont {L.~S.}\ \bibnamefont {{Wang}}}, \bibinfo {author}
  {\bibfnamefont {R.}~\bibnamefont {{Kato}}}, \ and\ \bibinfo {author}
  {\bibfnamefont {S.~Y.}\ \bibnamefont {{Li}}},\ }\bibfield  {title} {\enquote
  {\bibinfo {title} {{Absence of magnetic thermal conductivity in the quantum
  spin liquid candidate EtMe$_3$Sb[Pd(dmit)$_2$]$_2$ -- revisited}},}\ }\href@noop {} {\
  ,\ \bibinfo {pages} {arXiv:1904.10395}}\BibitemShut {NoStop}%
\bibitem [{\citenamefont {Helton}\ \emph {et~al.}(2007)\citenamefont {Helton},
  \citenamefont {Matan}, \citenamefont {Shores}, \citenamefont {Nytko},
  \citenamefont {Bartlett}, \citenamefont {Yoshida}, \citenamefont {Takano},
  \citenamefont {Suslov}, \citenamefont {Qiu}, \citenamefont {Chung},
  \citenamefont {Nocera},\ and\ \citenamefont {Lee}}]{heltonPRL07}%
  \BibitemOpen
  \bibfield  {author} {\bibinfo {author} {\bibfnamefont {J.~S.}\ \bibnamefont
  {Helton}}, \bibinfo {author} {\bibfnamefont {K.}~\bibnamefont {Matan}},
  \bibinfo {author} {\bibfnamefont {M.~P.}\ \bibnamefont {Shores}}, \bibinfo
  {author} {\bibfnamefont {E.~A.}\ \bibnamefont {Nytko}}, \bibinfo {author}
  {\bibfnamefont {B.~M.}\ \bibnamefont {Bartlett}}, \bibinfo {author}
  {\bibfnamefont {Y.}~\bibnamefont {Yoshida}}, \bibinfo {author} {\bibfnamefont
  {Y.}~\bibnamefont {Takano}}, \bibinfo {author} {\bibfnamefont
  {A.}~\bibnamefont {Suslov}}, \bibinfo {author} {\bibfnamefont
  {Y.}~\bibnamefont {Qiu}}, \bibinfo {author} {\bibfnamefont {J.-H.}\
  \bibnamefont {Chung}}, \bibinfo {author} {\bibfnamefont {D.~G.}\ \bibnamefont
  {Nocera}}, \ and\ \bibinfo {author} {\bibfnamefont {Y.~S.}\ \bibnamefont
  {Lee}},\ }\bibfield  {title} {\enquote {\bibinfo {title} {Spin dynamics of
  the spin-$1/2$ kagome lattice antiferromagnet
  ${\mathrm{ZnCu}}_{3}(\mathrm{OH}{)}_{6}{\mathrm{Cl}}_{2}$},}\ }\href
  {\doibase 10.1103/PhysRevLett.98.107204} {\bibfield  {journal} {\bibinfo
  {journal} {Phys. Rev. Lett.}\ }\textbf {\bibinfo {volume} {98}},\ \bibinfo
  {pages} {107204} (\bibinfo {year} {2007})}\BibitemShut {NoStop}%
\bibitem [{\citenamefont {de~Vries}\ \emph {et~al.}(2009)\citenamefont
  {de~Vries}, \citenamefont {Stewart}, \citenamefont {Deen}, \citenamefont
  {Piatek}, \citenamefont {Nilsen}, \citenamefont {R\o{}nnow},\ and\
  \citenamefont {Harrison}}]{devriesPRL09}%
  \BibitemOpen
  \bibfield  {author} {\bibinfo {author} {\bibfnamefont {M.~A.}\ \bibnamefont
  {de~Vries}}, \bibinfo {author} {\bibfnamefont {J.~R.}\ \bibnamefont
  {Stewart}}, \bibinfo {author} {\bibfnamefont {P.~P.}\ \bibnamefont {Deen}},
  \bibinfo {author} {\bibfnamefont {J.~O.}\ \bibnamefont {Piatek}}, \bibinfo
  {author} {\bibfnamefont {G.~J.}\ \bibnamefont {Nilsen}}, \bibinfo {author}
  {\bibfnamefont {H.~M.}\ \bibnamefont {R\o{}nnow}}, \ and\ \bibinfo {author}
  {\bibfnamefont {A.}~\bibnamefont {Harrison}},\ }\bibfield  {title} {\enquote
  {\bibinfo {title} {Scale-free antiferromagnetic fluctuations in the $s=1/2$
  kagome antiferromagnet herbertsmithite},}\ }\href {\doibase
  10.1103/PhysRevLett.103.237201} {\bibfield  {journal} {\bibinfo  {journal}
  {Phys. Rev. Lett.}\ }\textbf {\bibinfo {volume} {103}},\ \bibinfo {pages}
  {237201} (\bibinfo {year} {2009})}\BibitemShut {NoStop}%
\bibitem [{\citenamefont {Helton}\ \emph {et~al.}(2010)\citenamefont {Helton},
  \citenamefont {Matan}, \citenamefont {Shores}, \citenamefont {Nytko},
  \citenamefont {Bartlett}, \citenamefont {Qiu}, \citenamefont {Nocera},\ and\
  \citenamefont {Lee}}]{heltonPRL10}%
  \BibitemOpen
  \bibfield  {author} {\bibinfo {author} {\bibfnamefont {J.~S.}\ \bibnamefont
  {Helton}}, \bibinfo {author} {\bibfnamefont {K.}~\bibnamefont {Matan}},
  \bibinfo {author} {\bibfnamefont {M.~P.}\ \bibnamefont {Shores}}, \bibinfo
  {author} {\bibfnamefont {E.~A.}\ \bibnamefont {Nytko}}, \bibinfo {author}
  {\bibfnamefont {B.~M.}\ \bibnamefont {Bartlett}}, \bibinfo {author}
  {\bibfnamefont {Y.}~\bibnamefont {Qiu}}, \bibinfo {author} {\bibfnamefont
  {D.~G.}\ \bibnamefont {Nocera}}, \ and\ \bibinfo {author} {\bibfnamefont
  {Y.~S.}\ \bibnamefont {Lee}},\ }\bibfield  {title} {\enquote {\bibinfo
  {title} {Dynamic scaling in the susceptibility of the spin-$\frac{1}{2}$
  kagome lattice antiferromagnet herbertsmithite},}\ }\href {\doibase
  10.1103/PhysRevLett.104.147201} {\bibfield  {journal} {\bibinfo  {journal}
  {Phys. Rev. Lett.}\ }\textbf {\bibinfo {volume} {104}},\ \bibinfo {pages}
  {147201} (\bibinfo {year} {2010})}\BibitemShut {NoStop}%
\bibitem [{\citenamefont {Banerjee}\ \emph {et~al.}(2016)\citenamefont
  {Banerjee}, \citenamefont {Bridges}, \citenamefont {Yan}, \citenamefont
  {Aczel}, \citenamefont {Li}, \citenamefont {Stone}, \citenamefont {Granroth},
  \citenamefont {Lumsden}, \citenamefont {Yiu}, \citenamefont {Knolle},
  \citenamefont {Bhattacharjee}, \citenamefont {Kovrizhin}, \citenamefont
  {Moessner}, \citenamefont {Tennant}, \citenamefont {Mandrus},\ and\
  \citenamefont {Nagler}}]{banerjeeNATM16}%
  \BibitemOpen
  \bibfield  {author} {\bibinfo {author} {\bibfnamefont {A.}~\bibnamefont
  {Banerjee}}, \bibinfo {author} {\bibfnamefont {C.~A.}\ \bibnamefont
  {Bridges}}, \bibinfo {author} {\bibfnamefont {J.~Q.}\ \bibnamefont {Yan}},
  \bibinfo {author} {\bibfnamefont {A.~A.}\ \bibnamefont {Aczel}}, \bibinfo
  {author} {\bibfnamefont {L.}~\bibnamefont {Li}}, \bibinfo {author}
  {\bibfnamefont {M.~B.}\ \bibnamefont {Stone}}, \bibinfo {author}
  {\bibfnamefont {G.~E.}\ \bibnamefont {Granroth}}, \bibinfo {author}
  {\bibfnamefont {M.~D.}\ \bibnamefont {Lumsden}}, \bibinfo {author}
  {\bibfnamefont {Y.}~\bibnamefont {Yiu}}, \bibinfo {author} {\bibfnamefont
  {J.}~\bibnamefont {Knolle}}, \bibinfo {author} {\bibfnamefont
  {S.}~\bibnamefont {Bhattacharjee}}, \bibinfo {author} {\bibfnamefont {D.~L.}\
  \bibnamefont {Kovrizhin}}, \bibinfo {author} {\bibfnamefont {R.}~\bibnamefont
  {Moessner}}, \bibinfo {author} {\bibfnamefont {D.~A.}\ \bibnamefont
  {Tennant}}, \bibinfo {author} {\bibfnamefont {D.~G.}\ \bibnamefont
  {Mandrus}}, \ and\ \bibinfo {author} {\bibfnamefont {S.~E.}\ \bibnamefont
  {Nagler}},\ }\bibfield  {title} {\enquote {\bibinfo {title} {Proximate kitaev
  quantum spin liquid behaviour in a honeycomb magnet},}\ }\href
  {https://doi.org/10.1038/nmat4604} {\bibfield  {journal} {\bibinfo  {journal}
  {Nature Mater.}\ }\textbf {\bibinfo {volume} {15}},\ \bibinfo {pages} {733 EP
  --} (\bibinfo {year} {2016})}\BibitemShut {NoStop}%
\bibitem [{\citenamefont {Sachdev}(1992)}]{sachdevPRB92}%
  \BibitemOpen
  \bibfield  {author} {\bibinfo {author} {\bibfnamefont {Subir}\ \bibnamefont
  {Sachdev}},\ }\bibfield  {title} {\enquote {\bibinfo {title} {Kagom\'e- and
  triangular-lattice heisenberg antiferromagnets: Ordering from quantum
  fluctuations and quantum-disordered ground states with unconfined bosonic
  spinons},}\ }\href {\doibase 10.1103/PhysRevB.45.12377} {\bibfield  {journal}
  {\bibinfo  {journal} {Phys. Rev. B}\ }\textbf {\bibinfo {volume} {45}},\
  \bibinfo {pages} {12377--12396} (\bibinfo {year} {1992})}\BibitemShut
  {NoStop}%
\bibitem [{\citenamefont {Yildirim}\ and\ \citenamefont
  {Harris}(2006)}]{yildirimPRB06}%
  \BibitemOpen
  \bibfield  {author} {\bibinfo {author} {\bibfnamefont {T.}~\bibnamefont
  {Yildirim}}\ and\ \bibinfo {author} {\bibfnamefont {A.~B.}\ \bibnamefont
  {Harris}},\ }\bibfield  {title} {\enquote {\bibinfo {title} {Magnetic
  structure and spin waves in the kagom\'e jarosite compound
  $\mathrm{K}{\mathrm{Fe}}_{3}{(\mathrm{S}{\mathrm{O}}_{4})}_{2}{(\mathrm{O}\mathrm{H})}_{6}$},}\
  }\href {\doibase 10.1103/PhysRevB.73.214446} {\bibfield  {journal} {\bibinfo
  {journal} {Phys. Rev. B}\ }\textbf {\bibinfo {volume} {73}},\ \bibinfo
  {pages} {214446} (\bibinfo {year} {2006})}\BibitemShut {NoStop}%
\bibitem [{\citenamefont {Yan}\ \emph {et~al.}(2011)\citenamefont {Yan},
  \citenamefont {Huse},\ and\ \citenamefont {White}}]{yanSCI11}%
  \BibitemOpen
  \bibfield  {author} {\bibinfo {author} {\bibfnamefont {Simeng}\ \bibnamefont
  {Yan}}, \bibinfo {author} {\bibfnamefont {David~A.}\ \bibnamefont {Huse}}, \
  and\ \bibinfo {author} {\bibfnamefont {Steven~R.}\ \bibnamefont {White}},\
  }\bibfield  {title} {\enquote {\bibinfo {title} {Spin-liquid ground state of
  the s = 1/2 kagome heisenberg antiferromagnet},}\ }\href {\doibase
  10.1126/science.1201080} {\bibfield  {journal} {\bibinfo  {journal}
  {Science}\ }\textbf {\bibinfo {volume} {332}},\ \bibinfo {pages} {1173--1176}
  (\bibinfo {year} {2011})}\BibitemShut {NoStop}%
\bibitem [{\citenamefont {Jeschke}\ \emph {et~al.}(2013)\citenamefont
  {Jeschke}, \citenamefont {Salvat-Pujol},\ and\ \citenamefont
  {Valent\'{\i}}}]{jeschkePRB13}%
  \BibitemOpen
  \bibfield  {author} {\bibinfo {author} {\bibfnamefont {Harald~O.}\
  \bibnamefont {Jeschke}}, \bibinfo {author} {\bibfnamefont {Francesc}\
  \bibnamefont {Salvat-Pujol}}, \ and\ \bibinfo {author} {\bibfnamefont
  {Roser}\ \bibnamefont {Valent\'{\i}}},\ }\bibfield  {title} {\enquote
  {\bibinfo {title} {First-principles determination of heisenberg hamiltonian
  parameters for the spin-$\frac{1}{2}$ kagome antiferromagnet
  ZnCu${}_{3}$(OH)${}_{6}$Cl${}_{2}$},}\ }\href {\doibase
  10.1103/PhysRevB.88.075106} {\bibfield  {journal} {\bibinfo  {journal} {Phys.
  Rev. B}\ }\textbf {\bibinfo {volume} {88}},\ \bibinfo {pages} {075106}
  (\bibinfo {year} {2013})}\BibitemShut {NoStop}%
\bibitem [{\citenamefont {Motrunich}(2005)}]{motrunichPRB05}%
  \BibitemOpen
  \bibfield  {author} {\bibinfo {author} {\bibfnamefont {Olexei~I.}\
  \bibnamefont {Motrunich}},\ }\bibfield  {title} {\enquote {\bibinfo {title}
  {Variational study of triangular lattice spin-$1/2$ model with ring
  exchanges and spin liquid state in
  $\ensuremath{\kappa}\text{-}{(\mathrm{ET})}_{2}{\mathrm{Cu}}_{2}{(\mathrm{CN})}_{3}$},}\
  }\href {\doibase 10.1103/PhysRevB.72.045105} {\bibfield  {journal} {\bibinfo
  {journal} {Phys. Rev. B}\ }\textbf {\bibinfo {volume} {72}},\ \bibinfo
  {pages} {045105} (\bibinfo {year} {2005})}\BibitemShut {NoStop}%
\bibitem [{\citenamefont {Lee}\ and\ \citenamefont {Lee}(2005)}]{leePRL05}%
  \BibitemOpen
  \bibfield  {author} {\bibinfo {author} {\bibfnamefont {Sung-Sik}\
  \bibnamefont {Lee}}\ and\ \bibinfo {author} {\bibfnamefont {Patrick~A.}\
  \bibnamefont {Lee}},\ }\bibfield  {title} {\enquote {\bibinfo {title} {U(1)
  gauge theory of the hubbard model: Spin liquid states and possible
  application to
  $\ensuremath{\kappa}\mathrm{\text{-}}(\mathrm{BEDT}\mathrm{\text{-}}\mathrm{TTF}{)}_{2}{\mathrm{Cu}}_{2}(\mathrm{CN}{)}_{3}$},}\
  }\href {\doibase 10.1103/PhysRevLett.95.036403} {\bibfield  {journal}
  {\bibinfo  {journal} {Phys. Rev. Lett.}\ }\textbf {\bibinfo {volume} {95}},\
  \bibinfo {pages} {036403} (\bibinfo {year} {2005})}\BibitemShut {NoStop}%
\bibitem [{\citenamefont {Nave}\ and\ \citenamefont {Lee}(2007)}]{navePRB07}%
  \BibitemOpen
  \bibfield  {author} {\bibinfo {author} {\bibfnamefont {Cody~P.}\ \bibnamefont
  {Nave}}\ and\ \bibinfo {author} {\bibfnamefont {Patrick~A.}\ \bibnamefont
  {Lee}},\ }\bibfield  {title} {\enquote {\bibinfo {title} {Transport
  properties of a spinon fermi surface coupled to a U(1) gauge field},}\ }\href
  {\doibase 10.1103/PhysRevB.76.235124} {\bibfield  {journal} {\bibinfo
  {journal} {Phys. Rev. B}\ }\textbf {\bibinfo {volume} {76}},\ \bibinfo
  {pages} {235124} (\bibinfo {year} {2007})}\BibitemShut {NoStop}%
\bibitem [{\citenamefont {Lee}(2008)}]{leePRB08}%
  \BibitemOpen
  \bibfield  {author} {\bibinfo {author} {\bibfnamefont {Sung-Sik}\
  \bibnamefont {Lee}},\ }\bibfield  {title} {\enquote {\bibinfo {title}
  {Stability of the U(1) spin liquid with a spinon fermi surface in $2+1$
  dimensions},}\ }\href {\doibase 10.1103/PhysRevB.78.085129} {\bibfield
  {journal} {\bibinfo  {journal} {Phys. Rev. B}\ }\textbf {\bibinfo {volume}
  {78}},\ \bibinfo {pages} {085129} (\bibinfo {year} {2008})}\BibitemShut
  {NoStop}%
\bibitem [{\citenamefont {Lee}(2009)}]{leePRB09}%
  \BibitemOpen
  \bibfield  {author} {\bibinfo {author} {\bibfnamefont {Sung-Sik}\
  \bibnamefont {Lee}},\ }\bibfield  {title} {\enquote {\bibinfo {title}
  {Low-energy effective theory of fermi surface coupled with U(1) gauge field
  in $2+1$ dimensions},}\ }\href {\doibase 10.1103/PhysRevB.80.165102}
  {\bibfield  {journal} {\bibinfo  {journal} {Phys. Rev. B}\ }\textbf {\bibinfo
  {volume} {80}},\ \bibinfo {pages} {165102} (\bibinfo {year}
  {2009})}\BibitemShut {NoStop}%
\bibitem [{\citenamefont {Sanyal}\ \emph {et~al.}(2019)\citenamefont {Sanyal},
  \citenamefont {Dhochak},\ and\ \citenamefont {Bhattacharjee}}]{sanyalPRB19}%
  \BibitemOpen
  \bibfield  {author} {\bibinfo {author} {\bibfnamefont {Sambuddha}\
  \bibnamefont {Sanyal}}, \bibinfo {author} {\bibfnamefont {Kusum}\
  \bibnamefont {Dhochak}}, \ and\ \bibinfo {author} {\bibfnamefont {Subhro}\
  \bibnamefont {Bhattacharjee}},\ }\bibfield  {title} {\enquote {\bibinfo
  {title} {Interplay of uniform U(1) quantum spin liquid and magnetic phases
  in rare-earth pyrochlore magnets: A fermionic parton approach},}\ }\href
  {\doibase 10.1103/PhysRevB.99.134425} {\bibfield  {journal} {\bibinfo
  {journal} {Phys. Rev. B}\ }\textbf {\bibinfo {volume} {99}},\ \bibinfo
  {pages} {134425} (\bibinfo {year} {2019})}\BibitemShut {NoStop}%
\bibitem [{\citenamefont {Kitaev}(2006)}]{kitaevAP06}%
  \BibitemOpen
  \bibfield  {author} {\bibinfo {author} {\bibfnamefont {Alexei}\ \bibnamefont
  {Kitaev}},\ }\bibfield  {title} {\enquote {\bibinfo {title} {Anyons in an
  exactly solved model and beyond},}\ }\href@noop {} {\bibfield  {journal}
  {\bibinfo  {journal} {Annals of Physics}\ }\textbf {\bibinfo {volume}
  {321}},\ \bibinfo {pages} {2 -- 111} (\bibinfo {year} {2006})}\BibitemShut
  {NoStop}%
\bibitem [{\citenamefont {Jackeli}\ and\ \citenamefont
  {Khaliullin}(2009)}]{jackeliPRL09}%
  \BibitemOpen
  \bibfield  {author} {\bibinfo {author} {\bibfnamefont {G.}~\bibnamefont
  {Jackeli}}\ and\ \bibinfo {author} {\bibfnamefont {G.}~\bibnamefont
  {Khaliullin}},\ }\bibfield  {title} {\enquote {\bibinfo {title} {Mott
  insulators in the strong spin-orbit coupling limit: From heisenberg to a
  quantum compass and kitaev models},}\ }\href {\doibase
  10.1103/PhysRevLett.102.017205} {\bibfield  {journal} {\bibinfo  {journal}
  {Phys. Rev. Lett.}\ }\textbf {\bibinfo {volume} {102}},\ \bibinfo {pages}
  {017205} (\bibinfo {year} {2009})}\BibitemShut {NoStop}%
\bibitem [{\citenamefont {Knolle}\ \emph {et~al.}(2014)\citenamefont {Knolle},
  \citenamefont {Kovrizhin}, \citenamefont {Chalker},\ and\ \citenamefont
  {Moessner}}]{knollePRL14}%
  \BibitemOpen
  \bibfield  {author} {\bibinfo {author} {\bibfnamefont {J.}~\bibnamefont
  {Knolle}}, \bibinfo {author} {\bibfnamefont {D.~L.}\ \bibnamefont
  {Kovrizhin}}, \bibinfo {author} {\bibfnamefont {J.~T.}\ \bibnamefont
  {Chalker}}, \ and\ \bibinfo {author} {\bibfnamefont {R.}~\bibnamefont
  {Moessner}},\ }\bibfield  {title} {\enquote {\bibinfo {title} {Dynamics of a
  two-dimensional quantum spin liquid: Signatures of emergent majorana fermions
  and fluxes},}\ }\href {\doibase 10.1103/PhysRevLett.112.207203} {\bibfield
  {journal} {\bibinfo  {journal} {Phys. Rev. Lett.}\ }\textbf {\bibinfo
  {volume} {112}},\ \bibinfo {pages} {207203} (\bibinfo {year}
  {2014})}\BibitemShut {NoStop}%
\bibitem [{\citenamefont {Mandal}\ and\ \citenamefont
  {Surendran}(2009)}]{mandalPRB09}%
  \BibitemOpen
  \bibfield  {author} {\bibinfo {author} {\bibfnamefont {Saptarshi}\
  \bibnamefont {Mandal}}\ and\ \bibinfo {author} {\bibfnamefont {Naveen}\
  \bibnamefont {Surendran}},\ }\bibfield  {title} {\enquote {\bibinfo {title}
  {Exactly solvable kitaev model in three dimensions},}\ }\href {\doibase
  10.1103/PhysRevB.79.024426} {\bibfield  {journal} {\bibinfo  {journal} {Phys.
  Rev. B}\ }\textbf {\bibinfo {volume} {79}},\ \bibinfo {pages} {024426}
  (\bibinfo {year} {2009})}\BibitemShut {NoStop}%
\bibitem [{\citenamefont {Nasu}\ \emph {et~al.}(2014)\citenamefont {Nasu},
  \citenamefont {Udagawa},\ and\ \citenamefont {Motome}}]{nasuPRL14}%
  \BibitemOpen
  \bibfield  {author} {\bibinfo {author} {\bibfnamefont {Joji}\ \bibnamefont
  {Nasu}}, \bibinfo {author} {\bibfnamefont {Masafumi}\ \bibnamefont
  {Udagawa}}, \ and\ \bibinfo {author} {\bibfnamefont {Yukitoshi}\ \bibnamefont
  {Motome}},\ }\bibfield  {title} {\enquote {\bibinfo {title} {Vaporization of
  kitaev spin liquids},}\ }\href {\doibase 10.1103/PhysRevLett.113.197205}
  {\bibfield  {journal} {\bibinfo  {journal} {Phys. Rev. Lett.}\ }\textbf
  {\bibinfo {volume} {113}},\ \bibinfo {pages} {197205} (\bibinfo {year}
  {2014})}\BibitemShut {NoStop}%
\bibitem [{\citenamefont {Knolle}\ and\ \citenamefont
  {Moessner}(2019)}]{knolleAR19}%
  \BibitemOpen
  \bibfield  {author} {\bibinfo {author} {\bibfnamefont {J.}~\bibnamefont
  {Knolle}}\ and\ \bibinfo {author} {\bibfnamefont {R.}~\bibnamefont
  {Moessner}},\ }\bibfield  {title} {\enquote {\bibinfo {title} {A field guide
  to spin liquids},}\ }\href@noop {} {\bibfield  {journal} {\bibinfo  {journal}
  {Annual Review of Condensed Matter Physics}\ }\textbf {\bibinfo {volume}
  {10}},\ \bibinfo {pages} {451--472} (\bibinfo {year} {2019})}\BibitemShut
  {NoStop}%
\bibitem [{\citenamefont {Saitoh}\ \emph {et~al.}(2006)\citenamefont {Saitoh},
  \citenamefont {Ueda}, \citenamefont {Miyajima},\ and\ \citenamefont
  {Tatara}}]{saitohAPL06}%
  \BibitemOpen
  \bibfield  {author} {\bibinfo {author} {\bibfnamefont {E.}~\bibnamefont
  {Saitoh}}, \bibinfo {author} {\bibfnamefont {M.}~\bibnamefont {Ueda}},
  \bibinfo {author} {\bibfnamefont {H.}~\bibnamefont {Miyajima}}, \ and\
  \bibinfo {author} {\bibfnamefont {G.}~\bibnamefont {Tatara}},\ }\bibfield
  {title} {\enquote {\bibinfo {title} {Conversion of spin current into charge
  current at room temperature: Inverse spin-hall effect},}\ }\href@noop {}
  {\bibfield  {journal} {\bibinfo  {journal} {Appl. Phys. Lett.}\ }\textbf
  {\bibinfo {volume} {88}},\ \bibinfo {eid} {182509} (\bibinfo {year}
  {2006})}\BibitemShut {NoStop}%
\bibitem [{\citenamefont {Sinova}\ \emph {et~al.}(2015)\citenamefont {Sinova},
  \citenamefont {Valenzuela}, \citenamefont {Wunderlich}, \citenamefont
  {Back},\ and\ \citenamefont {Jungwirth}}]{sinovaRMP15}%
  \BibitemOpen
  \bibfield  {author} {\bibinfo {author} {\bibfnamefont {Jairo}\ \bibnamefont
  {Sinova}}, \bibinfo {author} {\bibfnamefont {Sergio~O.}\ \bibnamefont
  {Valenzuela}}, \bibinfo {author} {\bibfnamefont {J.}~\bibnamefont
  {Wunderlich}}, \bibinfo {author} {\bibfnamefont {C.~H.}\ \bibnamefont
  {Back}}, \ and\ \bibinfo {author} {\bibfnamefont {T.}~\bibnamefont
  {Jungwirth}},\ }\bibfield  {title} {\enquote {\bibinfo {title} {Spin hall
  effects},}\ }\href {\doibase 10.1103/RevModPhys.87.1213} {\bibfield
  {journal} {\bibinfo  {journal} {Rev. Mod. Phys.}\ }\textbf {\bibinfo {volume}
  {87}},\ \bibinfo {pages} {1213--1260} (\bibinfo {year} {2015})}\BibitemShut
  {NoStop}%
\bibitem [{\citenamefont {Uchida}\ \emph {et~al.}(2008)\citenamefont {Uchida},
  \citenamefont {Takahashi}, \citenamefont {Harii}, \citenamefont {Ieda},
  \citenamefont {Koshibae}, \citenamefont {Ando}, \citenamefont {Maekawa},\
  and\ \citenamefont {Saitoh}}]{uchidaNAT08}%
  \BibitemOpen
  \bibfield  {author} {\bibinfo {author} {\bibfnamefont {K.}~\bibnamefont
  {Uchida}}, \bibinfo {author} {\bibfnamefont {S.}~\bibnamefont {Takahashi}},
  \bibinfo {author} {\bibfnamefont {K.}~\bibnamefont {Harii}}, \bibinfo
  {author} {\bibfnamefont {J.}~\bibnamefont {Ieda}}, \bibinfo {author}
  {\bibfnamefont {W.}~\bibnamefont {Koshibae}}, \bibinfo {author}
  {\bibfnamefont {K.}~\bibnamefont {Ando}}, \bibinfo {author} {\bibfnamefont
  {S.}~\bibnamefont {Maekawa}}, \ and\ \bibinfo {author} {\bibfnamefont
  {E.}~\bibnamefont {Saitoh}},\ }\bibfield  {title} {\enquote {\bibinfo {title}
  {Observation of the spin seebeck effect},}\ }\href@noop {} {\bibfield
  {journal} {\bibinfo  {journal} {Nature}\ }\textbf {\bibinfo {volume} {455}},\
  \bibinfo {pages} {778--781} (\bibinfo {year} {2008})}\BibitemShut {NoStop}%
\bibitem [{\citenamefont {Uchida}\ \emph {et~al.}(2010)\citenamefont {Uchida},
  \citenamefont {Adachi}, \citenamefont {Ota}, \citenamefont {Nakayama},
  \citenamefont {Maekawa},\ and\ \citenamefont {Saitoh}}]{uchidaAPL10}%
  \BibitemOpen
  \bibfield  {author} {\bibinfo {author} {\bibfnamefont {Kenichi}\ \bibnamefont
  {Uchida}}, \bibinfo {author} {\bibfnamefont {Hiroto}\ \bibnamefont {Adachi}},
  \bibinfo {author} {\bibfnamefont {Takeru}\ \bibnamefont {Ota}}, \bibinfo
  {author} {\bibfnamefont {Hiroyasu}\ \bibnamefont {Nakayama}}, \bibinfo
  {author} {\bibfnamefont {Sadamichi}\ \bibnamefont {Maekawa}}, \ and\ \bibinfo
  {author} {\bibfnamefont {Eiji}\ \bibnamefont {Saitoh}},\ }\bibfield  {title}
  {\enquote {\bibinfo {title} {Observation of longitudinal spin-seebeck effect
  in magnetic insulators},}\ }\href {\doibase 10.1063/1.3507386} {\bibfield
  {journal} {\bibinfo  {journal} {Appl. Phys. Lett.}\ }\textbf {\bibinfo
  {volume} {97}},\ \bibinfo {eid} {172505} (\bibinfo {year}
  {2010})}\BibitemShut {NoStop}%
\bibitem [{\citenamefont {Hirobe}\ \emph {et~al.}(2017)\citenamefont {Hirobe},
  \citenamefont {Sato}, \citenamefont {Kawamata}, \citenamefont {Shiomi},
  \citenamefont {Uchida}, \citenamefont {Iguchi}, \citenamefont {Koike},
  \citenamefont {Maekawa},\ and\ \citenamefont {Saitoh}}]{hirobeNATP16}%
  \BibitemOpen
  \bibfield  {author} {\bibinfo {author} {\bibfnamefont {Daichi}\ \bibnamefont
  {Hirobe}}, \bibinfo {author} {\bibfnamefont {Masahiro}\ \bibnamefont {Sato}},
  \bibinfo {author} {\bibfnamefont {Takayuki}\ \bibnamefont {Kawamata}},
  \bibinfo {author} {\bibfnamefont {Yuki}\ \bibnamefont {Shiomi}}, \bibinfo
  {author} {\bibfnamefont {Ken-ichi}\ \bibnamefont {Uchida}}, \bibinfo {author}
  {\bibfnamefont {Ryo}\ \bibnamefont {Iguchi}}, \bibinfo {author}
  {\bibfnamefont {Yoji}\ \bibnamefont {Koike}}, \bibinfo {author}
  {\bibfnamefont {Sadamichi}\ \bibnamefont {Maekawa}}, \ and\ \bibinfo {author}
  {\bibfnamefont {Eiji}\ \bibnamefont {Saitoh}},\ }\bibfield  {title} {\enquote
  {\bibinfo {title} {One-dimensional spinon spin currents},}\ }\href@noop {}
  {\bibfield  {journal} {\bibinfo  {journal} {Nature Phys.}\ }\textbf {\bibinfo
  {volume} {13}},\ \bibinfo {pages} {30--34} (\bibinfo {year}
  {2017})}\BibitemShut {NoStop}%
\bibitem [{\citenamefont {Kajiwara}\ \emph {et~al.}(2010)\citenamefont
  {Kajiwara}, \citenamefont {Harii}, \citenamefont {Takahashi}, \citenamefont
  {Ohe}, \citenamefont {Uchida}, \citenamefont {Mizuguchi}, \citenamefont
  {Umezawa}, \citenamefont {Kawai}, \citenamefont {Ando}, \citenamefont
  {Takanashi}, \citenamefont {Maekawa},\ and\ \citenamefont
  {Saitoh}}]{kajiwaraNAT10}%
  \BibitemOpen
  \bibfield  {author} {\bibinfo {author} {\bibfnamefont {Y.}~\bibnamefont
  {Kajiwara}}, \bibinfo {author} {\bibfnamefont {K.}~\bibnamefont {Harii}},
  \bibinfo {author} {\bibfnamefont {S.}~\bibnamefont {Takahashi}}, \bibinfo
  {author} {\bibfnamefont {J.}~\bibnamefont {Ohe}}, \bibinfo {author}
  {\bibfnamefont {K.}~\bibnamefont {Uchida}}, \bibinfo {author} {\bibfnamefont
  {M.}~\bibnamefont {Mizuguchi}}, \bibinfo {author} {\bibfnamefont
  {H.}~\bibnamefont {Umezawa}}, \bibinfo {author} {\bibfnamefont
  {H.}~\bibnamefont {Kawai}}, \bibinfo {author} {\bibfnamefont
  {K.}~\bibnamefont {Ando}}, \bibinfo {author} {\bibfnamefont {K.}~\bibnamefont
  {Takanashi}}, \bibinfo {author} {\bibfnamefont {S.}~\bibnamefont {Maekawa}},
  \ and\ \bibinfo {author} {\bibfnamefont {E.}~\bibnamefont {Saitoh}},\
  }\bibfield  {title} {\enquote {\bibinfo {title} {Transmission of electrical
  signals by spin-wave interconversion in a magnetic insulator},}\ }\href@noop
  {} {\bibfield  {journal} {\bibinfo  {journal} {Nature}\ }\textbf {\bibinfo
  {volume} {464}},\ \bibinfo {pages} {262--266} (\bibinfo {year}
  {2010})}\BibitemShut {NoStop}%
\bibitem [{\citenamefont {Cornelissen}\ \emph {et~al.}(2015)\citenamefont
  {Cornelissen}, \citenamefont {Liu}, \citenamefont {Duine}, \citenamefont
  {Youssef},\ and\ \citenamefont {van Wees}}]{cornelissenNATP15}%
  \BibitemOpen
  \bibfield  {author} {\bibinfo {author} {\bibfnamefont {L.~J.}\ \bibnamefont
  {Cornelissen}}, \bibinfo {author} {\bibfnamefont {J.}~\bibnamefont {Liu}},
  \bibinfo {author} {\bibfnamefont {R.~A.}\ \bibnamefont {Duine}}, \bibinfo
  {author} {\bibfnamefont {J.~Ben}\ \bibnamefont {Youssef}}, \ and\ \bibinfo
  {author} {\bibfnamefont {B.~J.}\ \bibnamefont {van Wees}},\ }\bibfield
  {title} {\enquote {\bibinfo {title} {Long-distance transport of magnon spin
  information in a magnetic insulator at room temperature},}\ }\href@noop {}
  {\bibfield  {journal} {\bibinfo  {journal} {Nature Phys.}\ }\textbf {\bibinfo
  {volume} {11}},\ \bibinfo {pages} {1022--1026} (\bibinfo {year}
  {2015})}\BibitemShut {NoStop}%
\bibitem [{\citenamefont {Goennenwein}\ \emph {et~al.}(2015)\citenamefont
  {Goennenwein}, \citenamefont {Schlitz}, \citenamefont {Pernpeintner},
  \citenamefont {Ganzhorn}, \citenamefont {Althammer}, \citenamefont {Gross},\
  and\ \citenamefont {Huebl}}]{goennenweinAPL15}%
  \BibitemOpen
  \bibfield  {author} {\bibinfo {author} {\bibfnamefont {Sebastian T.~B.}\
  \bibnamefont {Goennenwein}}, \bibinfo {author} {\bibfnamefont {Richard}\
  \bibnamefont {Schlitz}}, \bibinfo {author} {\bibfnamefont {Matthias}\
  \bibnamefont {Pernpeintner}}, \bibinfo {author} {\bibfnamefont {Kathrin}\
  \bibnamefont {Ganzhorn}}, \bibinfo {author} {\bibfnamefont {Matthias}\
  \bibnamefont {Althammer}}, \bibinfo {author} {\bibfnamefont {Rudolf}\
  \bibnamefont {Gross}}, \ and\ \bibinfo {author} {\bibfnamefont {Hans}\
  \bibnamefont {Huebl}},\ }\bibfield  {title} {\enquote {\bibinfo {title}
  {Non-local magnetoresistance in YIG/Pt nanostructures},}\ }\href {\doibase
  10.1063/1.4935074} {\bibfield  {journal} {\bibinfo  {journal} {Appl. Phys.
  Lett.}\ }\textbf {\bibinfo {volume} {107}},\ \bibinfo {pages} {172405}
  (\bibinfo {year} {2015})}\BibitemShut {NoStop}%
\bibitem [{\citenamefont {Li}\ \emph {et~al.}(2016{\natexlab{a}})\citenamefont
  {Li}, \citenamefont {Xu}, \citenamefont {Aldosary}, \citenamefont {Tang},
  \citenamefont {Lin}, \citenamefont {Zhang}, \citenamefont {Lake},\ and\
  \citenamefont {Shi}}]{liNATC16}%
  \BibitemOpen
  \bibfield  {author} {\bibinfo {author} {\bibfnamefont {Junxue}\ \bibnamefont
  {Li}}, \bibinfo {author} {\bibfnamefont {Yadong}\ \bibnamefont {Xu}},
  \bibinfo {author} {\bibfnamefont {Mohammed}\ \bibnamefont {Aldosary}},
  \bibinfo {author} {\bibfnamefont {Chi}\ \bibnamefont {Tang}}, \bibinfo
  {author} {\bibfnamefont {Zhisheng}\ \bibnamefont {Lin}}, \bibinfo {author}
  {\bibfnamefont {Shufeng}\ \bibnamefont {Zhang}}, \bibinfo {author}
  {\bibfnamefont {Roger}\ \bibnamefont {Lake}}, \ and\ \bibinfo {author}
  {\bibfnamefont {Jing}\ \bibnamefont {Shi}},\ }\bibfield  {title} {\enquote
  {\bibinfo {title} {Observation of magnon-mediated current drag in Pt/yttrium
  iron garnet/Pt(Ta) trilayers},}\ }\href
  {http://dx.doi.org/10.1038/ncomms10858} {\bibfield  {journal} {\bibinfo
  {journal} {Nature Comm.}\ }\textbf {\bibinfo {volume} {7}} (\bibinfo {year}
  {2016}{\natexlab{a}})}\BibitemShut {NoStop}%
\bibitem [{\citenamefont {Han}\ \emph {et~al.}(2019)\citenamefont {Han},
  \citenamefont {Maekawa},\ and\ \citenamefont {Xie}}]{hanNATM19}%
  \BibitemOpen
  \bibfield  {author} {\bibinfo {author} {\bibfnamefont {Wei}\ \bibnamefont
  {Han}}, \bibinfo {author} {\bibfnamefont {Sadamichi}\ \bibnamefont
  {Maekawa}}, \ and\ \bibinfo {author} {\bibfnamefont {Xin-Cheng}\ \bibnamefont
  {Xie}},\ }\bibfield  {title} {\enquote {\bibinfo {title} {Spin current as a
  probe of quantum materials},}\ }\href {\doibase 10.1038/s41563-019-0456-7}
  {\bibfield  {journal} {\bibinfo  {journal} {Nature Mater.}\ } (\bibinfo
  {year} {2019}),\ 10.1038/s41563-019-0456-7}\BibitemShut {NoStop}%
\bibitem [{\citenamefont {Callen}\ and\ \citenamefont
  {Welton}(1951)}]{callenPR51}%
  \BibitemOpen
  \bibfield  {author} {\bibinfo {author} {\bibfnamefont {Herbert~B.}\
  \bibnamefont {Callen}}\ and\ \bibinfo {author} {\bibfnamefont {Theodore~A.}\
  \bibnamefont {Welton}},\ }\bibfield  {title} {\enquote {\bibinfo {title}
  {Irreversibility and generalized noise},}\ }\href {\doibase
  10.1103/PhysRev.83.34} {\bibfield  {journal} {\bibinfo  {journal} {Phys.
  Rev.}\ }\textbf {\bibinfo {volume} {83}},\ \bibinfo {pages} {34--40}
  (\bibinfo {year} {1951})}\BibitemShut {NoStop}%
\bibitem [{\citenamefont {Kubo}(1966)}]{kuboRPP66}%
  \BibitemOpen
  \bibfield  {author} {\bibinfo {author} {\bibfnamefont {R}~\bibnamefont
  {Kubo}},\ }\bibfield  {title} {\enquote {\bibinfo {title} {The
  fluctuation-dissipation theorem},}\ }\href {\doibase
  10.1088/0034-4885/29/1/306} {\bibfield  {journal} {\bibinfo  {journal}
  {Reports on Progress in Physics}\ }\textbf {\bibinfo {volume} {29}},\
  \bibinfo {pages} {255--284} (\bibinfo {year} {1966})}\BibitemShut {NoStop}%
\bibitem [{\citenamefont {Kamra}\ and\ \citenamefont
  {Belzig}(2016)}]{kamraPRB16}%
  \BibitemOpen
  \bibfield  {author} {\bibinfo {author} {\bibfnamefont {Akashdeep}\
  \bibnamefont {Kamra}}\ and\ \bibinfo {author} {\bibfnamefont {Wolfgang}\
  \bibnamefont {Belzig}},\ }\bibfield  {title} {\enquote {\bibinfo {title}
  {Magnon-mediated spin current noise in ferromagnet $|$ nonmagnetic conductor
  hybrids},}\ }\href {\doibase 10.1103/PhysRevB.94.014419} {\bibfield
  {journal} {\bibinfo  {journal} {Phys. Rev. B}\ }\textbf {\bibinfo {volume}
  {94}},\ \bibinfo {pages} {014419} (\bibinfo {year} {2016})}\BibitemShut
  {NoStop}%
\bibitem [{\citenamefont {Clerk}\ \emph {et~al.}(2010)\citenamefont {Clerk},
  \citenamefont {Devoret}, \citenamefont {Girvin}, \citenamefont {Marquardt},\
  and\ \citenamefont {Schoelkopf}}]{clerkRMP10}%
  \BibitemOpen
  \bibfield  {author} {\bibinfo {author} {\bibfnamefont {A.~A.}\ \bibnamefont
  {Clerk}}, \bibinfo {author} {\bibfnamefont {M.~H.}\ \bibnamefont {Devoret}},
  \bibinfo {author} {\bibfnamefont {S.~M.}\ \bibnamefont {Girvin}}, \bibinfo
  {author} {\bibfnamefont {Florian}\ \bibnamefont {Marquardt}}, \ and\ \bibinfo
  {author} {\bibfnamefont {R.~J.}\ \bibnamefont {Schoelkopf}},\ }\bibfield
  {title} {\enquote {\bibinfo {title} {Introduction to quantum noise,
  measurement, and amplification},}\ }\href {\doibase
  10.1103/RevModPhys.82.1155} {\bibfield  {journal} {\bibinfo  {journal} {Rev.
  Mod. Phys.}\ }\textbf {\bibinfo {volume} {82}},\ \bibinfo {pages}
  {1155--1208} (\bibinfo {year} {2010})}\BibitemShut {NoStop}%
\bibitem [{\citenamefont {Florens}\ and\ \citenamefont
  {Georges}(2004)}]{florensPRB04}%
  \BibitemOpen
  \bibfield  {author} {\bibinfo {author} {\bibfnamefont {Serge}\ \bibnamefont
  {Florens}}\ and\ \bibinfo {author} {\bibfnamefont {Antoine}\ \bibnamefont
  {Georges}},\ }\bibfield  {title} {\enquote {\bibinfo {title} {Slave-rotor
  mean-field theories of strongly correlated systems and the mott transition in
  finite dimensions},}\ }\href {\doibase 10.1103/PhysRevB.70.035114} {\bibfield
   {journal} {\bibinfo  {journal} {Phys. Rev. B}\ }\textbf {\bibinfo {volume}
  {70}},\ \bibinfo {pages} {035114} (\bibinfo {year} {2004})}\BibitemShut
  {NoStop}%
\bibitem [{\citenamefont {Nasu}\ \emph {et~al.}(2015)\citenamefont {Nasu},
  \citenamefont {Udagawa},\ and\ \citenamefont {Motome}}]{nasuPRB15}%
  \BibitemOpen
  \bibfield  {author} {\bibinfo {author} {\bibfnamefont {Joji}\ \bibnamefont
  {Nasu}}, \bibinfo {author} {\bibfnamefont {Masafumi}\ \bibnamefont
  {Udagawa}}, \ and\ \bibinfo {author} {\bibfnamefont {Yukitoshi}\ \bibnamefont
  {Motome}},\ }\bibfield  {title} {\enquote {\bibinfo {title} {Thermal
  fractionalization of quantum spins in a kitaev model: Temperature-linear
  specific heat and coherent transport of majorana fermions},}\ }\href
  {\doibase 10.1103/PhysRevB.92.115122} {\bibfield  {journal} {\bibinfo
  {journal} {Phys. Rev. B}\ }\textbf {\bibinfo {volume} {92}},\ \bibinfo
  {pages} {115122} (\bibinfo {year} {2015})}\BibitemShut {NoStop}%
\bibitem [{\citenamefont {Rousochatzakis}\ \emph {et~al.}(2019)\citenamefont
  {Rousochatzakis}, \citenamefont {Kourtis}, \citenamefont {Knolle},
  \citenamefont {Moessner},\ and\ \citenamefont
  {Perkins}}]{rousochatzakisPRB19}%
  \BibitemOpen
  \bibfield  {author} {\bibinfo {author} {\bibfnamefont {I.}~\bibnamefont
  {Rousochatzakis}}, \bibinfo {author} {\bibfnamefont {S.}~\bibnamefont
  {Kourtis}}, \bibinfo {author} {\bibfnamefont {J.}~\bibnamefont {Knolle}},
  \bibinfo {author} {\bibfnamefont {R.}~\bibnamefont {Moessner}}, \ and\
  \bibinfo {author} {\bibfnamefont {N.~B.}\ \bibnamefont {Perkins}},\
  }\bibfield  {title} {\enquote {\bibinfo {title} {Quantum spin liquid at
  finite temperature: Proximate dynamics and persistent typicality},}\ }\href
  {\doibase 10.1103/PhysRevB.100.045117} {\bibfield  {journal} {\bibinfo
  {journal} {Phys. Rev. B}\ }\textbf {\bibinfo {volume} {100}},\ \bibinfo
  {pages} {045117} (\bibinfo {year} {2019})}\BibitemShut {NoStop}%
\bibitem [{\citenamefont {Hall}(1964)}]{hallNBS68}%
  \BibitemOpen
  \bibfield  {author} {\bibinfo {author} {\bibfnamefont {L.~A.}\ \bibnamefont
  {Hall}},\ }\href@noop {} {\emph {\bibinfo {title} {Survey of Electrical
  Resistivity Measurements on 16 Pure Metals in the Temperature Range 0 to
  273$^\circ$ K}}},\ \bibinfo {organization} {National Bureau of Standards}
  (\bibinfo {year} {1964})\BibitemShut {NoStop}%
\bibitem [{\citenamefont {Desai}\ \emph {et~al.}(1984)\citenamefont {Desai},
  \citenamefont {Chu}, \citenamefont {James},\ and\ \citenamefont
  {Ho}}]{desaiJPC84}%
  \BibitemOpen
  \bibfield  {author} {\bibinfo {author} {\bibfnamefont {P.~D.}\ \bibnamefont
  {Desai}}, \bibinfo {author} {\bibfnamefont {T.~K.}\ \bibnamefont {Chu}},
  \bibinfo {author} {\bibfnamefont {H.~M.}\ \bibnamefont {James}}, \ and\
  \bibinfo {author} {\bibfnamefont {C.~Y.}\ \bibnamefont {Ho}},\ }\bibfield
  {title} {\enquote {\bibinfo {title} {Electrical resistivity of selected
  elements},}\ }\href@noop {} {\bibfield  {journal} {\bibinfo  {journal} {J.
  Phys. Chem. Ref. Data}\ }\textbf {\bibinfo {volume} {13}},\ \bibinfo {pages}
  {1069--1096} (\bibinfo {year} {1984})}\BibitemShut {NoStop}%
\bibitem [{\citenamefont {Joshi}\ \emph {et~al.}(2018)\citenamefont {Joshi},
  \citenamefont {Schnyder},\ and\ \citenamefont {Takei}}]{joshiPRB18}%
  \BibitemOpen
  \bibfield  {author} {\bibinfo {author} {\bibfnamefont {Darshan~G.}\
  \bibnamefont {Joshi}}, \bibinfo {author} {\bibfnamefont {Andreas~P.}\
  \bibnamefont {Schnyder}}, \ and\ \bibinfo {author} {\bibfnamefont
  {So}~\bibnamefont {Takei}},\ }\bibfield  {title} {\enquote {\bibinfo {title}
  {Detecting end states of topological quantum paramagnets via spin hall noise
  spectroscopy},}\ }\href {\doibase 10.1103/PhysRevB.98.064401} {\bibfield
  {journal} {\bibinfo  {journal} {Phys. Rev. B}\ }\textbf {\bibinfo {volume}
  {98}},\ \bibinfo {pages} {064401} (\bibinfo {year} {2018})}\BibitemShut
  {NoStop}%
\bibitem [{\citenamefont {Kamra}\ \emph {et~al.}(2014)\citenamefont {Kamra},
  \citenamefont {Witek}, \citenamefont {Meyer}, \citenamefont {Huebl},
  \citenamefont {Gepr\"ags}, \citenamefont {Gross}, \citenamefont {Bauer},\
  and\ \citenamefont {Goennenwein}}]{kamraPRB14}%
  \BibitemOpen
  \bibfield  {author} {\bibinfo {author} {\bibfnamefont {Akashdeep}\
  \bibnamefont {Kamra}}, \bibinfo {author} {\bibfnamefont {Friedrich~P.}\
  \bibnamefont {Witek}}, \bibinfo {author} {\bibfnamefont {Sibylle}\
  \bibnamefont {Meyer}}, \bibinfo {author} {\bibfnamefont {Hans}\ \bibnamefont
  {Huebl}}, \bibinfo {author} {\bibfnamefont {Stephan}\ \bibnamefont
  {Gepr\"ags}}, \bibinfo {author} {\bibfnamefont {Rudolf}\ \bibnamefont
  {Gross}}, \bibinfo {author} {\bibfnamefont {Gerrit E.~W.}\ \bibnamefont
  {Bauer}}, \ and\ \bibinfo {author} {\bibfnamefont {Sebastian T.~B.}\
  \bibnamefont {Goennenwein}},\ }\bibfield  {title} {\enquote {\bibinfo {title}
  {Spin hall noise},}\ }\href {\doibase 10.1103/PhysRevB.90.214419} {\bibfield
  {journal} {\bibinfo  {journal} {Phys. Rev. B}\ }\textbf {\bibinfo {volume}
  {90}},\ \bibinfo {pages} {214419} (\bibinfo {year} {2014})}\BibitemShut
  {NoStop}%
\bibitem [{\citenamefont {Wang}\ and\ \citenamefont
  {Vishwanath}(2006)}]{wangPRB06}%
  \BibitemOpen
  \bibfield  {author} {\bibinfo {author} {\bibfnamefont {Fa}~\bibnamefont
  {Wang}}\ and\ \bibinfo {author} {\bibfnamefont {Ashvin}\ \bibnamefont
  {Vishwanath}},\ }\bibfield  {title} {\enquote {\bibinfo {title} {Spin-liquid
  states on the triangular and kagom\'e lattices: A projective-symmetry-group
  analysis of schwinger boson states},}\ }\href {\doibase
  10.1103/PhysRevB.74.174423} {\bibfield  {journal} {\bibinfo  {journal} {Phys.
  Rev. B}\ }\textbf {\bibinfo {volume} {74}},\ \bibinfo {pages} {174423}
  (\bibinfo {year} {2006})}\BibitemShut {NoStop}%
\bibitem [{\citenamefont {Punk}\ \emph {et~al.}(2014)\citenamefont {Punk},
  \citenamefont {Chowdhury},\ and\ \citenamefont {Sachdev}}]{punkNATP14}%
  \BibitemOpen
  \bibfield  {author} {\bibinfo {author} {\bibfnamefont {Matthias}\
  \bibnamefont {Punk}}, \bibinfo {author} {\bibfnamefont {Debanjan}\
  \bibnamefont {Chowdhury}}, \ and\ \bibinfo {author} {\bibfnamefont {Subir}\
  \bibnamefont {Sachdev}},\ }\bibfield  {title} {\enquote {\bibinfo {title}
  {Topological excitations and the dynamic structure factor of spin liquids on
  the kagome lattice},}\ }\href {https://doi.org/10.1038/nphys2887} {\bibfield
  {journal} {\bibinfo  {journal} {Nature Phys.}\ }\textbf {\bibinfo {volume}
  {10}},\ \bibinfo {pages} {289 EP --} (\bibinfo {year} {2014})}\BibitemShut
  {NoStop}%
\bibitem [{\citenamefont {Lu}\ \emph {et~al.}(2011)\citenamefont {Lu},
  \citenamefont {Ran},\ and\ \citenamefont {Lee}}]{luPRB11hc}%
  \BibitemOpen
  \bibfield  {author} {\bibinfo {author} {\bibfnamefont {Yuan-Ming}\
  \bibnamefont {Lu}}, \bibinfo {author} {\bibfnamefont {Ying}\ \bibnamefont
  {Ran}}, \ and\ \bibinfo {author} {\bibfnamefont {Patrick~A.}\ \bibnamefont
  {Lee}},\ }\bibfield  {title} {\enquote {\bibinfo {title} {${\mathbb{z}}_{2}$
  spin liquids in the $s=\frac{1}{2}$ heisenberg model on the kagome lattice: A
  projective symmetry-group study of schwinger fermion mean-field states},}\
  }\href {\doibase 10.1103/PhysRevB.83.224413} {\bibfield  {journal} {\bibinfo
  {journal} {Phys. Rev. B}\ }\textbf {\bibinfo {volume} {83}},\ \bibinfo
  {pages} {224413} (\bibinfo {year} {2011})}\BibitemShut {NoStop}%
\bibitem [{\citenamefont {Huh}\ \emph {et~al.}(2010)\citenamefont {Huh},
  \citenamefont {Fritz},\ and\ \citenamefont {Sachdev}}]{huhPRB10}%
  \BibitemOpen
  \bibfield  {author} {\bibinfo {author} {\bibfnamefont {Yejin}\ \bibnamefont
  {Huh}}, \bibinfo {author} {\bibfnamefont {Lars}\ \bibnamefont {Fritz}}, \
  and\ \bibinfo {author} {\bibfnamefont {Subir}\ \bibnamefont {Sachdev}},\
  }\bibfield  {title} {\enquote {\bibinfo {title} {Quantum criticality of the
  kagome antiferromagnet with dzyaloshinskii-moriya interactions},}\ }\href
  {\doibase 10.1103/PhysRevB.81.144432} {\bibfield  {journal} {\bibinfo
  {journal} {Phys. Rev. B}\ }\textbf {\bibinfo {volume} {81}},\ \bibinfo
  {pages} {144432} (\bibinfo {year} {2010})}\BibitemShut {NoStop}%
\bibitem [{\citenamefont {Chatterjee}\ and\ \citenamefont
  {Sachdev}(2015)}]{chatterjeePRB15}%
  \BibitemOpen
  \bibfield  {author} {\bibinfo {author} {\bibfnamefont {Shubhayu}\
  \bibnamefont {Chatterjee}}\ and\ \bibinfo {author} {\bibfnamefont {Subir}\
  \bibnamefont {Sachdev}},\ }\bibfield  {title} {\enquote {\bibinfo {title}
  {Probing excitations in insulators via injection of spin currents},}\ }\href
  {\doibase 10.1103/PhysRevB.92.165113} {\bibfield  {journal} {\bibinfo
  {journal} {Phys. Rev. B}\ }\textbf {\bibinfo {volume} {92}},\ \bibinfo
  {pages} {165113} (\bibinfo {year} {2015})}\BibitemShut {NoStop}%
\bibitem [{\citenamefont {Li}\ \emph {et~al.}(2016{\natexlab{b}})\citenamefont
  {Li}, \citenamefont {Wang},\ and\ \citenamefont {Chen}}]{liPRB16}%
  \BibitemOpen
  \bibfield  {author} {\bibinfo {author} {\bibfnamefont {Yao-Dong}\
  \bibnamefont {Li}}, \bibinfo {author} {\bibfnamefont {Xiaoqun}\ \bibnamefont
  {Wang}}, \ and\ \bibinfo {author} {\bibfnamefont {Gang}\ \bibnamefont
  {Chen}},\ }\bibfield  {title} {\enquote {\bibinfo {title} {Anisotropic spin
  model of strong spin-orbit-coupled triangular antiferromagnets},}\ }\href
  {\doibase 10.1103/PhysRevB.94.035107} {\bibfield  {journal} {\bibinfo
  {journal} {Phys. Rev. B}\ }\textbf {\bibinfo {volume} {94}},\ \bibinfo
  {pages} {035107} (\bibinfo {year} {2016}{\natexlab{b}})}\BibitemShut
  {NoStop}%
\bibitem [{\citenamefont {Liu}\ \emph {et~al.}(2016)\citenamefont {Liu},
  \citenamefont {Yu},\ and\ \citenamefont {Wang}}]{liuPRB16}%
  \BibitemOpen
  \bibfield  {author} {\bibinfo {author} {\bibfnamefont {Changle}\ \bibnamefont
  {Liu}}, \bibinfo {author} {\bibfnamefont {Rong}\ \bibnamefont {Yu}}, \ and\
  \bibinfo {author} {\bibfnamefont {Xiaoqun}\ \bibnamefont {Wang}},\ }\bibfield
   {title} {\enquote {\bibinfo {title} {Semiclassical ground-state phase
  diagram and $\text{multi-}q$ phase of a spin-orbit-coupled model on
  triangular lattice},}\ }\href {\doibase 10.1103/PhysRevB.94.174424}
  {\bibfield  {journal} {\bibinfo  {journal} {Phys. Rev. B}\ }\textbf {\bibinfo
  {volume} {94}},\ \bibinfo {pages} {174424} (\bibinfo {year}
  {2016})}\BibitemShut {NoStop}%
\bibitem [{\citenamefont {Li}\ \emph {et~al.}(2017)\citenamefont {Li},
  \citenamefont {Lu},\ and\ \citenamefont {Chen}}]{liPRB17}%
  \BibitemOpen
  \bibfield  {author} {\bibinfo {author} {\bibfnamefont {Yao-Dong}\
  \bibnamefont {Li}}, \bibinfo {author} {\bibfnamefont {Yuan-Ming}\
  \bibnamefont {Lu}}, \ and\ \bibinfo {author} {\bibfnamefont {Gang}\
  \bibnamefont {Chen}},\ }\bibfield  {title} {\enquote {\bibinfo {title}
  {Spinon fermi surface $u(1)$ spin liquid in the spin-orbit-coupled
  triangular-lattice mott insulator ${\mathrm{YbMgGaO}}_{4}$},}\ }\href
  {\doibase 10.1103/PhysRevB.96.054445} {\bibfield  {journal} {\bibinfo
  {journal} {Phys. Rev. B}\ }\textbf {\bibinfo {volume} {96}},\ \bibinfo
  {pages} {054445} (\bibinfo {year} {2017})}\BibitemShut {NoStop}%
\bibitem [{\citenamefont {Lee}\ and\ \citenamefont {Nagaosa}(1992)}]{leePRB92}%
  \BibitemOpen
  \bibfield  {author} {\bibinfo {author} {\bibfnamefont {Patrick~A.}\
  \bibnamefont {Lee}}\ and\ \bibinfo {author} {\bibfnamefont {Naoto}\
  \bibnamefont {Nagaosa}},\ }\bibfield  {title} {\enquote {\bibinfo {title}
  {Gauge theory of the normal state of high-${\mathit{t}}_{\mathit{c}}$
  superconductors},}\ }\href {\doibase 10.1103/PhysRevB.46.5621} {\bibfield
  {journal} {\bibinfo  {journal} {Phys. Rev. B}\ }\textbf {\bibinfo {volume}
  {46}},\ \bibinfo {pages} {5621--5639} (\bibinfo {year} {1992})}\BibitemShut
  {NoStop}%
\bibitem [{\citenamefont {Polchinski}(1994)}]{polchinskiNPB94}%
  \BibitemOpen
  \bibfield  {author} {\bibinfo {author} {\bibfnamefont {Joseph}\ \bibnamefont
  {Polchinski}},\ }\bibfield  {title} {\enquote {\bibinfo {title} {Low-energy
  dynamics of the spinon-gauge system},}\ }\href {\doibase
  https://doi.org/10.1016/0550-3213(94)90449-9} {\bibfield  {journal} {\bibinfo
   {journal} {Nuclear Physics B}\ }\textbf {\bibinfo {volume} {422}},\ \bibinfo
  {pages} {617 -- 633} (\bibinfo {year} {1994})}\BibitemShut {NoStop}%
\bibitem [{\citenamefont {Kim}\ \emph {et~al.}(1994)\citenamefont {Kim},
  \citenamefont {Furusaki}, \citenamefont {Wen},\ and\ \citenamefont
  {Lee}}]{kimPRB94}%
  \BibitemOpen
  \bibfield  {author} {\bibinfo {author} {\bibfnamefont {Yong~Baek}\
  \bibnamefont {Kim}}, \bibinfo {author} {\bibfnamefont {Akira}\ \bibnamefont
  {Furusaki}}, \bibinfo {author} {\bibfnamefont {Xiao-Gang}\ \bibnamefont
  {Wen}}, \ and\ \bibinfo {author} {\bibfnamefont {Patrick~A.}\ \bibnamefont
  {Lee}},\ }\bibfield  {title} {\enquote {\bibinfo {title} {Gauge-invariant
  response functions of fermions coupled to a gauge field},}\ }\href {\doibase
  10.1103/PhysRevB.50.17917} {\bibfield  {journal} {\bibinfo  {journal} {Phys.
  Rev. B}\ }\textbf {\bibinfo {volume} {50}},\ \bibinfo {pages} {17917--17932}
  (\bibinfo {year} {1994})}\BibitemShut {NoStop}%
\bibitem [{\citenamefont {{Balents}}\ and\ \citenamefont
  {{Starykh}}()}]{balentsCM19}%
  \BibitemOpen
  \bibfield  {author} {\bibinfo {author} {\bibfnamefont {Leon}\ \bibnamefont
  {{Balents}}}\ and\ \bibinfo {author} {\bibfnamefont {Oleg~A.}\ \bibnamefont
  {{Starykh}}},\ }\bibfield  {title} {\enquote {\bibinfo {title} {{Spinon waves
  in magnetized spin liquids}},}\ }\href@noop {} {\bibinfo  {journal}
  {arXiv:1904.02117}\ }\BibitemShut {NoStop}%
\bibitem [{\citenamefont {Knolle}\ \emph {et~al.}(2015)\citenamefont {Knolle},
  \citenamefont {Kovrizhin}, \citenamefont {Chalker},\ and\ \citenamefont
  {Moessner}}]{knollePRB15}%
  \BibitemOpen
\bibfield  {journal} {  }\bibfield  {author} {\bibinfo {author} {\bibfnamefont
  {J.}~\bibnamefont {Knolle}}, \bibinfo {author} {\bibfnamefont {D.~L.}\
  \bibnamefont {Kovrizhin}}, \bibinfo {author} {\bibfnamefont {J.~T.}\
  \bibnamefont {Chalker}}, \ and\ \bibinfo {author} {\bibfnamefont
  {R.}~\bibnamefont {Moessner}},\ }\bibfield  {title} {\enquote {\bibinfo
  {title} {Dynamics of fractionalization in quantum spin liquids},}\ }\href
  {\doibase 10.1103/PhysRevB.92.115127} {\bibfield  {journal} {\bibinfo
  {journal} {Phys. Rev. B}\ }\textbf {\bibinfo {volume} {92}},\ \bibinfo
  {pages} {115127} (\bibinfo {year} {2015})}\BibitemShut {NoStop}%
\bibitem [{\citenamefont {Hermanns}\ \emph {et~al.}(2018)\citenamefont
  {Hermanns}, \citenamefont {Kimchi},\ and\ \citenamefont
  {Knolle}}]{hermannsAR18}%
  \BibitemOpen
  \bibfield  {author} {\bibinfo {author} {\bibfnamefont {M.}~\bibnamefont
  {Hermanns}}, \bibinfo {author} {\bibfnamefont {I.}~\bibnamefont {Kimchi}}, \
  and\ \bibinfo {author} {\bibfnamefont {J.}~\bibnamefont {Knolle}},\
  }\bibfield  {title} {\enquote {\bibinfo {title} {Physics of the kitaev model:
  Fractionalization, dynamic correlations, and material connections},}\
  }\href@noop {} {\bibfield  {journal} {\bibinfo  {journal} {Annual Review of
  Condensed Matter Physics}\ }\textbf {\bibinfo {volume} {9}},\ \bibinfo
  {pages} {17--33} (\bibinfo {year} {2018})}\BibitemShut {NoStop}%
\bibitem [{\citenamefont {Knolle}\ \emph {et~al.}(2018)\citenamefont {Knolle},
  \citenamefont {Bhattacharjee},\ and\ \citenamefont {Moessner}}]{knollePRB18}%
  \BibitemOpen
  \bibfield  {author} {\bibinfo {author} {\bibfnamefont {Johannes}\
  \bibnamefont {Knolle}}, \bibinfo {author} {\bibfnamefont {Subhro}\
  \bibnamefont {Bhattacharjee}}, \ and\ \bibinfo {author} {\bibfnamefont
  {Roderich}\ \bibnamefont {Moessner}},\ }\bibfield  {title} {\enquote
  {\bibinfo {title} {Dynamics of a quantum spin liquid beyond integrability:
  The Kitaev-Heisenberg-$\mathrm{\ensuremath{\Gamma}}$ model in an augmented
  parton mean-field theory},}\ }\href {\doibase 10.1103/PhysRevB.97.134432}
  {\bibfield  {journal} {\bibinfo  {journal} {Phys. Rev. B}\ }\textbf {\bibinfo
  {volume} {97}},\ \bibinfo {pages} {134432} (\bibinfo {year}
  {2018})}\BibitemShut {NoStop}%
\bibitem [{\citenamefont {Lieb}(1994)}]{liebPRL94}%
  \BibitemOpen
  \bibfield  {author} {\bibinfo {author} {\bibfnamefont {Elliott~H.}\
  \bibnamefont {Lieb}},\ }\bibfield  {title} {\enquote {\bibinfo {title} {Flux
  phase of the half-filled band},}\ }\href {\doibase
  10.1103/PhysRevLett.73.2158} {\bibfield  {journal} {\bibinfo  {journal}
  {Phys. Rev. Lett.}\ }\textbf {\bibinfo {volume} {73}},\ \bibinfo {pages}
  {2158--2161} (\bibinfo {year} {1994})}\BibitemShut {NoStop}%
\bibitem [{\citenamefont {Baskaran}\ \emph {et~al.}(2007)\citenamefont
  {Baskaran}, \citenamefont {Mandal},\ and\ \citenamefont
  {Shankar}}]{baskaranPRL07}%
  \BibitemOpen
  \bibfield  {author} {\bibinfo {author} {\bibfnamefont {G.}~\bibnamefont
  {Baskaran}}, \bibinfo {author} {\bibfnamefont {Saptarshi}\ \bibnamefont
  {Mandal}}, \ and\ \bibinfo {author} {\bibfnamefont {R.}~\bibnamefont
  {Shankar}},\ }\bibfield  {title} {\enquote {\bibinfo {title} {Exact results
  for spin dynamics and fractionalization in the kitaev model},}\ }\href
  {\doibase 10.1103/PhysRevLett.98.247201} {\bibfield  {journal} {\bibinfo
  {journal} {Phys. Rev. Lett.}\ }\textbf {\bibinfo {volume} {98}},\ \bibinfo
  {pages} {247201} (\bibinfo {year} {2007})}\BibitemShut {NoStop}%
\bibitem [{\citenamefont {Han}\ \emph {et~al.}(2016)\citenamefont {Han},
  \citenamefont {Norman}, \citenamefont {Wen}, \citenamefont
  {Rodriguez-Rivera}, \citenamefont {Helton}, \citenamefont {Broholm},\ and\
  \citenamefont {Lee}}]{hanPRB16}%
  \BibitemOpen
  \bibfield  {author} {\bibinfo {author} {\bibfnamefont {Tian-Heng}\
  \bibnamefont {Han}}, \bibinfo {author} {\bibfnamefont {M.~R.}\ \bibnamefont
  {Norman}}, \bibinfo {author} {\bibfnamefont {J.-J.}\ \bibnamefont {Wen}},
  \bibinfo {author} {\bibfnamefont {Jose~A.}\ \bibnamefont {Rodriguez-Rivera}},
  \bibinfo {author} {\bibfnamefont {Joel~S.}\ \bibnamefont {Helton}}, \bibinfo
  {author} {\bibfnamefont {Collin}\ \bibnamefont {Broholm}}, \ and\ \bibinfo
  {author} {\bibfnamefont {Young~S.}\ \bibnamefont {Lee}},\ }\bibfield  {title}
  {\enquote {\bibinfo {title} {Correlated impurities and intrinsic spin-liquid
  physics in the kagome material herbertsmithite},}\ }\href {\doibase
  10.1103/PhysRevB.94.060409} {\bibfield  {journal} {\bibinfo  {journal} {Phys.
  Rev. B}\ }\textbf {\bibinfo {volume} {94}},\ \bibinfo {pages} {060409}
  (\bibinfo {year} {2016})}\BibitemShut {NoStop}%
\bibitem [{\citenamefont {Freedman}\ \emph {et~al.}(2010)\citenamefont
  {Freedman}, \citenamefont {Han}, \citenamefont {Prodi}, \citenamefont
  {M{\"u}ller}, \citenamefont {Huang}, \citenamefont {Chen}, \citenamefont
  {Webb}, \citenamefont {Lee}, \citenamefont {McQueen},\ and\ \citenamefont
  {Nocera}}]{freedmanJACS10}%
  \BibitemOpen
  \bibfield  {author} {\bibinfo {author} {\bibfnamefont {Danna~E.}\
  \bibnamefont {Freedman}}, \bibinfo {author} {\bibfnamefont {Tianheng~H.}\
  \bibnamefont {Han}}, \bibinfo {author} {\bibfnamefont {Andrea}\ \bibnamefont
  {Prodi}}, \bibinfo {author} {\bibfnamefont {Peter}\ \bibnamefont
  {M{\"u}ller}}, \bibinfo {author} {\bibfnamefont {Qing-Zhen}\ \bibnamefont
  {Huang}}, \bibinfo {author} {\bibfnamefont {Yu-Sheng}\ \bibnamefont {Chen}},
  \bibinfo {author} {\bibfnamefont {Samuel~M.}\ \bibnamefont {Webb}}, \bibinfo
  {author} {\bibfnamefont {Young~S.}\ \bibnamefont {Lee}}, \bibinfo {author}
  {\bibfnamefont {Tyrel~M.}\ \bibnamefont {McQueen}}, \ and\ \bibinfo {author}
  {\bibfnamefont {Daniel~G.}\ \bibnamefont {Nocera}},\ }\bibfield  {title}
  {\enquote {\bibinfo {title} {Site specific x-ray anomalous dispersion of the
  geometrically frustrated kagom{\'e}magnet, herbertsmithite, ZnCu$_3$(OH)$_6$Cl$_2$},}\
  }\bibfield  {booktitle} {\emph {\bibinfo {booktitle} {Journal of the American
  Chemical Society}},\ }\href {\doibase 10.1021/ja1070398} {\bibfield
  {journal} {\bibinfo  {journal} {Journal of the American Chemical Society}\
  }\textbf {\bibinfo {volume} {132}},\ \bibinfo {pages} {16185--16190}
  (\bibinfo {year} {2010})}\BibitemShut {NoStop}%
\bibitem [{\citenamefont {Kimchi}\ \emph {et~al.}(2018)\citenamefont {Kimchi},
  \citenamefont {Nahum},\ and\ \citenamefont {Senthil}}]{kimchiPRX18}%
  \BibitemOpen
  \bibfield  {author} {\bibinfo {author} {\bibfnamefont {Itamar}\ \bibnamefont
  {Kimchi}}, \bibinfo {author} {\bibfnamefont {Adam}\ \bibnamefont {Nahum}}, \
  and\ \bibinfo {author} {\bibfnamefont {T.}~\bibnamefont {Senthil}},\
  }\bibfield  {title} {\enquote {\bibinfo {title} {Valence bonds in random
  quantum magnets: Theory and application to ${\mathrm{ybmggao}}_{4}$},}\
  }\href {\doibase 10.1103/PhysRevX.8.031028} {\bibfield  {journal} {\bibinfo
  {journal} {Phys. Rev. X}\ }\textbf {\bibinfo {volume} {8}},\ \bibinfo {pages}
  {031028} (\bibinfo {year} {2018})}\BibitemShut {NoStop}%
\bibitem [{\citenamefont {Ding}\ \emph {et~al.}(2019)\citenamefont {Ding},
  \citenamefont {Manuel}, \citenamefont {Bachus}, \citenamefont {Gru\ss{}ler},
  \citenamefont {Gegenwart}, \citenamefont {Singleton}, \citenamefont
  {Johnson}, \citenamefont {Walker}, \citenamefont {Adroja}, \citenamefont
  {Hillier},\ and\ \citenamefont {Tsirlin}}]{leiPRB19}%
  \BibitemOpen
  \bibfield  {author} {\bibinfo {author} {\bibfnamefont {Lei}\ \bibnamefont
  {Ding}}, \bibinfo {author} {\bibfnamefont {Pascal}\ \bibnamefont {Manuel}},
  \bibinfo {author} {\bibfnamefont {Sebastian}\ \bibnamefont {Bachus}},
  \bibinfo {author} {\bibfnamefont {Franziska}\ \bibnamefont {Gru\ss{}ler}},
  \bibinfo {author} {\bibfnamefont {Philipp}\ \bibnamefont {Gegenwart}},
  \bibinfo {author} {\bibfnamefont {John}\ \bibnamefont {Singleton}}, \bibinfo
  {author} {\bibfnamefont {Roger~D.}\ \bibnamefont {Johnson}}, \bibinfo
  {author} {\bibfnamefont {Helen~C.}\ \bibnamefont {Walker}}, \bibinfo {author}
  {\bibfnamefont {Devashibhai~T.}\ \bibnamefont {Adroja}}, \bibinfo {author}
  {\bibfnamefont {Adrian~D.}\ \bibnamefont {Hillier}}, \ and\ \bibinfo {author}
  {\bibfnamefont {Alexander~A.}\ \bibnamefont {Tsirlin}},\ }\bibfield  {title}
  {\enquote {\bibinfo {title} {Gapless spin-liquid state in the structurally
  disorder-free triangular antiferromagnet ${\mathrm{NaYbO}}_{2}$},}\ }\href
  {\doibase 10.1103/PhysRevB.100.144432} {\bibfield  {journal} {\bibinfo
  {journal} {Phys. Rev. B}\ }\textbf {\bibinfo {volume} {100}},\ \bibinfo
  {pages} {144432} (\bibinfo {year} {2019})}\BibitemShut {NoStop}%
\bibitem [{\citenamefont {Chaloupka}\ \emph {et~al.}(2013)\citenamefont
  {Chaloupka}, \citenamefont {Jackeli},\ and\ \citenamefont
  {Khaliullin}}]{chaloupkaPRL13}%
  \BibitemOpen
  \bibfield  {author} {\bibinfo {author} {\bibfnamefont {Ji\ifmmode
  \check{r}\else~\v{r}\fi{}\'{\i}}\ \bibnamefont {Chaloupka}}, \bibinfo
  {author} {\bibfnamefont {George}\ \bibnamefont {Jackeli}}, \ and\ \bibinfo
  {author} {\bibfnamefont {Giniyat}\ \bibnamefont {Khaliullin}},\ }\bibfield
  {title} {\enquote {\bibinfo {title} {Zigzag magnetic order in the iridium
  oxide ${\mathrm{Na}}_{2}{\mathrm{IrO}}_{3}$},}\ }\href {\doibase
  10.1103/PhysRevLett.110.097204} {\bibfield  {journal} {\bibinfo  {journal}
  {Phys. Rev. Lett.}\ }\textbf {\bibinfo {volume} {110}},\ \bibinfo {pages}
  {097204} (\bibinfo {year} {2013})}\BibitemShut {NoStop}%
\bibitem [{\citenamefont {Plumb}\ \emph {et~al.}(2014)\citenamefont {Plumb},
  \citenamefont {Clancy}, \citenamefont {Sandilands}, \citenamefont {Shankar},
  \citenamefont {Hu}, \citenamefont {Burch}, \citenamefont {Kee},\ and\
  \citenamefont {Kim}}]{plumbPRB14}%
  \BibitemOpen
  \bibfield  {author} {\bibinfo {author} {\bibfnamefont {K.~W.}\ \bibnamefont
  {Plumb}}, \bibinfo {author} {\bibfnamefont {J.~P.}\ \bibnamefont {Clancy}},
  \bibinfo {author} {\bibfnamefont {L.~J.}\ \bibnamefont {Sandilands}},
  \bibinfo {author} {\bibfnamefont {V.~Vijay}\ \bibnamefont {Shankar}},
  \bibinfo {author} {\bibfnamefont {Y.~F.}\ \bibnamefont {Hu}}, \bibinfo
  {author} {\bibfnamefont {K.~S.}\ \bibnamefont {Burch}}, \bibinfo {author}
  {\bibfnamefont {Hae-Young}\ \bibnamefont {Kee}}, \ and\ \bibinfo {author}
  {\bibfnamefont {Young-June}\ \bibnamefont {Kim}},\ }\bibfield  {title}
  {\enquote {\bibinfo {title}
  {$\ensuremath{\alpha}\ensuremath{-}{\mathrm{RuCl}}_{3}$: A spin-orbit
  assisted mott insulator on a honeycomb lattice},}\ }\href {\doibase
  10.1103/PhysRevB.90.041112} {\bibfield  {journal} {\bibinfo  {journal} {Phys.
  Rev. B}\ }\textbf {\bibinfo {volume} {90}},\ \bibinfo {pages} {041112}
  (\bibinfo {year} {2014})}\BibitemShut {NoStop}%
\bibitem [{\citenamefont {Liu}\ \emph {et~al.}(2011)\citenamefont {Liu},
  \citenamefont {Berlijn}, \citenamefont {Yin}, \citenamefont {Ku},
  \citenamefont {Tsvelik}, \citenamefont {Kim}, \citenamefont {Gretarsson},
  \citenamefont {Singh}, \citenamefont {Gegenwart},\ and\ \citenamefont
  {Hill}}]{liuPRB11}%
  \BibitemOpen
  \bibfield  {author} {\bibinfo {author} {\bibfnamefont {X.}~\bibnamefont
  {Liu}}, \bibinfo {author} {\bibfnamefont {T.}~\bibnamefont {Berlijn}},
  \bibinfo {author} {\bibfnamefont {W.-G.}\ \bibnamefont {Yin}}, \bibinfo
  {author} {\bibfnamefont {W.}~\bibnamefont {Ku}}, \bibinfo {author}
  {\bibfnamefont {A.}~\bibnamefont {Tsvelik}}, \bibinfo {author} {\bibfnamefont
  {Young-June}\ \bibnamefont {Kim}}, \bibinfo {author} {\bibfnamefont
  {H.}~\bibnamefont {Gretarsson}}, \bibinfo {author} {\bibfnamefont {Yogesh}\
  \bibnamefont {Singh}}, \bibinfo {author} {\bibfnamefont {P.}~\bibnamefont
  {Gegenwart}}, \ and\ \bibinfo {author} {\bibfnamefont {J.~P.}\ \bibnamefont
  {Hill}},\ }\bibfield  {title} {\enquote {\bibinfo {title} {Long-range
  magnetic ordering in Na${}_{2}$IrO${}_{3}$},}\ }\href {\doibase
  10.1103/PhysRevB.83.220403} {\bibfield  {journal} {\bibinfo  {journal} {Phys.
  Rev. B}\ }\textbf {\bibinfo {volume} {83}},\ \bibinfo {pages} {220403}
  (\bibinfo {year} {2011})}\BibitemShut {NoStop}%
\bibitem [{\citenamefont {Choi}\ \emph {et~al.}(2012)\citenamefont {Choi},
  \citenamefont {Coldea}, \citenamefont {Kolmogorov}, \citenamefont
  {Lancaster}, \citenamefont {Mazin}, \citenamefont {Blundell}, \citenamefont
  {Radaelli}, \citenamefont {Singh}, \citenamefont {Gegenwart}, \citenamefont
  {Choi}, \citenamefont {Cheong}, \citenamefont {Baker}, \citenamefont
  {Stock},\ and\ \citenamefont {Taylor}}]{choiPRL12}%
  \BibitemOpen
  \bibfield  {author} {\bibinfo {author} {\bibfnamefont {S.~K.}\ \bibnamefont
  {Choi}}, \bibinfo {author} {\bibfnamefont {R.}~\bibnamefont {Coldea}},
  \bibinfo {author} {\bibfnamefont {A.~N.}\ \bibnamefont {Kolmogorov}},
  \bibinfo {author} {\bibfnamefont {T.}~\bibnamefont {Lancaster}}, \bibinfo
  {author} {\bibfnamefont {I.~I.}\ \bibnamefont {Mazin}}, \bibinfo {author}
  {\bibfnamefont {S.~J.}\ \bibnamefont {Blundell}}, \bibinfo {author}
  {\bibfnamefont {P.~G.}\ \bibnamefont {Radaelli}}, \bibinfo {author}
  {\bibfnamefont {Yogesh}\ \bibnamefont {Singh}}, \bibinfo {author}
  {\bibfnamefont {P.}~\bibnamefont {Gegenwart}}, \bibinfo {author}
  {\bibfnamefont {K.~R.}\ \bibnamefont {Choi}}, \bibinfo {author}
  {\bibfnamefont {S.-W.}\ \bibnamefont {Cheong}}, \bibinfo {author}
  {\bibfnamefont {P.~J.}\ \bibnamefont {Baker}}, \bibinfo {author}
  {\bibfnamefont {C.}~\bibnamefont {Stock}}, \ and\ \bibinfo {author}
  {\bibfnamefont {J.}~\bibnamefont {Taylor}},\ }\bibfield  {title} {\enquote
  {\bibinfo {title} {Spin waves and revised crystal structure of honeycomb
  iridate ${\mathrm{Na}}_{2}{\mathrm{IrO}}_{3}$},}\ }\href {\doibase
  10.1103/PhysRevLett.108.127204} {\bibfield  {journal} {\bibinfo  {journal}
  {Phys. Rev. Lett.}\ }\textbf {\bibinfo {volume} {108}},\ \bibinfo {pages}
  {127204} (\bibinfo {year} {2012})}\BibitemShut {NoStop}%
\bibitem [{\citenamefont {Kubota}\ \emph {et~al.}(2015)\citenamefont {Kubota},
  \citenamefont {Tanaka}, \citenamefont {Ono}, \citenamefont {Narumi},\ and\
  \citenamefont {Kindo}}]{kubotaPRB15}%
  \BibitemOpen
  \bibfield  {author} {\bibinfo {author} {\bibfnamefont {Yumi}\ \bibnamefont
  {Kubota}}, \bibinfo {author} {\bibfnamefont {Hidekazu}\ \bibnamefont
  {Tanaka}}, \bibinfo {author} {\bibfnamefont {Toshio}\ \bibnamefont {Ono}},
  \bibinfo {author} {\bibfnamefont {Yasuo}\ \bibnamefont {Narumi}}, \ and\
  \bibinfo {author} {\bibfnamefont {Koichi}\ \bibnamefont {Kindo}},\ }\bibfield
   {title} {\enquote {\bibinfo {title} {Successive magnetic phase transitions
  in $\ensuremath{\alpha}\ensuremath{-}{\mathrm{RuCl}}_{3}$: Xy-like frustrated
  magnet on the honeycomb lattice},}\ }\href {\doibase
  10.1103/PhysRevB.91.094422} {\bibfield  {journal} {\bibinfo  {journal} {Phys.
  Rev. B}\ }\textbf {\bibinfo {volume} {91}},\ \bibinfo {pages} {094422}
  (\bibinfo {year} {2015})}\BibitemShut {NoStop}%
\bibitem [{\citenamefont {Hwan~Chun}\ \emph {et~al.}(2015)\citenamefont
  {Hwan~Chun}, \citenamefont {Kim}, \citenamefont {Kim}, \citenamefont {Zheng},
  \citenamefont {Stoumpos}, \citenamefont {Malliakas}, \citenamefont
  {Mitchell}, \citenamefont {Mehlawat}, \citenamefont {Singh}, \citenamefont
  {Choi}, \citenamefont {Gog}, \citenamefont {Al-Zein}, \citenamefont {Sala},
  \citenamefont {Krisch}, \citenamefont {Chaloupka}, \citenamefont {Jackeli},
  \citenamefont {Khaliullin},\ and\ \citenamefont {Kim}}]{chunNAT15}%
  \BibitemOpen
  \bibfield  {author} {\bibinfo {author} {\bibfnamefont {Sae}\ \bibnamefont
  {Hwan~Chun}}, \bibinfo {author} {\bibfnamefont {Jong-Woo}\ \bibnamefont
  {Kim}}, \bibinfo {author} {\bibfnamefont {Jungho}\ \bibnamefont {Kim}},
  \bibinfo {author} {\bibfnamefont {H.}~\bibnamefont {Zheng}}, \bibinfo
  {author} {\bibfnamefont {Constantinos~C.}\ \bibnamefont {Stoumpos}}, \bibinfo
  {author} {\bibfnamefont {C.~D.}\ \bibnamefont {Malliakas}}, \bibinfo {author}
  {\bibfnamefont {J.~F.}\ \bibnamefont {Mitchell}}, \bibinfo {author}
  {\bibfnamefont {Kavita}\ \bibnamefont {Mehlawat}}, \bibinfo {author}
  {\bibfnamefont {Yogesh}\ \bibnamefont {Singh}}, \bibinfo {author}
  {\bibfnamefont {Y.}~\bibnamefont {Choi}}, \bibinfo {author} {\bibfnamefont
  {T.}~\bibnamefont {Gog}}, \bibinfo {author} {\bibfnamefont {A.}~\bibnamefont
  {Al-Zein}}, \bibinfo {author} {\bibfnamefont {M.~Moretti}\ \bibnamefont
  {Sala}}, \bibinfo {author} {\bibfnamefont {M.}~\bibnamefont {Krisch}},
  \bibinfo {author} {\bibfnamefont {J.}~\bibnamefont {Chaloupka}}, \bibinfo
  {author} {\bibfnamefont {G.}~\bibnamefont {Jackeli}}, \bibinfo {author}
  {\bibfnamefont {G.}~\bibnamefont {Khaliullin}}, \ and\ \bibinfo {author}
  {\bibfnamefont {B.~J.}\ \bibnamefont {Kim}},\ }\bibfield  {title} {\enquote
  {\bibinfo {title} {Direct evidence for dominant bond-directional interactions
  in a honeycomb lattice iridate Na$_2$IrO$_3$},}\ }\href
  {https://doi.org/10.1038/nphys3322} {\bibfield  {journal} {\bibinfo
  {journal} {Nature Phys.}\ }\textbf {\bibinfo {volume} {11}},\ \bibinfo
  {pages} {462 EP --} (\bibinfo {year} {2015})}\BibitemShut {NoStop}%
\bibitem [{\citenamefont {Majumder}\ \emph {et~al.}(2015)\citenamefont
  {Majumder}, \citenamefont {Schmidt}, \citenamefont {Rosner}, \citenamefont
  {Tsirlin}, \citenamefont {Yasuoka},\ and\ \citenamefont
  {Baenitz}}]{majumderPRB15}%
  \BibitemOpen
  \bibfield  {author} {\bibinfo {author} {\bibfnamefont {M.}~\bibnamefont
  {Majumder}}, \bibinfo {author} {\bibfnamefont {M.}~\bibnamefont {Schmidt}},
  \bibinfo {author} {\bibfnamefont {H.}~\bibnamefont {Rosner}}, \bibinfo
  {author} {\bibfnamefont {A.~A.}\ \bibnamefont {Tsirlin}}, \bibinfo {author}
  {\bibfnamefont {H.}~\bibnamefont {Yasuoka}}, \ and\ \bibinfo {author}
  {\bibfnamefont {M.}~\bibnamefont {Baenitz}},\ }\bibfield  {title} {\enquote
  {\bibinfo {title} {Anisotropic ${\mathrm{Ru}}^{3+} 4{d}^{5}$ magnetism in the
  $\ensuremath{\alpha}\ensuremath{-}{\mathrm{RuCl}}_{3}$ honeycomb system:
  Susceptibility, specific heat, and zero-field nmr},}\ }\href {\doibase
  10.1103/PhysRevB.91.180401} {\bibfield  {journal} {\bibinfo  {journal} {Phys.
  Rev. B}\ }\textbf {\bibinfo {volume} {91}},\ \bibinfo {pages} {180401}
  (\bibinfo {year} {2015})}\BibitemShut {NoStop}%
\bibitem [{\citenamefont {Winter}\ \emph {et~al.}(2017)\citenamefont {Winter},
  \citenamefont {Riedl}, \citenamefont {Maksimov}, \citenamefont {Chernyshev},
  \citenamefont {Honecker},\ and\ \citenamefont {Valent{\'\i}}}]{winterNATC17}%
  \BibitemOpen
  \bibfield  {author} {\bibinfo {author} {\bibfnamefont {Stephen~M.}\
  \bibnamefont {Winter}}, \bibinfo {author} {\bibfnamefont {Kira}\ \bibnamefont
  {Riedl}}, \bibinfo {author} {\bibfnamefont {Pavel~A.}\ \bibnamefont
  {Maksimov}}, \bibinfo {author} {\bibfnamefont {Alexander~L.}\ \bibnamefont
  {Chernyshev}}, \bibinfo {author} {\bibfnamefont {Andreas}\ \bibnamefont
  {Honecker}}, \ and\ \bibinfo {author} {\bibfnamefont {Roser}\ \bibnamefont
  {Valent{\'\i}}},\ }\bibfield  {title} {\enquote {\bibinfo {title} {Breakdown
  of magnons in a strongly spin-orbital coupled magnet},}\ }\href {\doibase
  10.1038/s41467-017-01177-0} {\bibfield  {journal} {\bibinfo  {journal}
  {Nature Commun.}\ }\textbf {\bibinfo {volume} {8}},\ \bibinfo {pages} {1152}
  (\bibinfo {year} {2017})}\BibitemShut {NoStop}%
\bibitem [{\citenamefont {Niimi}\ \emph {et~al.}(2012)\citenamefont {Niimi},
  \citenamefont {Kawanishi}, \citenamefont {Wei}, \citenamefont {Deranlot},
  \citenamefont {Yang}, \citenamefont {Chshiev}, \citenamefont {Valet},
  \citenamefont {Fert},\ and\ \citenamefont {Otani}}]{niimiPRL12}%
  \BibitemOpen
  \bibfield  {author} {\bibinfo {author} {\bibfnamefont {Y.}~\bibnamefont
  {Niimi}}, \bibinfo {author} {\bibfnamefont {Y.}~\bibnamefont {Kawanishi}},
  \bibinfo {author} {\bibfnamefont {D.~H.}\ \bibnamefont {Wei}}, \bibinfo
  {author} {\bibfnamefont {C.}~\bibnamefont {Deranlot}}, \bibinfo {author}
  {\bibfnamefont {H.~X.}\ \bibnamefont {Yang}}, \bibinfo {author}
  {\bibfnamefont {M.}~\bibnamefont {Chshiev}}, \bibinfo {author} {\bibfnamefont
  {T.}~\bibnamefont {Valet}}, \bibinfo {author} {\bibfnamefont
  {A.}~\bibnamefont {Fert}}, \ and\ \bibinfo {author} {\bibfnamefont
  {Y.}~\bibnamefont {Otani}},\ }\bibfield  {title} {\enquote {\bibinfo {title}
  {Giant spin hall effect induced by skew scattering from bismuth impurities
  inside thin film cubi alloys},}\ }\href {\doibase
  10.1103/PhysRevLett.109.156602} {\bibfield  {journal} {\bibinfo  {journal}
  {Phys. Rev. Lett.}\ }\textbf {\bibinfo {volume} {109}},\ \bibinfo {pages}
  {156602} (\bibinfo {year} {2012})}\BibitemShut {NoStop}%
\bibitem [{\citenamefont {Kamenev}(2011)}]{kamenevBOOK11}%
  \BibitemOpen
  \bibfield  {author} {\bibinfo {author} {\bibfnamefont {Alex}\ \bibnamefont
  {Kamenev}},\ }\href@noop {} {\emph {\bibinfo {title} {Field Theory of
  Non-Equilibrium Systems}}}\ (\bibinfo  {publisher} {Cambridge University
  Press},\ \bibinfo {address} {Cambridge},\ \bibinfo {year} {2011})\BibitemShut
  {NoStop}%
\end{thebibliography}
\end{document}